\documentclass[12pt,letterpaper]{article}

\usepackage{amsmath,amssymb,calc}
\usepackage{graphicx}
\usepackage{color}

\newcommand\be{\begin{equation}}
\newcommand\ee{\end{equation}}
\newcommand{\bea}{\begin{eqnarray}}
\newcommand{\eea}{\end{eqnarray}}
\newcommand{\ben}{\begin{equation}}
\newcommand{\een}{\end{equation}}
\newcommand{\beqa}{\begin{eqnarray*}}
\newcommand{\eeqa}{\end{eqnarray*}}
\newcommand{\beqan}{\begin{eqnarray}}
\newcommand{\eeqan}{\end{eqnarray}}

\newcommand{\nn}{\nonumber}
\newcommand{\pd}{\partial}

\def\id{\protect{{1 \kern-.28em {\rm l}}}}

\font\mybbb=msbm10 at 8pt
\font\mybb=msbm10 at 12pt

\def\bbb#1{\hbox{\mybbb#1}}
\def\bb#1{\hbox{\mybb#1}}

\def\C{\bb{C}}

\def\id{\protect{{1 \kern-.28em {\rm l}}}}

\def\dd{\mathrm{d}}
\def\cM{ {\cal M}}

\def\R{\mathbb{R}}
\def\rS{\mathrm{S}}

\def\mD{\mathbb{D}}
\newcommand{\mA}{{\mathbb{A}}}

\DeclareSymbolFont{extraup}{U}{zavm}{m}{n}
\DeclareMathSymbol{\vardiamond}{\mathalpha}{extraup}{87}

\begin{document}

\begin{titlepage}
\begin{center}
\hfill \\
\vspace{2cm}
{\Large {\bf Two-field Cosmological $\alpha$-attractors with Noether Symmetry
\\[3mm] }}

\vskip 1.5cm
{\bf Lilia Anguelova${}^a$\footnote{anguelova@inrne.bas.bg}, Elena Mirela Babalic${}^b$\footnote{mbabalic@theory.nipne.ro} and Calin Iuliu Lazaroiu${}^c$\footnote{calin@ibs.re.kr}\\
\vskip 0.5cm  {\it ${}^a$ Institute for Nuclear Research and Nuclear Energy, BAS, Sofia, Bulgaria\\ ${}^b$ Horia Hulubei National Institute for Physics and Nuclear Engineering (IFIN-HH), Bucharest-Magurele, Romania\\
${}^c$ Center for Geometry and Physics, Institute for Basic Science, Pohang 37673, Republic of
Korea}
}

\vskip 6mm

\end{center}

\vskip .1in
\vspace{0.8cm}

\begin{center} {\bf Abstract}\end{center}

\vspace{-1cm}

\begin{quotation}\noindent

We study Noether symmetries in two-field cosmological
$\alpha$-attractors, investigating the case when the scalar manifold
is an elementary hyperbolic surface. This encompasses and generalizes
the case of the Poincar\'e disk. We solve the conditions for the existence
of a `separated' Noether symmetry and find the form of the scalar
potential compatible with such, for any elementary hyperbolic
surface. For this class of symmetries, we find that the 
$\alpha$-parameter must have a fixed value. Using those Noether symmetries, 
we also obtain many exact solutions of the equations of motion of these 
models, which were studied previously with numerical methods.

\end{quotation}

\end{titlepage}

\eject

\tableofcontents

\section{Introduction}

A period of an accelerated expansion in the Early Universe is thought
to be necessary for explaining the large-scale properties of the
present day Universe. The standard description of such an inflationary
stage is given by coupling the space-time metric to one or more
fundamental scalars, which have a nontrivial potential that
temporarily dominates the energy density of the Universe. There is, in
fact, a wide variety of such inflationary models. A particular class,
called $\alpha$-attractors \cite{KLR1,KLR2} (see also the earlier
related works \cite{KL,KL2}), stands out as being in an especially
good agreement with the current observational data.

This class of models has certain universal predictions for the
important cosmological observables $n_s$ (scalar spectral index) and
$r$ (tensor-to-scalar ratio). It has been understood that the key
reason for this is a specific property of the kinetic terms of the
scalars. More precisely, they are characterized by hyperbolic geometry
\cite{KL_hyp_g1,KL_hyp_g2}. In fact, the original works on
$\alpha$-attractors focused mostly on effectively single field
models.\footnote{By `effectively' single-field models we mean
  two-field models on the Poincar\'e disk, in which however one studies only
  radial trajectories. The importance of the hyperbolic geometry of
  the scalar manifold is much more manifest in the recent works
  \cite{AKLWW, LWWYA, DFRSW, AB, MM, CRS, GSRPR}, 
  which investigated novel behavior due to trajectories with nontrivial
  angular motion on the Poincar\'e disk. Note that this kind of 
  trajectories had already been considered in a much wider context in the 
  earlier references \cite{LS,BL1,BL2}.} The widest generalization in the context of
two-field models, which brings into sharp focus the essential role of
the hyperbolic geometry of the scalar kinetic terms and of
uniformization theory, was introduced in \cite{LS} and further
explored in \cite{BL1,BL2} by considering models whose scalar
manifolds are arbitrary hyperbolic surfaces, which can be much more
complicated than the Poincar\'e disk.

Although single-field inflationary models are the most studied, it is
quite natural to consider models with more than one scalar field. The
reason is that the underlying particle physics descriptions, including
string compactifications, usually contain many scalars. So it makes
sense to expect, in the context of a fundamental theory of matter and
gravity, that more than one field would play an important role during
an inflationary stage. In view of very recent developments in the
literature, there may also be another motivation to be interested in
multi-field cosmological models. Namely, it was conjectured in
\cite{OOSV} that quantum gravity requires the scalar potential to
satisfy a certain condition, which excludes dS minima and seems to be
in severe tension with single-field slow-roll inflationary models
\cite{GK,KVV}. It was argued in \cite{AP} that one can reconcile slow-roll
inflation with the conjecture of \cite{OOSV} by considering
multi-field models. One should note, however, that there are already
serious objections \cite{DHW,CdAMMQ,KT,AKLV,MYY} to that conjecture,
whose only motivation is that it is rather difficult to find
well-under-control stringy constructions that have (meta-)stable dS
minima.\footnote{The main conceptual objection can be summarized as
  follows. Effective field theory considerations clearly indicate the
  necessity to include quantum (in particular, non-perturbative)
  effects in order to obtain dS minima, while those string theory
  dS-related considerations which are sufficiently rigorous at present
  are essentially classical (relying on nontrivial background
  fluxes). So there should be no surprise at the difficulty, which can
  likely be overcome only upon developing a better non-perturbative
  understanding of string theory.} It could be helpful, in sorting out
arguments for or against the conjecture, to better understand
multi-field inflationary models and their embeddings in string
compactifications. Regardless of whether one is motivated by the
conjecture of \cite{OOSV} or by the general expectation that more than
one scalar field could play an important role for inflation, it is
natural to be interested in two-field models as the simplest case of
multi-field ones.

Most of the time, the equations of motion of two-field cosmological
models are solved numerically in the literature. See, in particular,
\cite{BL1,BL2,BL3} for such numerical investigations in two-field
$\alpha$-attractor models. Our goal here will be to find exact
solutions by imposing the requirement that the model possesses a
Noether symmetry. This method is well-known in the context of extended
theories of gravity, where it has long been used to find classes of
exact solutions \cite{CdeR,CMRS,CNP,EGM}. The basic idea is that the
presence of a Noether symmetry constrains the form of an otherwise
arbitrary function in the action (in our context, the scalar
potential) and allows one to simplify the equations of motion. In
general, this method does not give all solutions of the field
equations, but only a certain subset. However, having exact solutions
to analyze is often more informative conceptually than performing
numerical analysis. Furthermore, the relevant Noether symmetry may
have a deeper meaning, if the two-field models under consideration
could be embedded in some fundamental particle physics setup, like a
class of string theory compactifications.

The Noether symmetry method was already applied to one-field
$\alpha$-attractor models of inflation in \cite{KKC}. However, due to
the limitation to a single scalar field, that analysis could not
illustrate the essential role played by the hyperbolic geometry of the
scalar manifold. Here we will apply the Noether symmetry method to the
two-field generalized $\alpha$-attractors of \cite{LS,BL1,BL2}. A key
feature of this class of models is that the scalar manifold is a
hyperbolic surface. For a Riemannian 2-manifold, hyperbolicity amounts
to the condition that the Gaussian curvature is constant and
negative. In fact, it is inversely proportional to the
$\alpha$-parameter of these models. We will focus on the simplest
class of hyperbolic surfaces, called elementary, of which there are
three types: the Poincar\'e disk, the hyperbolic punctured disk and
the hyperbolic annuli (see, for example, \cite{BL1}). Using a
separation-of-variables Ansatz, we show that two-field
$\alpha$-attractor models, with scalar manifold given by {\it any}
elementary hyperbolic surface, have a `separated' Noether symmetry for
a certain form of the scalar potential. The existence of such a
symmetry requires a different form of the scalar potential for each of
the three types of elementary hyperbolic surface. The hyperbolic
geometry of the scalar kinetic terms will play an essential role in
this derivation.

It turns out that the special kind of Noether symmetry, which we find
using the separation of variables Ansatz, not only selects a
particular form of the scalar potential, but also fixes the value of
the otherwise arbitrary $\alpha$-parameter\footnote{This condition may
  be relaxed for more general Noether symmetries, which are not of the
  separation-of-variables type. We hope to say more on this in a
  future publication.}. That a specific value of the
$\alpha$-parameter is required for a separated Noether symmetry may
seem unexpected. However, it is also very intriguing. Recall that it
is not uncommon, especially in the context of string theory, to have
particular points in a certain parameter space, where an (enhanced)
symmetry occurs, although there is no such symmetry at generic points
of that parameter space. It would be very interesting to understand
whether this peculiar feature can help find specific embeddings of
two-field $\alpha$-attractor models with a separated Noether symmetry
in a more fundamental particle physics framework.

We also find many exact solutions of the equations of motion of
two-field $\alpha$-attractor models, which admit a separated Noether
symmetry. To achieve this, we transform the relevant Lagrangian to a
new system of generalized coordinates, which is adapted to the Noether
symmetry. We investigate each of the elementary hyperbolic surfaces in
detail and find a variety of exact solutions of the field equations in
each case.

The organization of the present paper is the following. In Section
\ref{Two_Field_Models}, we briefly review the action for the class of
cosmological models known as generalized two-field
$\alpha$-attractors. The two-dimensional scalar manifold of those
models is a hyperbolic surface. We write down the action for each
elementary hyperbolic surface, namely the Poincar\'e disk, the
hyperbolic punctured disk and the hyperbolic annuli. In Section
\ref{NoetherSym}, we write the cosmologically relevant point-particle
Lagrangian (the so-called `minisuperspace Lagrangian') and impose the
condition that it has a Noether symmetry. This leads to a coupled
system of seven PDEs. Using a separation-of-variables Ansatz, we find
solutions of that system for each elementary hyperbolic surface, in
particular determining the form of the scalar potential which is
compatible with the separated Noether symmetry. In Section
\ref{Sec:NewVar}, we find new generalized coordinates that are adapted
to this Noether symmetry. In Sections \ref{Sec:Disk},
\ref{Sec:PuncDisk} and \ref{Sec:Annulus}, we investigate the equations
of motion of the two-field $\alpha$-attractor Lagrangian in the new
coordinate system for the Poincar\'e disk, hyperbolic punctured disk
and hyperbolic annuli respectively. We find many exact solutions in
each of the three cases. Section \ref{Sec:Discussion} summarizes our
results and briefly mentions some directions for further
research. Appendix \ref{ElHypSs} recalls the basic definitions and
properties of elementary hyperbolic surfaces (whose geometry is
described in detail in reference \cite{BL1}). Appendix \ref{App:Traj_m0} 
illustrates some of the new exact solutions.

\section{Two-field cosmological $\alpha$-attractor models} \label{Two_Field_Models}
\setcounter{equation}{0}

Generalized two-field $\alpha$-attractors are a class of inflationary models obtained from Einstein gravity coupled to a non-linear sigma-model with two real scalar fields, whose target space (known as the {\em scalar manifold}) is a hyperbolic surface. This system is described by the action
\be \label{Action}
S = \int d^4 x \,\sqrt{- g} \left[ \,\frac{R}{2} - \frac{1}{2} \,G_{IJ} (\phi) \,\pd \phi^I \pd \phi^J - V ( \phi) \,\right] \, ,
\ee
where $R$ is the scalar curvature of the 4d space-time metric $g_{\mu \nu}$, the fields $\phi^I$ with $I=1,2$ are two real scalars and the non-linear sigma-model metric $G_{IJ} (\phi)$ is a complete hyperbolic metric, i.e. a complete metric of constant negative Gaussian curvature $K$ $\!${}\footnote{It is convenient to write the Gaussian curvature as $K = -\frac{const}{\alpha}$ in terms of an arbitrary positive parameter $\alpha$. (There are differing conventions in the literature, namely: either $K = - \frac{1}{3 \alpha}$ \,, \,$K = - \frac{2}{3 \alpha}$ or $K = - \frac{1}{2 \alpha}$.) It was shown in \cite{LS} that such models have universality properties similar to those of \cite{KLR1, KLR2}, hence the name `$\alpha$-attractors'.}. For brevity, we use the notation $\pd \phi^I \pd \phi^J \equiv g^{\mu \nu} \pd_{\mu} \phi^I \pd_{\nu} \phi^J$.

The simplest example is obtained by taking the scalar manifold to be the Poincar\'e disk $\mD$. In this case, using polar coordinates on $\mD$ and considering only radial trajectories, one recovers the original one-field $\alpha$-attractors of \cite{KLR1,KLR2}. It was understood in \cite{KL_hyp_g1,KL_hyp_g2} that the universal properties of the latter arise from the hyperbolic geometry of the Poincar\'e disk. Later, reference \cite{LS} considered a very wide generalization of the Poincar\'e disk models, obtained by taking the scalar manifold to be an arbitrary hyperbolic surface and showed that the universal properties of the original one-field $\alpha$-attractors persist under certain conditions. Specific examples of generalized two-field $\alpha$-attractors were explored in more detail in \cite{BL1,BL2,BL3}. In particular, \cite{BL1} studied $\alpha$-attractors whose scalar manifold is an elementary hyperbolic surface, i.e. the Poincar\'e disk, the punctured hyperbolic disk or a hyperbolic annulus. We briefly review their definitions and properties in Appendix \ref{ElHypSs}. 

Our goal here will be to show that, for each of the elementary hyperbolic surfaces, the cosmological model obtained from the action (\ref{Action}) possesses a Noether symmetry for a certain value of the parameter $\alpha$ and a particular form of the scalar potential $V (\phi)$. To achieve this goal, it will be useful to rewrite (\ref{Action}) in the form:
\be \label{S_alpha}
S = \int \!d^4x \,\sqrt{-g} \left[ \frac{R}{2} - \frac{1}{2} (\pd \varphi)^2 - \frac{f(\varphi)}{2} (\pd \theta)^2 - V (\varphi,\theta) \right] \, ,
\ee
where now the two real scalars are $\varphi$ and $\theta$ and all the information about the hyperbolic geometry of the sigma-model metric is contained in the function $f(\varphi)$. Such a rewriting can be achieved for any metric $G_{IJ}$, which admits a $U(1)$ isometry parameterized by $\theta$ and so, in particular, for any of the elementary hyperbolic surfaces. Namely:

\vspace{0.4cm}
\noindent
$\bullet$ \hspace*{0.03cm}{\bf Poincar\'e disk:}

\vspace{0.2cm}
\noindent
When $G_{IJ}$ is the metric on the hyperbolic disk $\bb{D}$, the action (\ref{Action}) can be written as:
\be \label{S_Pd}
S_{\bbb{D}} = \int \!d^4x \,\sqrt{-g} \left[ \frac{R}{2} - 3 \alpha \frac{\pd Z \pd \bar{Z}}{(1-Z \bar{Z})^2} - V (Z) \right] \, ,
\ee
in terms of a complex scalar $Z = \phi^1 + i \phi^2$. Writing the latter as:
\be
Z = \rho e^{i \theta}
\ee
and performing the field redefinition:
\be \label{rho_redef}
\rho = \tanh \!\left( \frac{\varphi}{\sqrt{6 \alpha}} \right) \, ,
\ee
we find that (\ref{S_Pd}) acquires the form (\ref{S_alpha}) with the following function $f(\varphi)$:
\be \label{fD}
f_{\bbb{D}} (\varphi) = \frac{3 \alpha}{2} \sinh^2 \!\left( \sqrt{\frac{2}{3 \alpha}} \,\varphi \right) \,\,\, .
\ee

\vspace{0.4cm}
\noindent
$\bullet$ \hspace*{0.03cm}{\bf Hyperbolic punctured disk:}

\vspace{0.2cm}
\noindent
For $G_{IJ}$ the metric on the hyperbolic punctured disk $\mD^*$, the action (\ref{Action}) can be written as:
\be \label{S_Punc_d}
S_{\bbb{D}^*} = \int \!d^4x \,\sqrt{-g} \left[ \frac{R}{2} - \frac{\alpha}{( \rho \ln \rho )^2} \left\{ (\pd \rho )^2 + \rho^2 (\pd \theta )^2 \right\} - V (\rho,\theta) \right] \, .
\ee
Hence, the field redefinition:
\be \label{rho_redef_D*}
\varphi = \sqrt{2\alpha} \,\ln (|\ln \!\rho|)
\ee
transforms it into (\ref{S_alpha}), where now the function $f(\varphi)$ is:
\be \label{fD*}
f_{\bbb{D}^*} (\varphi) = 2 \alpha \,\exp \!\left( - \sqrt{\frac{2}{\alpha}} \,\varphi \right) \,\,\, .
\ee


\vspace{0.4cm}
\noindent
$\bullet$ \hspace*{0.03cm}{\bf Hyperbolic annulus:}

\vspace{0.2cm}
\noindent
When $G_{IJ}$ is the metric on a hyperbolic annulus $\mA$, (\ref{Action}) acquires the form:
\be \label{S_Annuli}
S_{\bbb{A}} = \int \!d^4x \,\sqrt{-g} \left[ \frac{R}{2} - \frac{\alpha C_R^2}{[ \rho \cos ( C_R \ln \!\rho ) ]^2} \left\{ (\pd \rho )^2 + \rho^2 (\pd \theta )^2 \right\} - V (\rho,\theta) \right] \, ,
\ee
where $C_R \equiv \frac{\pi}{2 \ln \!\hat{R}}$. This can be transformed to the expression in (\ref{S_alpha}) by the redefinition:
\be \label{rho_redef_A}
\varphi = \sqrt{2\alpha} \,\ln \!\left[ \frac{1+ \sin (C_R \ln \!\rho \,)}{\cos (C_R \ln \!\rho \,)} \right] \, ,
\ee
which leads to the following function $f (\varphi)$:
\be \label{fA}
f_{\bbb{A}} (\varphi) = 2 \alpha C_R^2 \,\cosh^2 \!\left( \frac{\varphi}{\sqrt{2 \alpha}} \right) \,\,\, .
\ee

\section{Noether symmetries in two-field $\alpha$-attractors} \label{NoetherSym}
\setcounter{equation}{0}

We now investigate under what conditions the action (\ref{S_alpha}), namely:
\be \label{S_alpha_p}
S = \int \!d^4x \,\sqrt{-g} \left[ \frac{R}{2} - \frac{1}{2} (\pd \varphi)^2 - \frac{f(\varphi)}{2} (\pd \theta)^2 - V (\varphi,\theta) \right] \, ,
\ee
has a Noether symmetry. As usual, we will consider the following Ansatz for the four-dimensional inflationary metric:
\be \label{Infl_M}
ds_4^2 = - dt^2 + a^2(t) d \vec{x}^2 \, ,
\ee 
as well as spatially-homogeneous scalar fields $\varphi (x^{\mu}) = \varphi (t)$ and $\theta (x^{\mu}) = \theta (t)$. Substituting these in (\ref{S_alpha_p}), we obtain:
\be \label{S_ddot_a}
S = \int \!d^4x \, a^3 \!\left[ \frac{3 (\dot{a}^2 + a \ddot{a})}{a^2} + \frac{\dot{\varphi}^2}{2} + \frac{f(\varphi)}{2} \dot{\theta}^2 - V (\varphi, \theta) \right] \, .
\ee
Note that, since here $a$, $\varphi$ and $\theta$ depend only on time, the action per unit spatial volume in (\ref{S_ddot_a}) can be viewed as the classical action of a mechanical system with three degrees of freedom. 

To use the Noether method, we have to rewrite the Lagrangian in (\ref{S_ddot_a}) in canonical form, namely as ${\cal L} (q^i,\dot{q}^i)$ in terms of some generalized configuration space coordinates $q^i$ and the corresponding generalized velocities $\dot{q}^i$. To achieve this, we use integration by parts in the $\ddot{a}$ term in (\ref{S_ddot_a}). This allows us to write the action per unit spatial volume in (\ref{S_ddot_a}) as $\int \!d t \,{\cal L}$\,, with the following Lagrangian density:
\be \label{Lagr_alpha}
{\cal L} = - 3 a \dot{a}^2 + \frac{a^3 \dot{\varphi}^2}{2} + \frac{a^3 f(\varphi) \,\dot{\theta}^2}{2} - a^3 V(\varphi, \theta) \,\,\, .
\ee
In this point-like Lagrangian, we can view $\{a,\varphi,\theta\}$ as generalized coordinates on the configuration space $\cM=\R^2\times \rS^1$. Then $\{a,\dot{a},\varphi,\dot{\varphi},\theta,\dot{\theta}\}$ provide coordinates on the corresponding tangent bundle $T\cM$. Let us now write down the conditions for (\ref{Lagr_alpha}) to have a Noether symmetry.

\subsection{The Noether system}

Recall that a symmetry generator is a vector field $X$ defined on $T\cM$, which preserves the Lagrangian:
\be \label{Lie_der}
L_X {\cal L} = 0~~,
\ee
where $L_X$ is the Lie derivative along $X$. In fact, to generate a Noether symmetry of ${\cal L}$, the vector field $X$ has to be of the specific form:
\be 
\label{sym_gen}
X = \lambda_a \frac{\pd}{\pd a} + \dot{\lambda}_a \frac{\pd}{\pd \dot{a}} + \lambda_{\varphi} \frac{\pd}{\pd \varphi} + \dot{\lambda}_{\varphi} \frac{\pd}{\pd \dot{\varphi}} + \lambda_{\theta} \frac{\pd}{\pd \theta} + \dot{\lambda}_{\theta} \frac{\pd}{\pd \dot{\theta}} \,\,\,\, ,
\ee
where the coefficients $\lambda_{a,\varphi,\theta}$ are functions of the configuration space coordinates $\{a,\varphi,\theta\}$. Hence, the condition (\ref{Lie_der}) becomes:
\be \label{SymCond}
\lambda_a \frac{\pd {\cal L}}{\pd a} + \dot{\lambda}_a \frac{\pd {\cal L}}{\pd \dot{a}} + \lambda_{\varphi} \frac{\pd {\cal L}}{\pd \varphi} + \dot{\lambda}_{\varphi} \frac{\pd {\cal L}}{\pd \dot{\varphi}} + \lambda_{\theta} \frac{\pd {\cal L}}{\pd \theta} + \dot{\lambda}_{\theta} \frac{\pd {\cal L}}{\pd \dot{\theta}} = 0 \,\,\, .
\ee

Let us now investigate the implications of this condition for the Lagrangian (\ref{Lagr_alpha}). First, note that all terms in (\ref{SymCond}) are either quadratic in the generalized velocities $\dot{a}$, $\dot{\varphi}$ and $\dot{\theta}$ or contain no velocity at all. So we can view the left-hand side of (\ref{SymCond}) as a second degree polynomial in the generalized velocities. Since we want to find functions $\lambda_{a,\varphi,\theta} (a,\varphi,\theta)$\,, for which the symmetry condition (\ref{SymCond}) is satisfied identically, we have to require that each coefficient of this polynomial vanishes separately. Therefore, computing the various terms in (\ref{SymCond}) for the Lagrangian (\ref{Lagr_alpha}), we find the following coupled system (where in brackets we indicate the corresponding coefficient of the velocity polynomial):
\begin{align} \label{Syst_alpha}
&{\rm (E1)} \qquad &&(\mathrm{coeff.~of}~\dot{a}^2) \hspace*{-0.6cm}&&:\qquad \hspace*{0.2cm}&&\lambda_a + 2 a \frac{\pd \lambda_a}{\pd a} = 0 \,\, , \nn \\
&{\rm (E2)} \qquad &&(\mathrm{coeff.~of}~\dot{\varphi}^2) \hspace*{-0.6cm}&&:\qquad \hspace*{0.2cm}&&\frac{3}{2} \lambda_a + a \frac{\pd \lambda_{\varphi}}{\pd \varphi} = 0 \,\, , \nn \\
&{\rm (E3)} \qquad &&(\mathrm{coeff.~of}~\dot{\theta}^2) \hspace*{-0.6cm}&&:\qquad \hspace*{0.2cm}&&\frac{3}{2} f(\varphi) \lambda_a + \frac{a}{2} (\pd_{\varphi} f) \lambda_{\varphi} + a \,f(\varphi) \frac{\pd \lambda_{\theta}}{\pd \theta} = 0 \,\, , \nn \\
&{\rm (E4)} \qquad &&(\mathrm{coeff.~of}~\dot{a}\dot{\varphi}) \hspace*{-0.6cm}&&:\qquad \hspace*{0.2cm}&&- 6 \frac{\pd \lambda_a}{\pd \varphi} + a^2 \frac{\pd \lambda_{\varphi}}{\pd a} = 0 \,\, , \nn \\
&{\rm (E5)} \qquad &&(\mathrm{coeff.~of}~\dot{a}\dot{\theta}) \hspace*{-0.6cm}&&:\qquad \hspace*{0.2cm}&&- 6 \frac{\pd \lambda_a}{\pd \theta} + a^2 f(\varphi) \frac{\pd \lambda_{\theta}}{\pd a} = 0 \,\, , \nn \\
&{\rm (E6)} \qquad &&(\mathrm{coeff.~of}~\dot{\varphi}\dot{\theta}) \hspace*{-0.6cm}&&:\qquad \hspace*{0.2cm}&&\frac{\pd \lambda_{\varphi}}{\pd \theta} + f(\varphi) \frac{\pd \lambda_{\theta}}{\pd \varphi} = 0 \,\, , \nn \\
&{\rm (E7)} \quad &&({\rm ind.\,\,of\,\,velocity}) \hspace*{-0.6cm}&&:\qquad \hspace*{0.2cm}&&3 V \lambda_a + a V_{\varphi} \lambda_{\varphi} + a V_{\theta} \lambda_{\theta} = 0 \,\, .
\end{align}

In the next subsections, we will show that equations (E1)-(E6) can be solved for any function $f(\varphi)$, such that the scalar manifold metric in (\ref{S_alpha_p}) is hyperbolic, i.e. with a constant negative Gaussian curvature. Then, equation (E7) determines a particular form of the scalar potential. As in \cite{KKC}, we will look for solutions with the following separation-of-variables Ansatze:
\bea \label{Lambda_ans}
\lambda_a (a, \varphi, \theta) &=& A_1 (a) \Phi_1 (\varphi) \Theta_1 (\theta) \quad , \nn \\
\lambda_{\varphi} (a, \varphi, \theta) &=& A_2 (a) \Phi_2 (\varphi) \Theta_2 (\theta) \quad , \nn \\
\lambda_{\theta} (a, \varphi, \theta) &=& A_3 (a) \Phi_3 (\varphi) \Theta_3 (\theta) \quad .
\eea
Let us begin by considering equations (E1), (E2) and (E4), which do not depend on $f(\varphi)$ and hence have the same form for any elementary hyperbolic surface.

\subsection{Solving equations (E1), (E2) and (E4)}

Substituting (\ref{Lambda_ans}) in equation (E1), we obtain the following first order ODE:
\be
A_1 (a) + 2 a \frac{d A_1}{d a} = 0 \,\, .
\ee
Its general solution is:
\be \label{A1}
A_1 (a) = \frac{A}{\sqrt{a}} \,\, ,
\ee
where $A$ is an arbitrary integration constant.

Equating the expressions for $\lambda_a$ obtained from (E1) and (E2) in (\ref{Syst_alpha}), we have:
\be \label{Eq1Eq2}
\frac{\pd \lambda_a}{\pd a} = \frac{1}{3} \frac{\pd \lambda_{\varphi}}{\pd \varphi} \,\, .
\ee
Substituting (\ref{Lambda_ans}) in (\ref{Eq1Eq2}), we find the following set of equations:\footnote{For convenience, as well as for easier comparison with \cite{KKC}, we have assigned the $\frac{1}{3}$ coefficient in (\ref{Eq1Eq2}) to the first equation in (\ref{ThreeEqs}).}
\be \label{ThreeEqs}
\frac{d A_1}{d a} = \frac{A_2 (a)}{3} \quad , \quad \Phi_1 (\varphi) = k \frac{d \Phi_2}{d \varphi} \quad , \quad \Theta_1 (\theta) = \frac{1}{k} \Theta_2 (\theta) \,\, ,
\ee
where $k = const$. Using (\ref{A1}) in the first equation of (\ref{ThreeEqs}) gives:
\be \label{A2}
A_2 (a) = - \frac{3}{2} \frac{A}{a^{3/2}} \,\, .
\ee

Let us now consider equation (E4). Substituting (\ref{Lambda_ans}), (\ref{A1}) and (\ref{A2}) in this equation gives:
\be \label{Eq_phi_theta}
-8 \frac{d \Phi_1}{d \varphi} \Theta_1 (\theta) + 3 \Phi_2 (\varphi) \Theta_2 (\theta) = 0 \,\, .
\ee
This, together with the last relation in (\ref{ThreeEqs}), implies that:
\be \label{Phi2}
\Phi_2 = \frac{8}{3 k} \frac{d \Phi_1}{d \varphi}
\ee
Using the second equation of (\ref{ThreeEqs}) in (\ref{Phi2}), we end up with the following ODE:
\be
\frac{d^2 \Phi_2 (\varphi)}{d \varphi^2} - \frac{3}{8} \,\Phi_2 (\varphi) = 0 \,\, ,
\ee
whose general solution is:
\be
\Phi_2 (\varphi) = b_1 \sinh \!\left( \sqrt{\frac{3}{8}} \,\varphi \right) + b_2 \cosh \!\left( \sqrt{\frac{3}{8}} \,\varphi \right) \,\, ,
\ee
where $b_{1,2} = const$. Using this in (\ref{ThreeEqs}), we find:
\be \label{Phi1_gen}
\Phi_1 (\varphi) = k \sqrt{\frac{3}{8}} \left[ b_1 \cosh \!\left( \sqrt{\frac{3}{8}} \,\varphi \right) + b_2 \sinh \!\left( \sqrt{\frac{3}{8}} \,\varphi \right) \right] \,\, .
\ee
Note that our results above for $A_1 (a)$, $A_2 (a)$, $\Phi_1 (\varphi)$ and $\Phi_2 (\varphi)$ are consistent with those of \cite{KKC}, except that $b_1$ was set to zero in that work.

\subsection{Solving equations (E5) and (E6)}

Now we turn to equations (E5) and (E6) of (\ref{Syst_alpha}). We will see below that, for an arbitrary function $f(\varphi)$, the (E5)-(E6) system does not have a solution compatible with (\ref{Phi1_gen}). However, recall that we are only interested in functions $f$, such that the sigma-model metric in (\ref{S_alpha_p}), namely the metric $ds^2 = d\varphi^2 + f(\varphi) d\theta^2$, is hyperbolic. We will show now that, for any such $f (\varphi)$, equations (E5) and (E6) can be solved in a manner compatible with (\ref{Phi1_gen}).

Let us begin by substituting (\ref{Lambda_ans}) in (E5). This gives
\be \label{Th3}
\Theta_3 (\theta) = c \frac{d \Theta_1}{d \theta} \qquad {\rm with} \qquad c = const 
\ee
and
\be \label{A3_eq}
A_1 (a) - \beta a^2 \frac{d A_3}{d a} = 0 \qquad {\rm with} \qquad \beta = const \,\, ,
\ee
as well as an equation for $\Phi_3 (\varphi)$ which we will write down shortly. Using (\ref{A1}) allows us to solve (\ref{A3_eq}) as:
\be \label{A3}
A_3 (a) = - \frac{2}{3} \frac{A}{\beta a^{3/2}} \,\, ,
\ee
where we have set an additive integration constant to zero in order to ensure that $\frac{d \Phi_3}{d \varphi} \neq 0$ and $\frac{d \Theta_2}{d \theta} \neq 0$.\footnote{The sixth equation in (\ref{Syst_alpha}) implies that $\frac{d \Phi_3}{d \varphi} = 0$ and $\frac{d \Theta_2}{d \theta} = 0$ if there is a non-vanishing additive constant in (\ref{A3}).} Upon using (\ref{Th3}) and (\ref{A3}), equation (E5) reduces to the following algebraic relation:
\be \label{Eq5_f}
\Phi_3 (\varphi) = \frac{6 \beta}{c} \frac{1}{f (\varphi)} \,\Phi_1 (\varphi) \,\,\, .
\ee

Let us now consider equation (E6) of (\ref{Syst_alpha}). Substituting (\ref{A2}) and (\ref{A3}), one finds that the $a$-dependence factors out of this equation. Then, using the third relation of (\ref{ThreeEqs}) together with (\ref{Phi2}), equation (E6) reduces to:
\be
\frac{d \Phi_3}{d \varphi} = - \frac{6 \beta}{c} \frac{1}{f(\varphi)} \frac{d \Phi_1}{d \varphi} \,\,\, .
\ee
Comparing the last relation with (\ref{Eq5_f}), we conclude that
\be
\frac{1}{\Phi_3} \frac{d \Phi_3}{d \varphi} = - \frac{1}{\Phi_1} \frac{d \Phi_1}{d \varphi} \,\,\, ,
\ee
which implies:
\be \label{Ph3_inv_Ph1}
\Phi_3 (\varphi) = \frac{\Phi_0}{\Phi_1 (\varphi)} \,\,\, ,
\ee
where $\Phi_0 = const$. Substituting (\ref{Ph3_inv_Ph1}) in (\ref{Eq5_f}), we obtain:
\be \label{Ph1_f}
\Phi_1 (\varphi) = \sqrt{\frac{c \,\Phi_0}{6 \beta}} \,\sqrt{f (\varphi)}
\ee
and thus
\be \label{Ph3_f}
\Phi_3 (\varphi) = \sqrt{\frac{6 \beta \Phi_0}{c}} \frac{1}{\sqrt{f (\varphi)}} \,\,\, .
\ee

Clearly, for arbitrary $f(\varphi)$, the expression in (\ref{Ph1_f}) is not compatible with the $\Phi_1 (\varphi)$ solution found in (\ref{Phi1_gen}). However, we are interested only in functions $f(\varphi)$, for which the scalar manifold metric in (\ref{S_alpha_p}) is hyperbolic. In other words, we are only considering $f(\varphi)$ such that the Gaussian curvature $K$ of the metric $ds^2 = d \varphi^2 + f(\varphi) d \theta^2$ is constant and negative. This restricts the form of the function $f$. To see how, let us compute the Gaussian curvature in question:
\be \label{GausK}
K = - \frac{1}{4} \frac{(2 f f'' - f'^2)}{f^2} \,\,\, ,
\ee
where $f'\equiv \pd_{\varphi} f$. Imposing the condition that $K = const < 0$\,, we can view (\ref{GausK}) as an ODE for $f (\varphi)$. Solving it, we obtain:
\be \label{f_hyp}
f (\varphi) = \left[ C_1^{\varphi} \cosh \!\left( \sqrt{|K|}\,\varphi \right) + C_2^{\varphi} \sinh \!\left( \sqrt{|K|}\,\varphi \right) \right]^2 \qquad {\rm with} \qquad C_{1,2}^{\varphi} = const \,\,\, .
\ee
Substituting (\ref{f_hyp}) in (\ref{Ph1_f}), we find that the result has the same form as (\ref{Phi1_gen}). To completely match the two expressions for $\Phi_1 (\varphi)$, we have to take
\be
|K| = \frac{3}{8} \,\,\, . 
\ee
Note that this will restrict the value of the $\alpha$-parameter in each of the three cases with $f$ given by (\ref{fD}), (\ref{fD*}) and (\ref{fA}), as we will see shortly.\footnote{In \cite{LS}, the normalization $K = - \frac{1}{3 \alpha}$ was imposed for any hyperbolic surface. In the present work, however, the coefficients of proportionality between $K$ and $\frac{1}{\alpha}$ are different for each of the elementary hyperbolic surfaces. This follows from writing the relevant kinetic terms with the normalizations given in (\ref{S_Pd}), (\ref{S_Punc_d}) and (\ref{S_Annuli}), which is convenient for easier comparison with most of the  literature.}

Let us now compare in more detail the solution (\ref{Ph1_f}), with $f$ given by (\ref{f_hyp}), to the expression in (\ref{Phi1_gen}), for each elementary hyperbolic surface. The general form of $f$ in (\ref{f_hyp}) reduces to the specific form, in each of the three cases listed in equations (\ref{fD}), (\ref{fD*}) and (\ref{fA}), for the following respective choices of the integration constants:
\begin{align}
&f = f_{\bbb{D}} \hspace*{-1.1cm}&&: &&\hspace*{0.55cm}C_1^{\varphi} = 0 \hspace*{-0.55cm}&&{\rm and} &&\hspace*{-0.55cm}\left( C_2^{\varphi} \right)^2 = \frac{3 \alpha}{2} \quad , \nn \\
&f = f_{\bbb{D}^*} \hspace*{-1.1cm}&&: &&\left( C_1^{\varphi} \right)^2 = 2 \alpha \hspace*{-0.55cm}&&{\rm and} &&C_2^{\varphi} = - C_1^{\varphi} \quad , \nn \\
&f = f_{\bbb{A}} \hspace*{-1.1cm}&&: &&\left( C_1^{\varphi} \right)^2 = 2 \alpha C_R^2 \hspace*{-0.55cm}&&{\rm and} &&C_2^{\varphi} = 0 \quad .
\end{align}
Substituting these three cases for $f(\varphi)$ in relation (\ref{Ph1_f}) and comparing with (\ref{Phi1_gen}) gives the following conditions for the existence of a solution:  
\begin{align} \label{Constants}
\hspace*{0.2cm}&\bb{D} \hspace*{-0.4cm}&&: \qquad b_1 = 0 &&, \qquad b_2 = \frac{1}{k} \sqrt{\frac{2 \alpha c \,\Phi_0}{3 \beta}} &&, \qquad \alpha = \frac{16}{9} \quad , \nn \\
\hspace*{0.2cm}&\bb{D}^* \hspace*{-0.4cm}&&: \qquad b_1 = \frac{2}{3k} \sqrt{\frac{2 \alpha c \Phi_0}{\beta}} &&, \qquad b_2 = - b_1 &&, \qquad \alpha = \frac{4}{3} \quad , \nn \\
\hspace*{0.2cm}&\bb{A} \hspace*{-0.4cm}&&: \qquad b_1 = \frac{2C_R}{3k} \sqrt{\frac{2 \alpha c \Phi_0}{\beta}} &&, \qquad b_2 = 0 &&, \qquad \alpha = \frac{4}{3} \quad . 
\end{align}

To recapitulate, we have shown that, upon choosing integration constants satisfying the constraints (\ref{Constants}), the solutions of equations (E5) and (E6) are compatible with those of (E1), (E2) and (E4). More explicitly, the solutions for the functions $\Phi_{1,2} (\varphi)$ in the three cases of interest have the form:
\bea \label{Phi_solutions}
\bb{D} : &&\Phi_1 (\varphi) = k b_2 \sqrt{\frac{3}{8}} \,\sinh \!\left( \sqrt{\frac{3}{8}} \,\varphi \right) \quad , \quad \Phi_2 (\varphi) = b_2 \cosh \!\left( \sqrt{\frac{3}{8}} \,\varphi \right) \,\,\, , \nn \\
\bb{D}^* : &&\Phi_1 (\varphi) = k b_1 \sqrt{\frac{3}{8}} \,\exp \!\left( - \sqrt{\frac{3}{8}} \,\varphi \right) \quad , \quad \Phi_2 (\varphi) = - b_1 \exp \!\left( - \sqrt{\frac{3}{8}} \,\varphi \right) \,\,\, , \nn \\
\bb{A} : &&\Phi_1 (\varphi) = k b_1 \sqrt{\frac{3}{8}} \,\cosh \!\left( \sqrt{\frac{3}{8}} \,\varphi \right) \quad , \quad \Phi_2 (\varphi) = b_1 \sinh \!\left( \sqrt{\frac{3}{8}} \,\varphi \right) \,\,\, , 
\eea
while $\Phi_3 (\varphi)$ is given by (\ref{Ph3_inv_Ph1}) in all three cases. Also, for any function $f$, the solutions for $A_{1,2,3} (a)$ are:
\be \label{A123_sols}
A_1 (a) = \frac{A}{a^{1/2}} \,\,\,\quad , \,\,\,\quad A_2 (a) = - \frac{3}{2} \frac{A}{a^{3/2}} \,\,\,\quad , \,\,\,\quad A_3 (a) = - \frac{2}{3} \frac{A}{\beta a^{3/2}} \,\,\,\quad ,
\ee
as can be seen in (\ref{A1}), (\ref{A2}) and (\ref{A3}).

\subsection{Solving equation (E3)}

Next, we consider equation (E3) of the system (\ref{Syst_alpha}). Substituting the solutions for $A_{1,2,3}(a)$ given in (\ref{A123_sols}), we find that the $a$-dependence drops out from (E3). Then, using the third relation in (\ref{ThreeEqs}), as well as (\ref{Phi2}) and (\ref{Th3}), we find that (E3) reduces to:
\be \label{EQphth_f}
3 \left[ f (\varphi) \Phi_1 (\varphi) - \frac{4}{3} f' \Phi_1' \right] \Theta_1 (\theta) - 8 \Phi_1 (\varphi) \Theta_1'' (\theta) = 0 \,\,\, .
\ee
Substituting (\ref{Ph1_f}) in (\ref{EQphth_f}) gives:
\be \label{EQth_f}
\left( 3 f^2 - 2 f'^2 \right) \Theta_1 (\theta) - 8 f (\varphi) \Theta_1'' (\theta) = 0 \,\,\, .
\ee
Since the form of $f (\varphi)$ is fixed for each elementary hyperbolic surface, equation (\ref{EQth_f}) is an ODE for the function $\Theta_1 (\theta)$. This ODE admits solutions if and only if the following condition is satisfied:
\be \label{Th_eq}
\frac{\Theta_1'' (\theta)}{\Theta_1 (\theta)} = \frac{3 f^2 - 2 f'^2}{8 f} = const \equiv q \,\,\, .
\ee

It is easy to check that the expression $\frac{3 f^2 - 2 f'^2}{8 f}$ is indeed constant in each of the three cases of interest, namely the Poincar\'e disk, the punctured hyperbolic disk and the hyperbolic annuli. More precisely, substituting $f$ respectively from (\ref{fD}), (\ref{fD*}) and (\ref{fA}) gives: 
\be \label{q_value}
q_{\bbb{D}} = -1 \qquad , \qquad q_{\bbb{D}^*} = 0 \qquad , \qquad q_{\bbb{A}} = C_R^2 \quad .
\ee
In fact, one can show directly that $\frac{3 f^2 - 2 f'^2}{8 f} = const$ for any $f(\varphi)$\,, such that the scalar manifold metric in (\ref{S_alpha_p}) is hyperbolic. Namely, using the form of  $f(\varphi)$ given in (\ref{f_hyp}) with $|K| = \frac{3}{8}$\,, we obtain:
\be \label{ffpr_const}
\frac{3 f^2 - 2 f'^2}{8 f} = \frac{3}{8} \left[ (C_1^{\varphi})^2 - (C_2^{\varphi})^2 \right] \,\,\, .
\ee

Let us now study equation in (\ref{Th_eq}) for each of the three values of $q$ given in (\ref{q_value}).

\vspace{0.4cm}
\noindent
$\bullet$ \hspace*{0.03cm}{\bf Poincar\'e disk:}

\vspace{0.2cm}
\noindent
For $q=-1$, relation (\ref{Th_eq}) gives:
\be
\Theta_1'' (\theta) + \Theta_1 (\theta) = 0 \,\,\, ,
\ee
with the obvious solution 
\be \label{Th_d_sol}
\Theta_1 (\theta) = C_1 \sin \theta + C_2 \cos \theta \,\, .
\ee
Then (\ref{ThreeEqs}) implies:
\be
\Theta_2 (\theta) = k \left( C_1 \sin \theta + C_2 \cos \theta \right) \,\, ,
\ee
whereas (\ref{Th3}) gives:
\be \label{Th3_d_sol}
\Theta_3 (\theta) = c \left( C_1 \cos \theta - C_2 \sin \theta \right) \,\, .
\ee

\vspace{0.4cm}
\noindent
$\bullet$ \hspace*{0.03cm}{\bf Hyperbolic punctured disk:}

\vspace{0.2cm}
\noindent
For $q=0$, equation (\ref{Th_eq}) becomes:
\be
\Theta_1'' (\theta) = 0 \,\,\, ,
\ee
whose solution can be written as:
\be \label{Th_1_d_star_sol}
\Theta_1 (\theta) = C_3 \theta + \theta_0 \qquad {\rm with} \qquad C_3 , \theta_0 = const \,\,\, .
\ee
Using (\ref{Th_1_d_star_sol}) as well as (\ref{ThreeEqs}) and (\ref{Th3}), we find:
\be \label{Th_2_3_d_star_sol}
\Theta_2 (\theta) = k \!\left( C_3 \theta + \theta_0 \right) \qquad {\rm and} \qquad \Theta_3 (\theta) = c \,C_3  \,\,\, .
\ee

\vspace{0.4cm}
\noindent
$\bullet$ \hspace*{0.03cm}{\bf Hyperbolic annulus:}

\vspace{0.2cm}
\noindent
For $q=C_R^2$, equation (\ref{Th_eq}) takes the form:
\be
\Theta_1'' (\theta) - C_R^2 \,\Theta_1 (\theta) = 0 \,\,\, ,
\ee
which has the general solution
\be \label{Th_1_A_sol}
\Theta_1 (\theta) = C_4 \cosh (C_R \theta) + C_5 \sinh (C_R \theta) \,\,\, .
\ee
Hence, (\ref{ThreeEqs}) and (\ref{Th3}) give:
\bea \label{Th_2_3_A_sol}
\Theta_2 (\theta) &=& k \left[ C_4 \cosh (C_R \theta) + C_5 \sinh (C_R \theta) \right] \,\,\, , \nn \\
\Theta_3 (\theta) &=& c \,C_R \left[ C_4 \sinh (C_R \theta) + C_5 \cosh (C_R \theta) \right] \,\,\, .
\eea

\subsection{Solving equation (E7): the scalar potential}

So far, we have found functions $\lambda_{a,\varphi,\theta} (a, \varphi, \theta)$\,, which solve equations (E1)-(E6) of the Noether system (\ref{Syst_alpha}). Now we will show that the last equation of that system, namely (E7), determines the scalar potential $V(\varphi , \theta)$, if the latter is assumed to have  the separation of variables form:
\be \label{Vans}
V (\varphi, \theta) = \tilde{V} (\varphi) \hat{V} (\theta) \,\,\, .
\ee

We begin by substituting (\ref{Lambda_ans}) and (\ref{Vans}) into (E7). Then, using the solutions for $A_{1,2,3} (a)$ given in (\ref{A123_sols}) as well as the last relation in (\ref{ThreeEqs}) (namely  $\Theta_2 = k \Theta_1$), we find that (E7) reduces to:
\be
3 \left[ \tilde{V} (\varphi) \Phi_1 (\varphi) - \frac{k}{2} \tilde{V}' (\varphi) \Phi_2 (\varphi) \right] \hat{V} (\theta) \Theta_1 (\theta) - \frac{2}{3\beta} \tilde{V} (\varphi) \Phi_3 (\varphi) \hat{V} (\theta) \Theta_3 (\theta) = 0 \,\,\, .
\ee
Note that here we have not used any particular form of the function $f$. Hence, for any $f(\varphi)$, and thus for any $\Phi_{1,2,3} (\varphi)$ and $\Theta_{1,2,3} (\theta)$, we have the pair of equations
\be \label{V_eqs}
\frac{3 \left[ \tilde{V} (\varphi) \Phi_1 (\varphi) - \frac{k}{2} \tilde{V}' (\varphi) \Phi_2 (\varphi) \right]}{\frac{2}{3 \beta} \tilde{V} (\varphi) \Phi_3 (\varphi)} = p = \frac{\hat{V}' (\theta) \,\Theta_3 (\theta)}{\hat{V} (\theta) \,\Theta_1 (\theta)} \,\,\, ,
\ee
where $p = const$. Clearly, then, one has two separate equations for the two functions $\tilde{V} (\varphi)$ and $\hat{V} (\theta)$. Let us now study these two equations for each of the three types of elementary hyperbolic surface.

\vspace{0.4cm}
\noindent
$\bullet$ \hspace*{0.03cm}{\bf Poincar\'e disk:}

\vspace{0.2cm}
\noindent
Substituting the $\bb{D}$ expressions from (\ref{Phi_solutions}) and (\ref{Constants}) into (\ref{V_eqs}), we find the following equation for $\tilde{V} (\varphi)$:
\be
\frac{d \tilde{V} (\varphi)}{d \varphi} + \frac{\sqrt{\frac{3}{8}} \left[ \frac{p}{c} - 2 \sinh^2 \!\left( \sqrt{\frac{3}{8}} \,\varphi \right) \right] }{\sinh \!\left( \sqrt{\frac{3}{8}} \,\varphi \right) \cosh \!\left( \sqrt{\frac{3}{8}} \,\varphi \right)} \,\tilde{V} (\varphi) = 0 \,\,\, .
\ee
Its general solution has the form:
\be \label{Vtilde_sol}
\tilde{V} (\varphi) = \tilde{V}_0 \cosh^2 \!\left( \sqrt{\frac{3}{8}} \,\varphi \right) \coth^{\frac{p}{c}} \!\left( \sqrt{\frac{3}{8}} \,\varphi \right) \,\,\, ,
\ee
where $\tilde{V}_0$ is an integration constant.

Using (\ref{Th_d_sol}) and (\ref{Th3_d_sol}) inside (\ref{V_eqs}), we obtain:
\be \label{Vhat_th_Eq}
\frac{d \hat{V} (\theta)}{d \theta} - \frac{p}{c} \frac{(C_1 \sin \theta + C_2 \cos \theta)}{(C_1 \cos \theta - C_2 \sin \theta)} \,\hat{V} (\theta) = 0 \,\,\, ,
\ee
whose solution is:
\be \label{Vtilde}
\hat{V} (\theta) = \hat{V}_0 \left[ C_1 \cos \theta - C_2 \sin \theta \right]^{-\frac{p}{c}}
\ee
with $\hat{V}_0 = const$.

Therefore, for the case of the hyperbolic disk, the form of the scalar potential, that is compatible with Noether's symmetry, is:
\be \label{Vpot_tot}
V (\varphi , \theta) = V_0 \cosh^2 \!\left( \sqrt{\frac{3}{8}} \,\varphi \right) \coth^{\frac{p}{c}} \!\left( \sqrt{\frac{3}{8}} \,\varphi \right) \left[ C_1 \cos \theta - C_2 \sin \theta \right]^{-\frac{p}{c}} \,\,\, ,
\ee
where $V_0 = const$. Note that this expression reduces to the single-field result of \cite{KKC} for $p=0$. It is also worth pointing out that the $\theta$-dependence in (\ref{Vtilde}) allows as a special case the particular form needed for natural inflation. Indeed, by taking $C_2 = 0$ and $\frac{p}{c} = - 2$, we have $\hat{V} (\theta) = const \times \cos^2 \!\theta$. In that regard, it may be interesting to make a connection to the recent considerations of \cite{LWWYA} on realizing natural inflation in two-field attractor models.

\eject

\vspace{0.4cm}
\noindent
$\bullet$ \hspace*{0.03cm}{\bf Hyperbolic punctured disk:}

\vspace{0.2cm}
\noindent
Using the $\bb{D}^*$ expressions from (\ref{Phi_solutions}) and (\ref{Constants}) in (\ref{V_eqs}), we have:
\be
\frac{d \tilde{V} (\varphi)}{d \varphi} + \sqrt{\frac{3}{2}} \left( 1 - \frac{p}{2 c} \,e^{\sqrt{\frac{3}{2}} \,\varphi} \right) \tilde{V} (\varphi) = 0 \,\,\, ,
\ee
whose solution is:
\be \label{Vtilde_D*}
\tilde{V} (\varphi) = \tilde{V}_0 \,\exp \!\left( \!- \sqrt{\frac{3}{2}} \,\varphi + \frac{p}{2c} e^{\sqrt{\frac{3}{2}} \varphi} \right) \,\,\, .
\ee

Now, substituting the solutions for $\Theta_{1,3}$ from (\ref{Th_1_d_star_sol}) and (\ref{Th_2_3_d_star_sol}) inside (\ref{V_eqs}), we end up with:
\be
\frac{d \hat{V} (\theta)}{d \theta} - \frac{p}{c} \!\left( \theta + \frac{\theta_0}{C_3} \right) \!\hat{V} (\theta) = 0 \,\,\, .
\ee
The solution of the last equation is:
\be \label{Vhat_D*}
\hat{V} (\theta) = \hat{V}_0 \exp \!\left[ {\frac{p \,\theta}{c} \!\left( \frac{\theta}{2} + \frac{\theta_0}{C_3} \right)} \!\right] \,\,\, .
\ee

Note that $p=0$ again gives a result independent of $\theta$ and thus leads to an effectively single-field system. It may be interesting to investigate this special case further and to see whether or how it differs from the single-field system studied in \cite{KKC} (which arises from taking $p=0$ for the Poincar\'e disk).

\vspace{0.4cm}
\noindent
$\bullet$ \hspace*{0.03cm}{\bf Hyperbolic annulus:}

\vspace{0.2cm}
\noindent
Finally, from the $\bb{A}$ expressions in (\ref{Phi_solutions}) and (\ref{Constants}), substituted in (\ref{V_eqs}), we obtain:
\be
\frac{d \tilde{V} (\varphi)}{d \varphi} + \frac{\sqrt{\frac{3}{8}} \left[ \frac{p}{c C_R^2} - 2 \cosh^2 \!\left( \sqrt{\frac{3}{8}} \,\varphi \right) \right] }{\sinh \!\left( \sqrt{\frac{3}{8}} \,\varphi \right) \cosh \!\left( \sqrt{\frac{3}{8}} \,\varphi \right)} \,\tilde{V} (\varphi) = 0 \,\,\, .
\ee
Hence, in this case the solution for $\tilde{V}$ is:
\be \label{Vtilde_A}
\tilde{V} (\varphi) = \tilde{V}_0 \sinh^2 \!\left( \sqrt{\frac{3}{8}} \,\varphi \right) \coth^{\frac{p}{c C_R^2}} \!\left( \sqrt{\frac{3}{8}} \,\varphi \right) \,\,\, .
\ee

Now substituting (\ref{Th_1_A_sol}) and (\ref{Th_2_3_A_sol}) in (\ref{V_eqs}), one finds the following equation for $\hat{V}$:
\be
\frac{d \hat{V} (\theta)}{d \theta} - \frac{p}{c C_R} \frac{\left[ C_4 \cosh (C_R \theta) + C_5 \sinh (C_R \theta) \right]}{\left[ C_4 \sinh (C_R \theta ) + C_5 \cosh (C_R \theta ) \right]} \,\hat{V} (\theta) = 0 \,\,\, ,
\ee
whose solution is given by:
\be \label{Vhat_A}
\hat{V} (\theta) = \hat{V}_0 \left[ C_4 \sinh (C_R \theta) + C_5 \cosh (C_R \theta) \right]^{\frac{p}{c C_R^2}} \,\,\, .
\ee

\section{New variables: cyclic coordinate} \label{Sec:NewVar}
\setcounter{equation}{0}

In this section, we will look for a suitable coordinate transformation $(a,\varphi,\theta) \rightarrow (u,v,w)$, such that $w$ is the cyclic coordinate corresponding to the symmetry with generator $X$ that we found above. This will be very useful for finding analytical solutions of the $\alpha$-attractor equations of motion for the following reason. In the new variables the symmetry generator will have the form $X = \frac{\pd }{\pd w}$ and thus the condition $L_X {\cal L} =0$ will become:
\be \label{L_w_zero}
\frac{\pd {\cal L}}{\pd w} = 0 \,\,\, .
\ee 
This will simplify the relevant equations of motion significantly, as we will see below. Note that, due to (\ref{L_w_zero}), the Euler-Lagrange equation for $w$ becomes:
\be
\frac{d}{d t} \,\frac{\pd {\cal L}}{\pd \dot{w}} = 0 \,\,\, ,
\ee
which shows that the generalized momentum $p_w \equiv \frac{\pd {\cal L}}{\pd \dot{w}}$ is conserved. 

To find such coordinates, we must solve the conditions $i_X \dd u = 0$\,, \,$i_X \dd v = 0$ \,and \,$i_X \dd w =1$, which amount to the system:
\bea \label{SystCyclic}
\lambda_a \frac{\pd u}{\pd a} + \lambda_{\varphi} \frac{\pd u}{\pd \varphi} + \lambda_{\theta} \frac{\pd u}{\pd \theta} = 0 \,\,\, , \nn \\
\lambda_a \frac{\pd v}{\pd a} + \lambda_{\varphi} \frac{\pd v}{\pd \varphi} + \lambda_{\theta} \frac{\pd v}{\pd \theta} = 0 \,\,\, , \nn \\
\lambda_a \frac{\pd w}{\pd a} + \lambda_{\varphi} \frac{\pd w}{\pd \varphi} + \lambda_{\theta} \frac{\pd w}{\pd \theta} = 1 \,\,\, ,
\eea
Since the first two equations in (\ref{SystCyclic}) are formally identical, the general solutions for $u (a, \varphi, \theta)$ and $v (a, \varphi, \theta)$ will have the same form. Ensuring different functions for $u$ and $v$ will be due to choosing different values for (some of) the constants that characterize this general form, as will become clear below. 

\subsection{Finding the coordinates $u$ and $v$}

In this subsection we consider the first equation in (\ref{SystCyclic}), namely
\be \label{EQu}
\lambda_a \frac{\pd u}{\pd a} + \lambda_{\varphi} \frac{\pd u}{\pd \varphi} + \lambda_{\theta} \frac{\pd u}{\pd \theta} = 0 \,\,\, .
\ee
As already pointed out, this will enable us to find not only $u$, but $v$ as well. 

We will look for solutions with the separation of variables Ansatz:
\be \label{u_ans}
u (a,\varphi,\theta) = A_u (a) \Phi_u (\varphi) \Theta_u (\theta) \,\,\, .
\ee
Using (\ref{Lambda_ans}), (\ref{u_ans}) and the last relation in (\ref{ThreeEqs}) (i.e. $\Theta_2 = k \Theta_1$), equation (\ref{EQu}) reduces to:
\be
\left[ A_1 \Phi_1 A_u' \Phi_u + k A_2 \Phi_2 A_u \Phi_u' \right] \Theta_1 \Theta_u = - A_3 \Phi_3  A_u \Phi_u \Theta_3 \Theta_u' \,\,\, .
\ee 
Separating out the $\theta$-dependence gives:
\be \label{Th_u}
\frac{\Theta_3 (\theta)}{\Theta_1 (\theta) \Theta_u (\theta)} \frac{d \Theta_u (\theta)}{d \theta} = c_{\theta}
\ee
and
\be \label{A_uPh_u}
A_1 \Phi_1 A_u' \Phi_u + k A_2 \Phi_2 A_u \Phi_u' + c_{\theta} A_3 \Phi_3  A_u \Phi_u = 0
\ee
for some $c_{\theta} = const$. Now, substituting $A_{1,2,3} (a)$ from (\ref{A123_sols}) in (\ref{A_uPh_u}), we find that the $a$-dependence factors out provided that:
\be
\frac{d A_u}{d a} = c_a \frac{A_u (a)}{a}
\ee
for some $c_a = const$. The last equation is solved by:
\be \label{A_u_sol}
A_u (a) = a^{c_a} \,\,\, ,
\ee
where for convenience we have set the overall multiplicative integration constant to one.\footnote{Note that, to preform a coordinate transformation $(a,\varphi,\theta) \rightarrow (u,v,w)$, we only need a {\it particular} solution of the system (\ref{SystCyclic}).} 

Substituting (\ref{A_u_sol}) and (\ref{A123_sols}) in (\ref{A_uPh_u}) gives:
\be \label{EQ_Ph_u}
\frac{3}{2} k \Phi_2 \frac{d \Phi_u (\varphi)}{d \varphi} + \left( \frac{2}{3} \frac{c_{\theta}}{\beta} \Phi_3 - c_a \Phi_1 \!\right) \!\Phi_u (\varphi) = 0 \,\,\, .
\ee
This equation has different coefficients for each elementary hyperbolic surface, since the functions $\Phi_{1,2,3} (\varphi)$ differ in each case (see equation (\ref{Phi_solutions})). Before specializing to the various cases, we can further simplify (\ref{EQ_Ph_u}) by using the expressions (\ref{Ph1_f})-(\ref{Ph3_f}) and (\ref{Phi2}) for $\Phi_{1,2,3}$ in terms of the function $f(\varphi)$. This allow us to bring (\ref{EQ_Ph_u}) to the form:
\be \label{EQ_phu_u_f}
\frac{d \Phi_u (\varphi)}{d \varphi} + \frac{\left[ \frac{2 c_{\theta}}{c} - \frac{c_a}{2} f (\varphi) \right]}{ f' (\varphi)} \,\Phi_u (\varphi) = 0 \,\,\, .
\ee

To recapitulate, the solution for $A_u (a)$ is independent of $f$ and is given by (\ref{A_u_sol}). On the other hand, the solutions for $\Phi_u (\varphi)$ and $\Theta_u (\theta)$ do depend on the form of the function $f$ and are determined by equations (\ref{Th_u}) and (\ref{EQ_phu_u_f}), respectively. Let us now find $\Phi_u$ and $\Theta_u$ for each type of elementary hyperbolic surface.

\eject

\vspace{0.4cm}
\noindent
$\bullet$ \hspace*{0.03cm}{\bf Poincar\'e disk:}

\vspace{0.2cm}
\noindent
For $f(\varphi)$ given in (\ref{fD}) with $\alpha = \frac{16}{9}$ (see the corresponding row in  (\ref{Constants})), we find that (\ref{EQ_phu_u_f}) acquires the form:
\be
\frac{d \Phi_u (\varphi)}{d \varphi} + \frac{\sqrt{\frac{3}{8}} \left[ \frac{c_{\theta}}{c} - \frac{2 c_a}{3} \sinh^2 \!\left( \sqrt{\frac{3}{8}} \,\varphi \right) \right] }{\sinh \!\left( \sqrt{\frac{3}{8}} \,\varphi \right) \cosh \!\left( \sqrt{\frac{3}{8}} \,\varphi \right)} \,\Phi_u (\varphi) = 0 \,\,\, .
\ee
This equation has the general solution:
\be \label{Ph_u_sol_gen}
\Phi_u (\varphi) = \left[ \coth \!\left( \sqrt{\frac{3}{8}} \,\varphi \right) \right]^{\frac{c_{\theta}}{c}} \left[ \cosh \!\left( \sqrt{\frac{3}{8}} \,\varphi \right) \right]^{\frac{2 c_a}{3}} \,\,\, ,
\ee
where we have again set the overall integration constant to one for convenience. Note that the single-field result for $\Phi_u (\varphi)$ in (\ref{Ph_u_sol_gen}) is obtained by taking $c_{\theta} = 0$. Then, setting $c_a = 3$, we find from (\ref{Ph_u_sol_gen}) and (\ref{A_u_sol}) the same particular solution for $u (a, \varphi) = A_u (a) \Phi_u (\varphi)$, as that in \cite{KKC}.

Now, substituting (\ref{Th_d_sol}) and (\ref{Th3_d_sol}) in (\ref{Th_u}), we find:
\be 
\frac{d \Theta_u (\theta)}{d \theta} - \frac{c_{\theta}}{c} \frac{(C_1 \sin \theta + C_2 \cos \theta)}{(C_1 \cos \theta - C_2 \sin \theta)} \,\Theta_u (\theta) = 0 \,\,\, ,
\ee
whose solution is:
\be \label{Th_u_sol}
\Theta_u (\theta) = \left( C_1 \cos \theta - C_2 \sin \theta \right)^{-\frac{c_{\theta}}{c}}
\ee
with the overall integration constant once again set to one.

\vspace{0.4cm}
\noindent
$\bullet$ \hspace*{0.03cm}{\bf Hyperbolic punctured disk:}

\vspace{0.2cm}
\noindent
Taking $f(\varphi)$ as in (\ref{fD*}) with $\alpha = \frac{4}{3}$ (in accordance with the $\bb{D}^*$ row of (\ref{Constants})), equation (\ref{EQ_phu_u_f}) becomes: 
\be
\frac{d \Phi_u (\varphi)}{d \varphi} + \sqrt{\frac{3}{8}} \left[ \frac{2 c_a}{3} - \frac{c_{\theta}}{c} \exp \!\left( \sqrt{\frac{3}{2}} \,\varphi \right) \right] \!\Phi_u (\varphi) = 0 \,\,\, .
\ee
This ODE is solved by
\be \label{Phi_u_D*}
\Phi_u (\varphi) = \exp \left[ - \frac{c_a}{\sqrt{6}} \,\varphi + \frac{c_{\theta}}{2 c} \,\exp \!\left( \sqrt{\frac{3}{2}} \,\varphi \right) \right] \,\, ,
\ee
where again the overall integration constant has been set to one.

The solution of (\ref{Th_u}), after substituting (\ref{Th_1_d_star_sol}) and (\ref{Th_2_3_d_star_sol}), is given by:
\be \label{Th_u_D*}
\Theta_u (\theta) = \exp \!\left[ {\frac{c_{\theta} \,\theta}{c} \!\left( \frac{\theta}{2} + \frac{\theta_0}{C_3} \right)} \!\right] \,\,\, ,
\ee
where the overall integration constant was set to one.

\vspace{0.4cm}
\noindent
$\bullet$ \hspace*{0.03cm}{\bf Hyperbolic annulus:}

\vspace{0.2cm}
\noindent
For $f(\varphi)$ given by (\ref{fA}) with $\alpha = \frac{4}{3}$ (as in the $\bb{A}$ line of (\ref{Constants})), equation (\ref{EQ_phu_u_f}) becomes:
\be
\frac{d \Phi_u (\varphi)}{d \varphi} + \frac{\sqrt{\frac{3}{8}} \left[ \frac{c_{\theta}}{c C_R^2} - \frac{2 c_a}{3} \cosh^2 \!\left( \sqrt{\frac{3}{8}} \,\varphi \right) \right] }{\sinh \!\left( \sqrt{\frac{3}{8}} \,\varphi \right) \cosh \!\left( \sqrt{\frac{3}{8}} \,\varphi \right)} \,\Phi_u (\varphi) = 0 \,\,\, .
\ee
Hence, the solution in this case is
\be \label{Phi_u_A}
\Phi_u (\varphi) = \left[ \coth \!\left( \sqrt{\frac{3}{8}} \,\varphi \right) \right]^{\frac{c_{\theta}}{c C_R^2}} \left[ \,\sinh \!\left( \sqrt{\frac{3}{8}} \,\varphi \right) \right]^{\frac{2 c_a}{3}} \,\,\, .
\ee

Finally, the solution of (\ref{Th_u}), after substituting (\ref{Th_1_A_sol}) and (\ref{Th_2_3_A_sol}), has the form:
\be \label{Th_u_A}
\Theta_u (\theta) = \left[ C_4 \sinh (C_R \theta) + C_5 \cosh (C_R \theta) \right]^{\frac{c_{\theta}}{c C_R^2}} \,\,\, .
\ee

\vspace{0.4cm}
\noindent
{\bf Remark on the coordinate $v$:}

\vspace{0.2cm}
\noindent
So far, we have found a function $u (a, \varphi, \theta) = A_u (a) \Phi_u (\varphi) \Theta_u (\theta)$, for each of the three cases under consideration, that solves the first equation in (\ref{SystCyclic}). As mentioned above, the second equation in (\ref{SystCyclic}) is then solved by a function $v(a , \varphi, \theta) = A_v (a) \Phi_v (\varphi) \Theta_v (\theta)$, such that $A_v$, $\Phi_v$ and $\Theta_v$ have the same general form as their $u$-indexed counterparts. To ensure that $v$ is a different function, one has to choose different values of the constants $c_a$ and $c_{\theta}$ than those taken for the function $u$.

\subsection{Finding the cyclic coordinate $w$}

Now we will consider the last equation in (\ref{SystCyclic}), namely:
\be \label{EQ_cycl_w}
\lambda_a \frac{\pd w}{\pd a} + \lambda_{\varphi} \frac{\pd w}{\pd \varphi} + \lambda_{\theta} \frac{\pd w}{\pd \theta} = 1 \,\,\, .
\ee
As usual, we will make the separation of variables Ansatz: 
\be \label{w_prod}
w (a ,\varphi, \theta) = A_w (a) \Phi_w (\varphi) \Theta_w (\theta) \,\,\, .
\ee
Substituting (\ref{w_prod}), (\ref{Lambda_ans}) and (\ref{A123_sols}) in (\ref{EQ_cycl_w}), it is easy to realize that the $a$-dependence can be canceled within each term by taking
\be \label{A_w_sol}
A_w (a) = \frac{1}{A} \,a^{3/2} \,\,\, .
\ee
Using (\ref{A_w_sol}) and the relation $\Theta_2 = k \Theta_1$ (see (\ref{ThreeEqs})), we find that (\ref{EQ_cycl_w}) acquires the form:
\be \label{EqPh_wTh_w}
\frac{3}{2} \left[ \Phi_1 \Phi_w - k \Phi_2 \Phi_w' \right] \Theta_1 \Theta_w - \frac{2}{3 \beta} \Phi_3 \Phi_w \Theta_3 \Theta_w' = 1 \,\,\, .
\ee

Now we will show that one can remove the $\varphi$-dependence in (\ref{EqPh_wTh_w}) by a suitable choice of the function $\Phi_w (\varphi)$. The result will be an equation for $\Theta_w (\theta)$. Indeed, let us take:
\be \label{Ph_w_sol}
\Phi_w (\varphi) = \frac{\phi_0}{\Phi_3 (\varphi)} \qquad {\rm with} \qquad \phi_0 = const \,\,\, .
\ee
Then, obviously, the second term in (\ref{EqPh_wTh_w}) becomes independent of $\varphi$. In addition, one can show that the combination $[\Phi_1 \Phi_w - k \Phi_2 \Phi_w']$ in the first term, with $\Phi_w$ given by (\ref{Ph_w_sol}), is a constant for each of the three cases in (\ref{Phi_solutions}). In fact, one can see directly that this combination is constant for any function $f(\varphi)$ compatible with the hyperbolic geometry of the scalar manifold. Indeed, using (\ref{Phi2}), (\ref{Ph_w_sol}), (\ref{Ph3_inv_Ph1}) and (\ref{Ph1_f}), we obtain:
\be \label{Combination}
\Phi_1 \Phi_w - k \Phi_2 \Phi_w' = \frac{\phi_0}{\Phi_0} \left[ \Phi_1^2 - \frac{8}{3} \,\Phi_1'^2 \right] = \frac{\phi_0 c}{6 \beta} \,\frac{\left( f^2 - \frac{2}{3} f'^2 \right)}{f} \,\,\, .
\ee
Now recall relation (\ref{ffpr_const}), which holds for any $f (\varphi)$ of the form (\ref{f_hyp}) with $|K| = \frac{3}{8}$. Using this relation, we find that (\ref{Combination}) implies:
\be \label{PhComb_q}
\Phi_1 \Phi_w - k \Phi_2 \Phi_w' = \frac{\phi_0 c}{6 \beta} \left[ \left( C_1^{\varphi} \right)^2 - \left( C_2^{\varphi} \right)^2 \right] = \frac{4}{9} \frac{\phi_0 c}{\beta} q \,\,\, ,
\ee
where for convenience we also wrote the result in terms of the constant $q$ defined in (\ref{Th_eq}).

We are finally ready to extract an ODE for $\Theta_w (\theta)$. Substituting (\ref{Ph_w_sol}) and (\ref{PhComb_q}) into (\ref{EqPh_wTh_w}) gives:
\be \label{EQTh_w}
\frac{2}{3} \frac{\phi_0}{\beta} \big[ c \,q \,\Theta_1 (\theta) \,\Theta_w (\theta) - \Theta_3 (\theta) \,\Theta_w' (\theta) \big] = 1 \,\,\, .
\ee
Then, using (\ref{Th3}) and setting
\be \label{phi_0}
\phi_0 = - \frac{3}{2} \frac{\beta}{c}
\ee
in order to simplify the equation, we obtain from (\ref{EQTh_w}):
\be \label{EQTh_w_f}
\Theta_1' (\theta) \,\Theta_w' (\theta) - q \,\Theta_1 (\theta) \,\Theta_w (\theta) - 1 = 0 \,\,\, .
\ee
Let us now solve the last equation for each type of elementary hyperbolic surface.

\vspace{0.4cm}
\noindent
$\bullet$ \hspace*{0.03cm}{\bf Poincar\'e disk:}

\vspace{0.2cm}
\noindent
In this case $q=-1$ (see (\ref{q_value})) and $\Theta_1 (\theta)$ is given by (\ref{Th_d_sol}). Therefore, (\ref{EQTh_w_f}) becomes:
\be
\left[ C_1 \cos \theta - C_2 \sin \theta \right] \frac{d \Theta_w}{d \theta} + \left[ C_1 \sin \theta + C_2 \cos \theta \right] \Theta_w (\theta) - 1 = 0 \,\,\, .
\ee
The general solution of the last equation can be written as:
\be \label{Th_w_sol}
\Theta_w (\theta) = \frac{\sin \theta}{C_1} + \hat{C}_{\theta} \left[ C_1 \cos \theta - C_2 \sin \theta \right] \qquad {\rm with} \qquad \hat{C}_{\theta} = const \,\,\, .
\ee

Note that, upon redefinition of the integration constant $\hat{C}_{\theta}$, the solution can also be written as:
\be \label{Th_w_sol_c}
\Theta_w (\theta) = \frac{\cos \theta}{C_2} + \hat{C}_{\theta} \left[ C_1 \cos \theta - C_2 \sin \theta \right]
\ee 
or as:
\be \label{Th_w_sol_s}
\Theta_w (\theta) = \frac{1}{2} \left( \frac{\sin \theta}{C_1} + \frac{\cos \theta}{C_2} \right) + \hat{C}_{\theta} \left[ C_1 \cos \theta - C_2 \sin \theta \right] \,\,\, .
\ee
The last form might seem preferable, since it is symmetric with respect to interchange of the trigonometric functions sin and cos. However, this form requires both $C_1 \neq 0$ and $C_2 \neq 0$. On the other hand, the forms (\ref{Th_w_sol}) and (\ref{Th_w_sol_c}) allow one to take respectively  the limits $C_2=0$ and $C_1 = 0$. Since we will be particularly interested in the limit $C_2 = 0$, we will use the form (\ref{Th_w_sol}) in what follows (although we will comment more on using (\ref{Th_w_sol_s}) below).

\vspace{0.4cm}
\noindent
$\bullet$ \hspace*{0.03cm}{\bf Hyperbolic punctured disk:}

\vspace{0.2cm}
\noindent
In this case $q=0$ (see (\ref{q_value})). Also, $\Theta_1 (\theta)$ has the form (\ref{Th_1_d_star_sol}). Substituting these in (\ref{EQTh_w_f}), we obtain the ODE:
\be
C_3 \,\Theta_w' (\theta) - 1 = 0 \,\,\, ,
\ee
which has the solution:
\be \label{Th_w_D*}
\Theta_w (\theta) = \frac{\theta}{C_3} + const \,\,\, .
\ee

\vspace{0.4cm}
\noindent
$\bullet$ \hspace*{0.03cm}{\bf Hyperbolic annulus:}

\vspace{0.2cm}
\noindent
In this case, relation (\ref{q_value}) gives $q = C_R^2$\,. Using this and the relevant $\Theta_1 (\theta)$ expression (\ref{Th_1_A_sol}), we find that (\ref{EQTh_w_f}) acquires the form:
\be \label{EQTh_w_A}
C_R \big[ C_4 \sinh (C_R \theta) + C_5 \cosh (C_R \theta) \big] \Theta_w' (\theta) - C_R^2 \big[ C_4 \cosh (C_R \theta) + C_5 \sinh (C_R \theta) \big] \Theta_w (\theta) = 1 \,\,\, .
\ee
Similarly to the $\bb{D}$ case above, the general solution of (\ref{EQTh_w_A}) can be written in three equivalent ways, namely:
\be \label{Th_w_sol_A}
\Theta_w (\theta) = \frac{\sinh (C_R \theta )}{C_R^2 C_5} + \tilde{C}_{\theta} \big[ C_4 \sinh (C_R \theta) + C_5 \cosh (C_R \theta) \big]
\ee
or
\be
\Theta_w (\theta) = - \frac{\cosh (C_R \theta )}{C_R^2 C_4} + \tilde{C}_{\theta} \big[ C_4 \sinh (C_R \theta) + C_5 \cosh (C_R \theta) \big]
\ee
or
\be
\Theta_w (\theta) = \frac{1}{2 C_R^2} \left( \frac{\sinh (C_R \theta )}{C_5} - \frac{\cosh (C_R \theta )}{C_4} \right) + \tilde{C}_{\theta} \big[ C_4 \sinh (C_R \theta) + C_5 \cosh (C_R \theta) \big] \,\,\, .
\ee

\section{Equations of motion for the Poincar\'e disk} \label{Sec:Disk}
\setcounter{equation}{0}

In this section our goal will be to find solutions to the equations of motion of the Lagrangian (\ref{Lagr_alpha}) for the case of the Poincar\'e disk. For that purpose, we will first rewrite the Lagrangian in terms of the new coordinates $(u,v,w)$ with the cyclic variable $w$. As already pointed out, this will lead to a significant simplification of the equations that will enable us to find analytical solutions.

Let us begin by summarizing the relevant results, which we have obtained so far for the two-field cosmological model based on the Poincar\'e disk. For $f(\varphi)$ given by (\ref{fD}) with $\alpha = \frac{16}{9}$ as in (\ref{Constants}), the Lagrangian (\ref{Lagr_alpha}) has the form:
\be \label{Lagr_D}
{\cal L} = - 3 a \dot{a}^2 + \frac{a^3 \dot{\varphi}^2}{2} + \frac{4}{3} \,a^3 \sinh^2 \!\left( \sqrt{\frac{3}{8}} \,\varphi \right) \!\dot{\theta}^2 - a^3 V(\varphi, \theta) \,\,\, .
\ee
We found that (\ref{Lagr_D}) has a certain Noether symmetry, when the scalar potential is of the form (\ref{Vpot_tot}), namely:
\be \label{V_D}
V (\varphi , \theta) = V_0 \cosh^2 \!\left( \sqrt{\frac{3}{8}} \,\varphi \right) \coth^{\frac{p}{c}} \!\left( \sqrt{\frac{3}{8}} \,\varphi \right) \left[ C_1 \cos \theta - C_2 \sin \theta \right]^{-\frac{p}{c}} \,\,\, .
\ee
Also, according to (\ref{A_u_sol}), (\ref{Ph_u_sol_gen}), (\ref{Th_u_sol}), (\ref{A_w_sol}), (\ref{Ph_w_sol}) and (\ref{Th_w_sol}), the general form of the new variables $u$, $v$ and $w$, with the latter being the cyclic coordinate corresponding to the Noether symmetry of Section \ref{NoetherSym}, is the following:
\bea \label{NewCoord}
u (a, \varphi, \theta) &=& a^{c_a^u} \left[ \cosh \!\left( \sqrt{\frac{3}{8}} \,\varphi \right) \right]^{\frac{2 c_a^u}{3}} \left[ \coth \!\left( \sqrt{\frac{3}{8}} \,\varphi \right) \right]^{\frac{c_{\theta}^u}{c}} \left( C_1 \cos \theta - C_2 \sin \theta \right)^{-\frac{c^u_{\theta}}{c}} \nn \\
v (a, \varphi, \theta) &=& a^{c_a^v} \left[ \cosh \!\left( \sqrt{\frac{3}{8}} \,\varphi \right) \right]^{\frac{2 c_a^v}{3}} \left[ \coth \!\left( \sqrt{\frac{3}{8}} \,\varphi \right) \right]^{\frac{c_{\theta}^v}{c}} \left( C_1 \cos \theta - C_2 \sin \theta \right)^{-\frac{c^v_{\theta}}{c}} \nn \\
w (a, \varphi, \theta) &=& C_w \,a^{3/2} \,\sinh \!\left( \sqrt{\frac{3}{8}} \,\varphi \right) \sin \theta \,\,\, ,
\eea
where for convenience we have denoted $C_w \equiv - \frac{1}{A C_1} \sqrt{\frac{\beta}{c \,\Phi_0}}$ and have taken $\hat{C}_{\theta} = 0$ in (\ref{Th_w_sol}). Note that, to obtain this expression for the coefficient $C_w$\,, one has to take into account (\ref{phi_0}), as well as the relevant coefficient for $\Phi_1 (\varphi)$ according to (\ref{Constants})-(\ref{Phi_solutions}). Finally, we have labeled the $c_a$ and $c_{\theta}$ constants, characterizing the functions $u (a, \varphi, \theta)$ and $v (a, \varphi, \theta)$\,, with upper $u$ and $v$ indices, respectively, to underline the fact that their values in the two cases are independent of each other.

Now we are ready to rewrite the Lagrangian in terms of the variables $(u,v,w)$ and to study the resulting equations of motion. An important remark is in order, though, before we embark on that investigation. Namely, the Lagrangian (\ref{Lagr_D}) is subject to the Hamiltonian constraint $E_{\cal L} = 0$, where 
\be \label{E_L_def}
E_{\cal L} = \frac{\pd {\cal L}}{\pd \dot{q}^i} \,\dot{q}^i - {\cal L}
\ee
is the energy function corresponding to any point particle Lagrangian ${\cal L} (q^i,\dot{q}^i)$ with generalized coordinates $q^i$. It is well-known that the Hamiltonian $E_{\cal L}$ is conserved on any solutions of the Euler-Lagrange equations, i.e. that for such solutions one has $E_{\cal L} = const$. So imposing the constraint $E_{\cal L} = 0$ (which is equivalent to the first order Einstein equation, often also called Friedman constraint) only results in a relation between the integration constants of the Euler-Lagrange equations; see for example \cite{CdeR}. Instead of just using the Hamiltonian constraint at the end of the computation, in order to eliminate one of the integration constants, it is tempting to try to utilize it from the start, in order to facilitate the search for solutions. However, since this constraint is generally (highly) non-linear, there is no guarantee that it will make a crucial difference for that purpose. In particular, for the cases that we will investigate below, it will turn out not to be useful in our search for analytical solutions.

\subsection{Lagrangian in the new variables}

To obtain the Lagrangian in terms of the new variables $u$, $v$ and $w$, we only need a particular coordinate transformation $(a, \varphi, \theta) \rightarrow (u,v,w)$. Hence, we can choose convenient values for the arbitrary constants $c_a^{u,v}$ and $c_{\theta}^{u,v}$ in (\ref{NewCoord}). Particularly simple (and convenient for comparison with \cite{KKC}) expressions are obtained for the following choices:
\be \label{Choices_consts}
c^u_{\theta} = 0 \quad , \quad c^u_a = 3 \qquad \,\,\, {\rm and} \qquad \,\,\, c^v_{\theta} = - c \,\quad , \quad c^v_a = \frac{3}{2} \qquad .
\ee
Substituting (\ref{Choices_consts}) in (\ref{NewCoord}) gives:\footnote{As mentioned earlier, here we use (\ref{Th_w_sol}), since we are interested in encompassing the special case with $C_2 = 0$. For a discussion of the coordinate transformation and resulting Lagrangian, when using the form of the $\Theta_w$ solution in (\ref{Th_w_sol_s}), see footnote \ref{fn_C1_C2} below.}
\bea \label{CoordTr}
u &=& a^3 \,\cosh^2 \!\left( \sqrt{\frac{3}{8}} \,\varphi \right) \,\,\, , \nn \\
v &=& a^{3/2} \,\sinh \!\left( \sqrt{\frac{3}{8}} \,\varphi \right) (C_1 \cos \theta - C_2 \sin \theta) \,\,\,\, , \nn \\
w &=& C_w \,a^{3/2} \,\sinh \!\left( \sqrt{\frac{3}{8}} \,\varphi \right) \sin \theta \,\,\,\, .
\eea
Note that here we need $C_1 \neq 0$, to ensure that $v$ and $w$ are independent variables. However, this was already tacitly assumed when using the $\Theta_w (\theta)$ solution (\ref{Th_w_sol}) in (\ref{NewCoord}); to allow for $C_1 = 0$, one would have to use the form (\ref{Th_w_sol_c}) instead. 

From now on, we will work with the coordinate transformation (\ref{CoordTr}), whose inverse transformation is:
\bea \label{InvCoordTr}
a &=& \left\{ u - \left[ \frac{1}{C_1^2} \left( \frac{v}{w} + \frac{C_2}{C_w} \right)^2 + \frac{1}{C_w^2} \right] \!w^2 \right\}^{1/3} \,\,\, , \nn \\
\varphi &=& 2 \sqrt{\frac{2}{3}} \,{\rm arccoth} \!\left( \sqrt{\frac{u}{w^2}} \left[ \frac{1}{C_1^2} \left( \frac{v}{w} + \frac{C_2}{C_w} \right)^2 + \frac{1}{C_w^2} \right]^{-1/2} \right) \,\,\, , \nn \\
\theta &=& {\rm arccot} \!\left[ \frac{C_w}{C_1} \!\left( \frac{v}{w} + \frac{C_2}{C_w} \right) \right] \,\,\, .
\eea
Note that, when $\theta = const$\,, the variables $v$ and $w$ coincide up to a constant and the resulting expressions in (\ref{CoordTr}) and (\ref{InvCoordTr}) are consistent with the single-field ones obtained in \cite{KKC}. 

Now, substituting (\ref{InvCoordTr}) in (\ref{Lagr_D})-(\ref{V_D}), we find that in the new variables the Lagrangian is:
\be \label{Lagr_uvw}
{\cal L} = - \frac{1}{3} \frac{\dot{u}^2}{u} + \frac{4}{3} \frac{1}{C_1^2} \,\dot{v}^2 +\frac{4}{3} \frac{1}{C_w^2} \left( 1 + \frac{C_2^2}{C_1^2} \right) \dot{w}^2 + \frac{8}{3} \frac{C_2}{C_1^2 C_w} \, \dot{v} \dot{w} - V_0 \frac{u^{\frac{p}{2c}+1}}{v^{\frac{p}{c}}} \,\,\, .
\ee
As already mentioned above, the single-field case is obtained for $w = const \times v$ and $p=0$. In that case, the Lagrangian (\ref{Lagr_uvw}) is consistent with that in \cite{KKC}. Also, note that the mixed term drops out for $C_2 = 0$, which is exactly the special case relevant for natural inflation as mentioned below (\ref{Vpot_tot}).\footnote{Note that, if we had used the $\Theta_w$ solution in (\ref{Th_w_sol_s}) (still with $\hat{C}_{\theta} = 0$), then the third line of (\ref{CoordTr}) would have been modified to $w = C_w \,a^{3/2} \,\sinh \!\left( \sqrt{\frac{3}{8}} \,\varphi \right) ( C_1 \cos \theta + C_2 \sin \theta )$ with $C_w = - \frac{1}{2 A C_1 C_2} \sqrt{\frac{\beta}{c \Phi_0}}$ and $C_{1,2} \neq 0$. Then, the inverse transformation would be:
\bea 
a &=& \left\{ u - \frac{\left[ C_1^2 \left( w - v C_w \right)^2 + C_2^2 \left( w + v C_w \right)^2 \right]}{4 C_1 C_2^2 C_w^2} \right\}^{1/3} \,\,\, , \nn \\
\varphi &=& 2 \sqrt{\frac{2}{3}} \,{\rm arccoth} \!\left( \frac{2 C_1 C_2 C_w \sqrt{u}}{\sqrt{ C_1^2 \left( w - v C_w \right)^2 + C_2^2 \left( w + v C_w \right)^2}} \right) \,\,\, , \nn \\
\theta &=& {\rm arccot} \!\left[ \frac{C_2 (w + v C_w)}{C_1 (w - v C_w)} \right] \,\,\, .
\eea
That would lead to the following Lagrangian: 
\be
{\cal L} = - \frac{1}{3} \frac{\dot{u}^2}{u} + \frac{1}{3} \frac{(C_1^2+C_2^2)}{C_1^2 C_2^2} \left( \dot{v}^2 + \frac{1}{C_w^2} \dot{w}^2 \right) + \frac{2}{3 C_w} \frac{(C_2^2 - C_1^2)}{C_1^2 C_2^2} \,\dot{v} \dot{w} - V_0 \frac{u^{\frac{p}{2c}+1}}{v^{p/c}} \,\,\, .
\ee
Clearly, in this case, the mixed $\dot{v} \dot{w}$ term would vanish for $C_1 = C_2$. \label{fn_C1_C2}} 

Before we begin looking for solutions, let us underline again that (\ref{Lagr_uvw}) is subject to the Hamiltonian constraint $E_{\cal L} = 0$, where
\be \label{E_L_D_uvw}
E_{\cal L} = \frac{\pd {\cal L}}{\pd \dot{u}} \,\dot{u} + \frac{\pd {\cal L}}{\pd \dot{v}} \,\dot{v} + \frac{\pd {\cal L}}{\pd \dot{w}} \,\dot{w} - {\cal L} \,\,\, ,
\ee
as discussed above.

\subsection{Solutions}

It is convenient to introduce the notation:
\be
m \equiv \frac{p}{c} \,\,\, .
\ee
Then the Euler-Lagrange equations of (\ref{Lagr_uvw}) are:
\bea \label{Syst_wvu}
\left( 1 + \frac{C_2^2}{C_1^2} \right) \ddot{w} + \frac{C_2 C_w}{C_1^2} \,\ddot{v} &=& 0 \,\,\, , \nn \\
V_0 \,m \,\frac{u^{\frac{m}{2}+1}}{v^{m+1}} - \frac{8}{3} \left( \frac{1}{C_1^2} \,\ddot{v} + \frac{C_2}{C_1^2 C_w} \,\ddot{w} \right) &=& 0 \,\,\ , \nn \\
2 u \ddot{u} -\dot{u}^2 - 3 V_0 \left(\frac{m}{2} + 1 \right) \frac{u^{\frac{m}{2}+2}}{v^m} &=& 0 \,\,\, .
\eea
Note that in the single-field limit, which for us is given by $p=0$ (equivalently, $m=0$) and $v = const \times w$, this system is in complete agreement with \cite{KKC}. 

To simplify the system (\ref{Syst_wvu}), let us express $\ddot{w}$ from the first equation, namely
\be \label{ddw}
\ddot{w} = - \frac{C_2 C_w}{C_1^2 + C_2^2} \,\ddot{v} \,\,\, ,
\ee
and introduce the function
\be \label{utilde_def}
\tilde{u} (t) \equiv \sqrt{u (t)} \,\,\, .
\ee
Substituting (\ref{ddw}) and (\ref{utilde_def}) in the second and third equations of (\ref{Syst_wvu}) gives:
\bea \label{Syst_utilde_v}
\ddot{v} - \frac{3}{8} V_0 C_0 \,m \,\frac{\tilde{u}^{m+2}}{v^{m+1}} &=& 0 \,\,\, , \nn \\
\ddot{\tilde{u}} - \frac{3}{8} V_0 (m+2) \frac{\tilde{u}^{m+1}}{v^m} &=& 0 \,\,\, ,
\eea
where for convenience we have denoted $C_0 \equiv C_1^2 + C_2^2$.

Before we begin solving (\ref{Syst_utilde_v}), let us make an important remark. Equation (\ref{ddw}) can be solved immediately for $w$ in terms of $v$. One of the integration constants in this solution is determined by the constant of motion $\Sigma_0$, that is due to the Noether symmetry. Indeed, in general, $\Sigma_0$ is given by:
\be \label{CoM}
\Sigma_0 \equiv \frac{\pd {\cal L}}{\pd \dot{w}} = \frac{8}{3} \frac{1}{C_w^2} \left( 1 + \frac{C_2^2}{C_1^2} \right) \dot{w} + \frac{8}{3} \frac{C_2}{C_1^2 C_w} \,\dot{v} \,\,\, .
\ee
The first equation in (\ref{Syst_wvu}) (equivalently, equation (\ref{ddw})) is precisely the time derivative of (\ref{CoM}), due to the fact that $w$ is a cyclic coordinate. So the general solution for $w$ is:
\be \label{w_D_sol}
w (t) = - \frac{C_2 C_w}{(C_1^2 + C_2^2)} \,v(t) + \hat{\Sigma}_0 \,t + C_0^w \,\,\, ,
\ee
where $\hat{\Sigma}_0 = \frac{3}{8} \frac{C_1^2 C_w^2}{(C_1^2 + C_2^2)} \Sigma_0$ and $C_0^w = const$\,. Hence, using (\ref{w_D_sol}) and (\ref{utilde_def}) in (\ref{E_L_D_uvw}), we have:
\be \label{E_L_D_ut_v}
E_{\cal L} = - \frac{4}{3} \,\dot{\tilde{u}}^2 + \frac{4}{3} \frac{\dot{v}^2}{(C_1^2+C_2^2)} + V_0 \frac{\tilde{u}^{m+2}}{v^m} + \frac{3}{16} \frac{C_1^2 C_w^2}{(C_1^2+C_2^2)} \Sigma_0^2 \,\,\, .
\ee
As alluded to earlier, the constraint $E_{\cal L} = 0$ is highly nonlinear and we have not found it helpful in looking for exact solutions. So we will utilize it only at the end, in order to fix one of the integration constants of the solutions of (\ref{Syst_utilde_v}) that we will manage to find. 

Now let us turn to solving the system (\ref{Syst_utilde_v}). It simplifies significantly for three special choices of $m$, namely $m = 0, -1, - 2$. We will begin by investigating these special cases in order of increasing complexity. 
Finally, we will address the generic case with $m \neq 0,-1,-2$.

\subsubsection{Special cases: $m=0, -1, -2$}

The simplest special cases are $m=0$ and $m=-2$. So we will consider them first, before turning to the $m=-1$ case.

\vspace{0.4cm}
\noindent
$\bullet$ \hspace*{0.03cm}{\bf $m=0$ case:}

\vspace{0.2cm}
\noindent
In this case, the system (\ref{Syst_utilde_v}) reduces to:
\bea
\ddot{v} &=& 0 \,\,\, , \nn \\
\ddot{\tilde{u}} - \frac{3}{4} V_0 \,\tilde{u} &=& 0 \,\,\, ,
\eea
with the general solution:
\bea \label{SingleF-D-sol}
v (t) &=& C_1^v \,t + C_2^v \,\,\, , \nn \\
\tilde{u} (t) &=& C^u_1 \cosh \!\left( \frac{1}{2} \sqrt{3 V_0} \,t \right) + C^u_2 \sinh \!\left( \frac{1}{2} \sqrt{3 V_0} \,t \right) \,\,\, ,
\eea
where $C_{1,2}^u = const$ and $C_{1,2}^v = const$. 

Substituting (\ref{SingleF-D-sol}) in (\ref{E_L_D_ut_v}) with $m=0$ gives:
\be
E_{\cal L} = \left[ (C_1^u)^2 - (C_2^u)^2 \right] V_0 + \frac{4}{3} \,\frac{(C_1^v)^2}{(C_1^2 + C_2^2)} + \frac{3}{16} \frac{C_1^2 C_w^2}{(C_1^2+C_2^2)} \Sigma_0^2 \,\,\, .
\ee
Hence, we can enforce the Hamiltonian constraint $E_{\cal L} = 0$, for example, by taking:
\be \label{HamCon_m0_D}
(C_1^v)^2 = \frac{3}{4} \,(C_1^2 + C_2^2) \left[ (C_2^u)^2 - (C_1^u)^2 \right] V_0 - \frac{9}{64} C_1^2 C_w^2 \Sigma_0^2 \,\,\, .
\ee 

Note that, depending on the choice of integration constants, these $m=0$ solutions can have either $w= const \times v$ (which is the single-field limit) or $w\neq const \times v$. In Appendix \ref{App:Disk} we illustrate genuine two-field trajectories obtained in the latter case for certain values of the integration constants.


\vspace{0.4cm}
\noindent
$\bullet$ \hspace*{0.03cm}{\bf $m=-2$ case:}

\vspace{0.2cm}
\noindent
In this case, (\ref{Syst_utilde_v}) acquires the form:
\bea
\ddot{v} + \frac{3}{4} V_0 C_0 \,v &=& 0 \,\,\, , \nn \\
\ddot{\tilde{u}} &=& 0 \,\,\, ,
\eea
whose general solution is:
\bea \label{uv_m_minus_2_sol}
v (t) &=& C_1^v \sin \!\left( \frac{1}{2} \sqrt{3 V_0 C_0} \,t \right) + C_2^v \cos \!\left( \frac{1}{2} \sqrt{3 V_0 C_0} \,t \right) \,\,\, , \nn \\
\tilde{u} (t) &=& C_1^u \,t + C_2^u \,\,\, .
\eea

Substituting (\ref{uv_m_minus_2_sol}) in (\ref{E_L_D_ut_v}) with $m=-2$, we find:
\be
E_{\cal L} = \left[ (C_1^v)^2 + (C_2^v)^2 \right] V_0 - \frac{4}{3} (C_1^u)^2 + \frac{3}{16} \,\frac{C_1^2 C_w^2}{(C_1^2 + C_2^2)} \,\Sigma_0^2 \,\,\, .
\ee
So, to ensure that $E_{\cal L} = 0$, we can take for instance:
\be
(C_1^u)^2 = \frac{3}{4} \left[ (C_1^v)^2 + (C_2^v)^2 \right] V_0 + \frac{9}{64} \,\frac{C_1^2 C_w^2}{(C_1^2 + C_2^2)} \,\Sigma_0^2 \,\,\, .
\ee

Note that, for $m=-2$ and $C_2 = 0$\,, our scalar potential is of the kind relevant for natural inflation, namely $V \sim \cos^2 \theta$\,. It would be interesting to compare the solution with Noether symmetry obtained here to the considerations of \cite{LWWYA}.


\vspace{0.4cm}
\noindent
$\bullet$ \hspace*{0.03cm}{\bf $m=-1$ case:}

\vspace{0.2cm}
\noindent
In this case, the system (\ref{Syst_utilde_v}) becomes:
\bea \label{Syst_uv_m_minus_one}
\ddot{v} + \frac{3}{8} V_0 C_0 \,\tilde{u} &=& 0 \,\,\, , \nn \\
\ddot{\tilde{u}} - \frac{3}{8} V_0 \,v &=& 0 \,\,\, .
\eea
Denoting for convenience 
\be
Q \equiv \frac{3}{8} V_0 \,\,\, ,
\ee
we find from the first equation:
\be \label{sq_u_v}
\tilde{u} = - \frac{\ddot{v}}{Q C_0} \,\,\, .
\ee
Differentiating (\ref{sq_u_v}) twice gives:
\be \label{udd}
\ddot{\tilde{u}} = - \frac{v^{(4)}}{Q C_0} \,\,\, ,
\ee 
where $v^{(4)} \equiv \frac{d^4 v}{d t^4}$. Substituting (\ref{udd}) in the second equation of (\ref{Syst_uv_m_minus_one}), we end up with:
\be \label{Eqv4}
v^{(4)} + Q^2 C_0 \,v = 0 \,\,\, .
\ee
Recall that $C_0 > 0$ by definition. So the general solution of (\ref{Eqv4}) has the form:
\be \label{v_sol_m_m1}
v (t) = C_1^v \cosh (\omega t) \cos (\omega t) + C_2^v \sinh (\omega t) \cos (\omega t) + C_3^v \cosh (\omega t) \sin (\omega t) + C_4^v \sinh (\omega t) \sin (\omega t)
\ee
with $\omega \equiv \sqrt{\frac{1}{2} Q} \,C_0^{1/4} = \frac{1}{4} \sqrt{3 V_0} \,C_0^{1/4}$\,. Hence, using (\ref{sq_u_v}), the solution for $\tilde{u}(t)$ is:
\be \label{u_sol_m_m1}
\tilde{u} = \tilde{C}_1^v \sinh (\omega t) \sin (\omega t) + \tilde{C}_2^v \cosh (\omega t) \sin (\omega t) - \tilde{C}_3^v \sinh (\omega t) \cos (\omega t) - \tilde{C}_4^v \cosh (\omega t) \cos (\omega t) ,
\ee
where $\tilde{C}_i^v = C_i^v / \sqrt{C_0}$ for $i=1,...,4$.

Using (\ref{v_sol_m_m1}) and (\ref{u_sol_m_m1}) in (\ref{E_L_D_ut_v}) with $m=-1$, we obtain:
\be \label{EL_m_m1}
E_{\cal L} = \frac{(C_2^v C_3^v - C_1^v C_4^v)}{\sqrt{C_1^2+C_2^2}} \,V_0 + \frac{3}{16} \,\frac{C_1^2 C_w^2}{(C_1^2 + C_2^2)} \,\Sigma_0^2 \,\,\, .
\ee
Clearly, we can ensure that $E_{\cal L} = 0$ by choosing appropriately any one of the integration constants $C_{1,...,4}^v$ in terms of the remaining constants in (\ref{EL_m_m1}).

\subsubsection{Generic case: $m \neq 0,-1,-2$} \label{Sec:D_gen_m}

For $m \neq 0,-2$ the two equations in (\ref{Syst_utilde_v}) are always coupled. In principle, one could use a procedure similar to that used for the $m=-1$ case above, in order to reduce the system to a single fourth order ODE. Namely, we can express $\tilde{u}$ from the first equation in terms of $v$ and $\ddot{v}$. Upon differentiating this expression twice, we would obtain $\ddot{\tilde{u}}$ in terms of $v$ and its derivatives up to and including $v^{(4)}$. Finally, substituting the results for $\tilde{u} (v,\ddot{v})$ and $\ddot{\tilde{u}} (v,...,v^{(4)})$ in the second equation of (\ref{Syst_utilde_v}), we would end up with a single 4th order ODE for $v(t)$. However, this equation is generally nonlinear and rather messy. Alternatively, one could substitute the expression for $\tilde{u} (v, \ddot{v})$, resulting from the first equation of (\ref{Syst_utilde_v}), into (\ref{E_L_D_ut_v}) in order to obtain a 3rd order ODE for $v (t)$ from the constraint $E_{\cal L} = 0$. This equation, however, is also highly nonlinear and quite messy. Despite not being able to solve (\ref{Syst_utilde_v}) analytically in full generality, we will nevertheless manage to find particular classes of solutions for any $m<-2$ or $m>0$.

For that purpose, let us first note that the two equations in (\ref{Syst_utilde_v}), together, imply the relation:
\be \label{Rel_tilde_u_v_m}
C_0 \,m \,\ddot{\tilde{u}} \,\tilde{u} = (m+2) \,\ddot{v} \,v \,\,\, .
\ee
So we can view (\ref{Rel_tilde_u_v_m}) and one of the equations in (\ref{Syst_utilde_v}) as the two independent equations to solve. An obvious Ansatz solving (\ref{Rel_tilde_u_v_m}) is 
\be \label{v_ans_u_tilde}
v = \pm \,C_0^{1/2} \sqrt{\frac{m}{m+2}} \,\tilde{u} \,\,\, .
\ee
However, notice that, in order to have real solutions with this Ansatz, we need to assume that $m<-2$ or $m>0$. Substituting (\ref{v_ans_u_tilde}) in any of the two equations of (\ref{Syst_utilde_v}), we end up with:
\be \label{u_D_eq_part}
\ddot{\tilde{u}} \,- \,\frac{3}{8} \,V_0 \,C_0^{-m/2} \,(\pm 1)^m \,(m+2) \left( \frac{m+2}{m} \right)^{\!\!\frac{m}{2}} \tilde{u} \,= \,0 \,\,\, .
\ee
Depending on the sign of the $\tilde{u}$ term\footnote{Note that this sign is correlated with the choice of sign in (\ref{v_ans_u_tilde}).}, the solutions of (\ref{u_D_eq_part}) are:
\be \label{uv_PartSol1}
\tilde{u} (t) = C_1^u \sinh \!\left( \tilde{\omega} t \right) + C_2^u \cosh \!\left( \tilde{\omega} t \right) \qquad {\rm for} \qquad (\pm 1)^m \,(m+2) > 0
\ee
and 
\be \label{uv_PartSol2}
\tilde{u} (t) = C_1^u \sin \!\left( \tilde{\omega} t \right) + C_2^u \cos \!\left( \tilde{\omega} t \right) \qquad {\rm for} \qquad (\pm 1)^m \,(m+2) < 0 \,\,\, ,
\ee
where 
\be
\tilde{\omega} = \frac{1}{2} \sqrt{\frac{3}{2} V_0 C_0^{-m/2} \,|\,(\pm 1)^m \,(m+2)| \left( \frac{m+2}{m} \right)^{\!\!\frac{m}{2}}} \,\,\, .
\ee

Substituting (\ref{v_ans_u_tilde}) and (\ref{uv_PartSol1}) in (\ref{E_L_D_ut_v}), we have:
\be \label{E_L_u_D_sol_Part_I}
E_{\cal L} = V_0 C_0^{-m/2} (\pm 1)^m \!\left( \frac{m+2}{m} \right)^{\!\!\frac{m}{2}} \!\left[ (C_2^u)^2 - (C_1^u)^2 \right] + \frac{3}{16} \frac{C_1^2 C_w^2}{(C_1^2+C_2^2)} \Sigma_0^2 \,\,\, ,
\ee
while using (\ref{v_ans_u_tilde}) and (\ref{uv_PartSol2}) inside (\ref{E_L_D_ut_v}) gives:
\be \label{E_L_u_D_sol_Part_II}
E_{\cal L} = V_0 C_0^{-m/2} (\pm 1)^m \!\left( \frac{m+2}{m} \right)^{\!\!\frac{m}{2}} \!\left[ (C_2^u)^2 + (C_1^u)^2 \right] + \frac{3}{16} \frac{C_1^2 C_w^2}{(C_1^2+C_2^2)} \Sigma_0^2 \,\,\, .
\ee
Clearly, one can always satisfy the Hamiltonian constraint $E_{\cal L} = 0$ for the expression (\ref{E_L_u_D_sol_Part_I}), upon fixing suitably an integration constant. On the other hand, (\ref{E_L_u_D_sol_Part_II}) can vanish only for the minus sign in $(\pm 1)^m$ together with $m$ being odd. In that case, the condition $(\pm 1)^m (m+2) < 0$, together with the earlier requirement $m\in (-\infty , -2) \cup (0, \infty)$, implies that $m>0$. Hence, the system (\ref{Syst_utilde_v}) has particular solutions with $\tilde{u} (t)$ of the form (\ref{uv_PartSol2}), only for the ``$-$" sign of the $v(t)$ expression in (\ref{v_ans_u_tilde}), as well as $m$ odd and positive. 

To illustrate the above considerations, let us write down, for example, the particular solutions for $m=1$:
\bea \label{uv_sol_part_m_eq_plus_1_a}
v (t) &=& \sqrt{\frac{C_0}{3}} \,\,\tilde{u} (t) \,\,\, , \nn \\
\tilde{u} (t) &=& C_1^u \sinh \!\left( \tilde{\omega} t \right) + C_2^u \cosh \!\left( \tilde{\omega} t \right)
\eea
and 
\bea \label{uv_sol_part_m_eq_plus_1_b}
v (t) &=& - \,\sqrt{\frac{C_0}{3}} \,\,\tilde{u} (t) \,\,\, , \nn \\
\tilde{u} (t) &=& C_1^u \sin \!\left( \tilde{\omega} t \right) + C_2^u \cos \!\left( \tilde{\omega} t \right) 
\eea
with \,$\tilde{\omega} = \frac{3}{2} \sqrt{\frac{3^{1/2}}{2} \,V_0 C_0^{-1/2}}$ \,and \,$(C_1^u)^2 = \frac{\sqrt{3}}{16} \frac{C_1^2 C_w^2}{V_0 C_0^{1/2}} \Sigma_0^2 \pm (C_2^u)^2$\,\,, where the ``+" corresponds to (\ref{uv_sol_part_m_eq_plus_1_a}) and the ``$-$" to (\ref{uv_sol_part_m_eq_plus_1_b}).

In view of the $m=-1$ case considered above, relation (\ref{Rel_tilde_u_v_m}) also seems to suggest looking for solutions with an Ansatz of the form $\tilde{u} = const \times \ddot{v}$. Unlike (\ref{v_ans_u_tilde}), however, such an Ansatz would lead to two independent equations for $v$ since it does not solve identically (\ref{Rel_tilde_u_v_m}), but instead brings it in the form $v^{(4)} - const_1 \times v = 0$. Indeed, substituting the same Ansatz $\tilde{u} = const \times \ddot{v}$ in any of the two equations in (\ref{Syst_utilde_v}), would lead to an equation of the form $\ddot{v} + const_2 \times v = 0$\,. Since in general $const_1 \neq const_2^2$\,, the two equations for $v(t)$ would be incompatible. One can ensure $const_1 = const_2^2$ by viewing it as a constraint relating $V_0$, $C_0$ and $m$ and then solving it for one of those constants in terms of the other two. In that case, one would still end up with a solution of the same kind as (\ref{uv_PartSol1}) or (\ref{uv_PartSol2}), but with at least one of the previously arbitrary integration constants now fixed.

\section{Equations of motion for the hyperbolic punctured disk} \label{Sec:PuncDisk}
\setcounter{equation}{0}

In this section, we will look for solutions of the equations of motion for the case of the hyperbolic punctured disk. Let us begin with a summary of the necessary results from the previous sections. 

For the hyperbolic punctured disk, we have $\alpha = \frac{4}{3}$ according to (\ref{Constants}). Hence, using (\ref{fD*}), the Lagrangian (\ref{Lagr_alpha}) acquires the form:
\be \label{Lagr_D*}
{\cal L} = - 3 a \dot{a}^2 + \frac{a^3 \dot{\varphi}^2}{2} + \frac{4}{3} \,a^3 \exp \!\left( -\sqrt{\frac{3}{2}} \,\varphi \right) \!\dot{\theta}^2 - a^3 V(\varphi, \theta) \,\,\, ,
\ee
where the potential is
\be \label{V_tot_D*}
V (\varphi , \theta) = V_0 \,\exp \!\left( \!- \sqrt{\frac{3}{2}} \,\varphi + \frac{p}{2c} e^{\sqrt{\frac{3}{2}} \varphi} \right) \exp \!\left( \,\frac{p}{2 c} \,\theta^2 \right)
\ee
in accord with (\ref{Vtilde_D*}) and (\ref{Vhat_D*}); note that, for technical simplicity, here and in the following we will only consider the $\theta_0 = 0$ case. In addition, from (\ref{A_u_sol}), (\ref{Phi_u_D*}), (\ref{Th_u_D*}), (\ref{A_w_sol}), (\ref{Ph_w_sol}) and (\ref{Th_w_D*}), we have:
\bea \label{NewCoord_D*}
u (a, \varphi, \theta) &=& a^{c_a^u} \exp \!\left[ - \frac{c_a^u}{\sqrt{6}} \,\varphi + \frac{c_{\theta}^u}{2 c} \,\exp \!\left( \sqrt{\frac{3}{2}} \,\varphi \right) \right] \,\exp \!\left( {\frac{c_{\theta}^u}{2c} \,\theta^2} \!\right) \nn \\
v (a, \varphi, \theta) &=& a^{c_a^v} \exp \!\left[ - \frac{c_a^v}{\sqrt{6}} \,\varphi + \frac{c_{\theta}^v}{2 c} \,\exp \!\left( \sqrt{\frac{3}{2}} \,\varphi \right) \right] \,\exp \!\left( {\frac{c_{\theta}^v}{2 c} \,\theta^2} \!\right) \nn \\
w (a, \varphi, \theta) &=& C_w \,a^{3/2} \,\theta \,\exp \!\left( -\sqrt{\frac{3}{8}} \,\varphi \right) \,\,\, ,
\eea
where we have denoted $C_w = - \frac{1}{A C_3} \sqrt{\frac{\beta}{c \Phi_0}}$ and have taken $const = 0$ in (\ref{Th_w_D*}) for convenience. Notice that we used (\ref{phi_0}), (\ref{Ph3_inv_Ph1}) and the $\bb{D}^*$ lines for $\Phi_1 (\varphi)$ in (\ref{Constants})-(\ref{Phi_solutions}), in order to obtain the expression for $C_w$ given above.

\subsection{Lagrangian in the new variables} \label{Subsec:newLagr}

To rewrite the Lagrangian (\ref{Lagr_D*}) in terms of $(u,v,w)$, let us first choose suitably the constants $c_a^{u,v}$ and $c_{\theta}^{u,v}$ in (\ref{NewCoord_D*}). It is convenient to take:
\be \label{Choices_consts_D*}
c^u_{\theta} = 0 \quad , \quad c^u_a = \frac{3}{2} \qquad \,\,\, {\rm and} \qquad \,\,\, c^v_{\theta} = c \,\quad , \quad c^v_a = 0 \qquad .
\ee
Using (\ref{Choices_consts_D*}), the coordinate transformation (\ref{NewCoord_D*}) becomes:
\bea \label{NewCoor_D*_part_c}
u &=& a^{3/2} \exp \!\left( -\sqrt{\frac{3}{8}} \,\varphi \right) \,\,\, , \nn \\
v &=& \exp \!\left( \frac{e^{\sqrt{\frac{3}{2}} \,\varphi} + \theta^2}{2} \right) \,\,\, , \nn \\
w &=& C_w \,a^{3/2} \,\theta \,\exp \!\left( -\sqrt{\frac{3}{8}} \,\varphi\right) \,\,\, ,
\eea
whose inverse transformation has the form:
\bea \label{InvCoordTr_D*}
a &=& u^{2/3} \left( \!2 \ln v - \frac{w^2}{u^2 C_w^2} \right)^{1/3} \,\,\, , \nn \\
\varphi &=& (2/3)^{1/2} \, \ln \!\left( \!2 \ln v - \frac{w^2}{u^2 C_w^2} \right) \,\,\, , \nn \\
\theta &=& \frac{w}{u C_w} \,\,\, .
\eea

Substituting (\ref{InvCoordTr_D*}) in (\ref{Lagr_D*})-(\ref{V_tot_D*}) gives:
\be \label{L_uvw_D*}
{\cal L} = - \frac{8}{3} \,\dot{u}^2 \ln v - \frac{8}{3} \,\dot{u} \,\dot{v} \,\frac{u}{v} + \frac{4}{3} \frac{1}{C_w^2} \,\dot{w}^2 - V_0 u^2 v^m \,\,\, ,
\ee
where for convenience we introduced the notation
\be
m \equiv \frac{p}{c} \,\,\, ,
\ee
as in the previous section. Notice that the Lagrangian (\ref{L_uvw_D*}) can be simplified upon exchanging $v$ for a new variable $\hat{v}$, defined through:
\be \label{v_new_var_hat}
\hat{v} = u \ln v \,\,\, .
\ee
Indeed, equation (\ref{v_new_var_hat}) implies that $v = e^{\hat{v}/u}$. Substituting this in (\ref{L_uvw_D*}), we find:
\be \label{L_u_vhat_w}
{\cal L} = - \frac{8}{3} \,\dot{u} \,\dot{\hat{v}} + \frac{4}{3} \frac{1}{C_w^2} \,\dot{w}^2 - V_0 \,u^2 e^{m \hat{v}/u} \,\,\, .
\ee
Recall also that (\ref{L_u_vhat_w}) is subject to the Hamiltonian constraint $E_{\cal L} = 0$, as discussed in the previous Section.

Before we begin looking for solutions of the equations of motion, it is worth making a couple of remarks. First, one can easily see from (\ref{v_new_var_hat}) that the expression for $\hat{v}$ is not of the form (\ref{u_ans}), and consequently not of the same form as the $u$ and $v$ solutions in (\ref{NewCoord_D*}). Note, however, that the separation of variables Ansatz (\ref{u_ans}) only enables us to find a particular class of solutions of (\ref{EQu}). Furthermore, for any $u$ and $v$ satisfying the latter equation, the expression $u f (v)$, where $f(v)$ is an arbitrary function of $v$, is clearly a solution of (\ref{EQu}) too. 

Finally, let us comment on the single-field case, which is again obtained for $m=0$. At first sight, it might seem that there is a problem, as (\ref{NewCoor_D*_part_c}) implies \,$w = const \times u$ \,for $\theta = const$, whereas the Lagrangian (\ref{L_u_vhat_w}) depends explicitly on $u$, and not on $\hat{v}$, after setting $m=0$. However, this is exactly the correct dependence, since the Lagrangian does not contain the usual kinetic terms for $u$ and $\hat{v}$, but only the mixed $\dot{u} \dot{\hat{v}}$ term. As a result, the $u$-variation gives the $v$-equation of motion and vice-versa. This will become apparent shortly.

\subsection{Solutions} \label{Sec:D*_Sols}

Let us now turn to investigating the equations of motion of the Lagrangian (\ref{L_u_vhat_w}). Clearly, the $w$-equation is:
\be \label{wEq_D*}
\ddot{w} = 0 \,\,\, ,
\ee 
Hence, we immediately have:
\be \label{Punct_d_sol_w}
w (t) = \Sigma_* t + C_0^w \,\,\, ,
\ee
where $C_0^w = const$ and $\Sigma_*$ is the Noether symmetry constant of motion, up to a numerical factor. Substituting (\ref{L_u_vhat_w}) and (\ref{Punct_d_sol_w}) in the general expression (\ref{E_L_def}), we find the Hamiltonian:
\be \label{E_L_D*}
E_{\cal L} = - \frac{8}{3} \,\dot{u} \,\dot{\hat{v}} + V_0 \,u^2 e^{m \hat{v}/u} + \frac{4}{3} \frac{\Sigma_*^2}{C_w^2} \,\,\,\, .
\ee
As in the previous Section, the constraint $E_{\cal L} = 0$ will not turn out to be helpful in finding new analytical solutions, due to its non-linearity. So we will use it only at the end, to fix one of the integration constants of the solutions of the Euler-Lagrange equations that we find.

The $u$ and $\hat{v}$ equations of motion, following from (\ref{L_u_vhat_w}), are:
\bea \label{EQs_u_v_hat}
\ddot{u} - \frac{3}{8} V_0 m \,u \,e^{m \hat{v} / u} &=& 0 \,\,\, , \nn \\
\ddot{\hat{v}} + \frac{3}{8} V_0 (m \hat{v} - 2 u) e^{m \hat{v} / u} &=& 0 \,\,\, .
\eea
Note that, due to the unusual kinetic term, the first equation in (\ref{EQs_u_v_hat}) arises from the $\hat{v}$-variation, i.e. from $\frac{\pd \cal L}{\pd \hat{v}} - \frac{d}{dt} \frac{\pd {\cal L}}{\pd \dot{\hat{v}}} = 0$, while the second one comes from the $u$-variation. Clearly, taking $m=0$ simplifies greatly the system (\ref{EQs_u_v_hat}). So let us consider this case first.

\vspace{0.4cm}
\noindent
$\bullet$ \hspace*{0.03cm}{\bf Special case: $m=0$}

\vspace{0.2cm}
\noindent
In this case, (\ref{EQs_u_v_hat}) acquires the form:
\bea \label{EQs_u_v_hat_m_0}
\ddot{u} &=& 0 \,\,\, , \nn \\
\ddot{\hat{v}} - \frac{3}{4} V_0 \,u &=& 0 \,\,\, .
\eea 
Recall that, as pointed out above, $m=0$ corresponds to the single-field limit. Obviously, in view of (\ref{wEq_D*}), the first equation in (\ref{EQs_u_v_hat_m_0}) is consistent with the single-field identification $w = const \times u$ when $\theta = const$, that we discussed in Subsection \ref{Subsec:newLagr}. The solutions of (\ref{EQs_u_v_hat_m_0}) are:
\bea \label{Punct_d_uv_sol_m_0}
u &=& C_1^* \,t + C_2^* \,\,\, , \nn \\
\hat{v} &=& \frac{1}{8} V_0 C_1^* \,t^3 + \frac{3}{8} V_0 C_2^* \,t^2 + C_3^* \,t + C_4^* \,\,\, ,
\eea
where $C_i^*$ with $i=1,...,4$ are integration constants. Note that this is quite different from the analogous solutions in the Poincar\'e disk case, given in (\ref{SingleF-D-sol}).
It may be worth exploring further what distinguishing features that may lead to for the punctured disk case, even with just one scalar field.

Now, substituting (\ref{Punct_d_uv_sol_m_0}) in (\ref{E_L_D*}) with $m=0$, we obtain:
\be
E_{\cal L} = (C_2^*)^2 \,V_0 - \frac{8}{3} \,C_1^* C_3^* + \frac{4}{3} \frac{\Sigma_*^2}{C_w^2} \,\,\, .
\ee
To ensure that $E_{\cal L} = 0$\,, we can take for example:
\be \label{Ham_Con_m0_D*}
(C_2^*)^2 = \frac{1}{V_0} \left[ \frac{8}{3} \,C_1^* C_3^* - \frac{4}{3} \frac{\Sigma_*^2}{C_w^2} \right] \,\,\, .
\ee

Note that the solutions above can have $w \neq const \times u$, even though $m=0$, depending on the choice of integration constants. In Appendix \ref{App:PuncD} we illustrate such two-field solutions for particular values of the constants.

\vspace{0.4cm}
\noindent
$\bullet$ \hspace*{0.03cm}{\bf Generic case: $m \neq 0$}

\vspace{0.2cm}
\noindent
Now let us consider the generic case with $m \neq 0$\,.\footnote{Note that in this case the local $\theta$-dependence solution (\ref{Th_1_d_star_sol}), (\ref{Th_2_3_d_star_sol}) of the symmetry condition (\ref{SymCond}), as well as the resulting potential (\ref{V_tot_D*}), can be extended globally in field space only on the universal cover of the punctured disk. Nevertheless, solutions on the punctured disk itself also can be physically meaningful, when they lie in a segment $\theta \in [0 , \theta_0]$ with $\theta_0 < 2\pi$. (As one can see in Appendix \ref{App:Traj_m0}, albeit for $m=0$, such solutions are a common occurrence.) Indeed, there could be circumstances, in which the physical configuration space of (\ref{Lagr_alpha}) could be a submanifold (determined by some constraint on $\theta$, for instance) of the manifold parametrized by $\{a, \varphi ,\theta \}$. Finally, even for solutions, which cannot be extended to $t \rightarrow \infty$ on the punctured disk itself (due to global issues in field space), parts with finite-time duration can still be of physical relevance, since the models under consideration are expected to provide appropriate effective descriptions only in a finite time interval. \label{GlobalFS}} Then, one could solve the first equation in (\ref{EQs_u_v_hat}) algebraically for $\hat{v}$, obtaining $\hat{v} = \frac{u}{m} \ln \!\left( \frac{8}{3 V_0 m} \frac{\ddot{u}}{u} \right)$\,. Substituting this expression in the second equation of (\ref{EQs_u_v_hat}), one would find a fourth order ODE for $u(t)$. However, the resulting equation is highly non-linear and thus cannot be solved analytically in full generality. Alternatively, one could substitute $\hat{v} = \frac{u}{m} \ln \!\left( \frac{8}{3 V_0 m} \frac{\ddot{u}}{u} \right)$ in the constraint $E_{\cal L} = 0$, in order to obtain a third order ODE for $u(t)$. This equation, though, is also highly nonlinear and unwieldy. So we will pursue a different route instead. Namely, we will use a certain Ansatz that will enable us to find particular analytical solutions for any $m > 0$\,. 

Notice that, from the first equation in (\ref{EQs_u_v_hat}), we have $e^{m\hat{v}/u} = \frac{8}{3 V_0 m} \frac{\ddot{u}}{u}$\,. Substituting this in the second equation of (\ref{EQs_u_v_hat}), we end up with:
\be \label{muv}
m ( u \ddot{\hat{v}} + \ddot{u} \hat{v} ) - 2 u \ddot{u} = 0 \,\,\, .
\ee
Clearly, one can view (\ref{muv}) and one of (\ref{EQs_u_v_hat}) as the two independent equations to solve. Therefore, an obvious Ansatz, that solves (\ref{muv}) identically, is:
\be \label{vhat_ans}
\hat{v} = \frac{u}{m} \,\,\, .
\ee
Substituting (\ref{vhat_ans}) in any of the two equations in (\ref{EQs_u_v_hat}), we obtain:
\be
\ddot{u}(t) \,- \,\frac{3}{8} V_0 m e \,\, u(t) \,= \,0 \,\,\, .
\ee
The solutions of this equation are:
\be \label{u_sol_Punc_Disk_gen_m_I}
u (t) = C_1^* \sinh (\omega_* t) + C_2^* \cosh (\omega_* t) \qquad {\rm for} \qquad m > 0
\ee
and 
\be \label{u_sol_Punc_Disk_gen_m_II}
u (t) = C_1^* \sin (\omega_* t) + C_2^* \cos (\omega_* t) \qquad {\rm for} \qquad m < 0 \,\,\, ,
\ee
where 
\be
\omega_* = \sqrt{\frac{3}{8} V_0 e |m|} \,\,\, .
\ee

Let us now impose the Hamiltonian constraint. Substituting (\ref{vhat_ans}) and (\ref{u_sol_Punc_Disk_gen_m_I}) in (\ref{E_L_D*}), we obtain:
\be \label{E_L_D*_part_I}
m > 0 \,: \qquad \,\, E_{\cal L} = e V_0 \left[ (C_2^*)^2 - (C_1^*)^2 \right] + \frac{4}{3} \frac{\Sigma_*^2}{C_w^2} \,\,\, ,
\ee
while substituting (\ref{vhat_ans}) and (\ref{u_sol_Punc_Disk_gen_m_II}) in (\ref{E_L_D*}) gives:
\be \label{E_L_D*_part_II}
m < 0 \,: \qquad \,\, E_{\cal L} = e V_0 \left[ (C_2^*)^2 + (C_1^*)^2 \right] + \frac{4}{3} \frac{\Sigma_*^2}{C_w^2} \,\,\, .
\ee
Clearly, one can ensure that (\ref{E_L_D*_part_I}) satisfies the constraint $E_{\cal L} = 0$ by fixing suitably $C_1^*$ or $C_2^*$. On the other hand, the expression (\ref{E_L_D*_part_II}), following from (\ref{u_sol_Punc_Disk_gen_m_II}), is incompatible with the Hamiltonian constraint. Hence, this constraint allows only particular solutions of the form (\ref{u_sol_Punc_Disk_gen_m_I}). 

Note that (\ref{vhat_ans}), together with (\ref{v_new_var_hat}), implies that $v = const$. Nevertheless, this is not a degenerate case, since from (\ref{InvCoordTr_D*}) we can see that all of $a(t)$, $\varphi (t)$ and $\theta (t)$ are nontrivial functions. For $v=const$, however, it is evident that $\varphi$ and $\theta$ become functionally dependent. So this particular solution corresponds to yet another effectively single-field system, although it has $m \neq 0$. It would be very interesting to understand whether there is a deeper reason for this outcome.

\section{Equations of motion for the hyperbolic annuli} \label{Sec:Annulus}
\setcounter{equation}{0}

In this section, we turn to finding analytical solutions of the equations of motion for the hyperbolic annuli case. As before, we begin by summarizing the relevant results from Sections \ref{NoetherSym} and \ref{Sec:NewVar}.

In the $\bb{A}$ case, we have from (\ref{Constants}) that $\alpha = \frac{4}{3}$\,. Using this and (\ref{fA}), we find that the Lagrangian (\ref{Lagr_alpha}) becomes:
\be \label{Lagr_A}
{\cal L} = - 3 a \dot{a}^2 + \frac{a^3 \dot{\varphi}^2}{2} + \frac{4}{3} \,a^3 C_R^2 \cosh^2 \!\left( \sqrt{\frac{3}{8}} \,\varphi \right) \!\dot{\theta}^2 - a^3 V(\varphi, \theta)
\ee
with potential given by:
\be \label{Vpot_A}
V (\varphi , \theta) = V_0 \sinh^2 \!\left( \sqrt{\frac{3}{8}} \,\varphi \right) \coth^{\frac{p}{c C_R^2}} \!\left( \sqrt{\frac{3}{8}} \,\varphi \right) \left[ C_4 \sinh (C_R \theta) + C_5 \cosh (C_R \theta) \right]^{\frac{p}{c C_R^2}} \, ,
\ee
according to (\ref{Vtilde_A}) and (\ref{Vhat_A}). In addition, the new variables $u$, $v$ and $w$, with $w$ being the cyclic coordinate, have the form:
\bea \label{NewCoord_A}
u (a,\varphi,\theta) \!\!\!&=& \!\!\!a^{c_a^u} \!\left[ \,\sinh \!\left( \!\sqrt{\frac{3}{8}} \,\varphi \!\right) \right]^{\frac{2 c_a^u}{3}} \left[ \coth \!\left( \!\sqrt{\frac{3}{8}} \,\varphi \!\right) \right]^{\frac{c_{\theta}^u}{c C_R^2}} \left[ C_4 \sinh (C_R \theta) + C_5 \cosh (C_R \theta) \right]^{\frac{c_{\theta}^u}{c C_R^2}} \nn \\
v (a,\varphi,\theta) \!\!\!&=& \!\!\!a^{c_a^v} \!\left[ \,\sinh \!\left( \!\sqrt{\frac{3}{8}} \,\varphi \!\right) \right]^{\frac{2 c_a^v}{3}} \left[ \coth \!\left( \!\sqrt{\frac{3}{8}} \,\varphi \!\right) \right]^{\frac{c_{\theta}^v}{c C_R^2}} \left[ C_4 \sinh (C_R \theta) + C_5 \cosh (C_R \theta) \right]^{\frac{c_{\theta}^v}{c C_R^2}} \nn \\
w (a,\varphi,\theta) \!\!\!&=& \!\!\!C_w \,a^{3/2} \cosh \!\left( \!\sqrt{\frac{3}{8}} \,\varphi \!\right) \sinh \!\left( C_R \theta \right) \,\,\, ,
\eea
where we have used (\ref{A_u_sol}), (\ref{Phi_u_A}), (\ref{Th_u_A}), (\ref{A_w_sol}), (\ref{Ph_w_sol}) and (\ref{Th_w_sol_A}). Also, for convenience we have denoted $C_w \equiv - \frac{1}{A C_R C_5} \sqrt{\frac{\beta}{c \Phi_0}}$ and have taken $\tilde{C}_{\theta} = 0$ in (\ref{Th_w_sol_A}). Finally, note that, similarly to Sections \ref{Sec:Disk} and \ref{Sec:PuncDisk}, we have obtained the expression for $C_w$ here by using (\ref{Ph3_inv_Ph1}), the $\bb{A}$ lines for $\Phi_1 (\varphi)$ in (\ref{Constants})-(\ref{Phi_solutions}), (\ref{Ph_w_sol}), (\ref{phi_0}) and (\ref{Th_w_sol_A}).

\subsection{Lagrangian in the new variables}

In the hyperbolic annuli case, it is convenient to choose the constants, defining the coordinate transformation $(a,\varphi,\theta) \rightarrow (u,v,w)$, as follows:
\be \label{Choices_consts_A}
c^u_{\theta} = 0 \,\quad , \quad c^u_a = \frac{3}{2} \qquad \,\,\, {\rm and} \qquad \,\,\, c^v_{\theta} = c \,C_R^2 \,\quad , \quad c^v_a = \frac{3}{2} \qquad .
\ee
Substituting (\ref{Choices_consts_A}) in (\ref{NewCoord_A}), we find:
\bea \label{NewCoor_A_part_c}
u &=& a^{3/2} \sinh \!\left( \sqrt{\frac{3}{8}} \,\varphi \right) \,\,\, , \nn \\
v &=& a^{3/2} \cosh \!\left( \sqrt{\frac{3}{8}} \,\varphi \right) \left[ C_4 \sinh (C_R \theta) + C_5 \cosh (C_R \theta) \right] \,\,\, , \nn \\
w &=& C_w \,a^{3/2} \cosh \!\left( \sqrt{\frac{3}{8}} \,\varphi \right) \sinh (C_R \theta) \,\,\, .
\eea
Note that, to have independent functions for $v$ and $w$, we need $C_5 \neq 0$ in (\ref{NewCoor_A_part_c}). However, we have already assumed that by choosing to use inside (\ref{NewCoord_A}) the form of the $\Theta_w (\theta)$ solution, given by (\ref{Th_w_sol_A}).

The inverse of the transformation (\ref{NewCoor_A_part_c}) is:
\bea \label{InvCoordTr_A}
a &=& \left\{ \left[ \frac{1}{C_5^2} \left( \frac{v}{w} - \frac{C_4}{C_w} \right)^2 - \frac{1}{C_w^2} \right] \!w^2 - u^2 \right\}^{1/3} \,\,\, , \nn \\
\varphi &=& 2 \sqrt{\frac{2}{3}} \,{\rm arccoth} \!\left( \frac{w}{u} \left[ \frac{1}{C_5^2} \left( \frac{v}{w} - \frac{C_4}{C_w} \right)^2 - \frac{1}{C_w^2} \right]^{1/2} \right) \,\,\, , \nn \\
\theta &=& \frac{1}{C_R} \,{\rm arccoth} \!\left[ \frac{C_w}{C_5} \!\left( \frac{v}{w} - \frac{C_4}{C_w} \right) \right] \,\,\, .
\eea
Using (\ref{InvCoordTr_A}) in (\ref{Lagr_A}) and (\ref{Vpot_A}), we obtain the following action:
\be \label{Lagr_uvw_A}
{\cal L} = \frac{4}{3} \,\dot{u}^2 - \frac{4}{3} \frac{1}{C_5^2} \,\dot{v}^2 +\frac{4}{3} \frac{1}{C_w^2} \left( 1 - \frac{C_4^2}{C_5^2} \right) \dot{w}^2 + \frac{8}{3} \frac{C_4}{C_5^2 C_w} \, \dot{v} \dot{w} - V_0 \frac{v^m}{u^{m-2}} \,\,\, ,
\ee
where for convenience we have denoted
\be
m = \frac{p}{c C_R^2} \,\,\, .
\ee
Note that for \,$C_4 = \pm C_5$ \,the $\dot{w}^2$ term in (\ref{Lagr_uvw_A}) drops out, whereas for $C_4 = 0$ the mixed $\dot{v} \dot{w}$ term vanishes. Finally, recall also that the Lagrangian (\ref{Lagr_uvw_A}) is subject to the Hamiltonian constraint $E_{\cal L} = 0$. Due to its non-linearity, this constraint again will be of practical use only for fixing an integration constant of the solutions of the Euler-Lagrange equations. 

\subsection{Solutions} \label{Subsec:Sols_A}

Let us now look for solutions of the equations of motion of (\ref{Lagr_uvw_A}). In order to keep the $\dot{w}^2$ term in the Lagrangian, we will assume that $C_4^2 \neq C_5^2$\,.\footnote{We will comment on the degenerate $C_4 = \pm C_5$ case in an appropriate place below.} Then the $w$-equation immediately gives:
\be \label{w_sol_A_C_hat_non_zero}
\ddot{w} = - \frac{C_4 C_w}{C_5^2 - C_4^2} \,\ddot{v} \,\,\, .
\ee
The solution of the latter is 
\be \label{w_D*_sol}
w (t) = - \frac{C_4 C_w}{(C_5^2 - C_4^2)} \,v(t) + \hat{\Sigma}_0 \,t + C_0^w \,\,\, ,
\ee
where $C_0^w = const$ and $\hat{\Sigma}_0 = \frac{3}{8} \frac{C_5^2 C_w^2}{(C_5^2 - C_4^2)} \Sigma_0$ with $\Sigma_0$ being the Noether symmetry constant of motion. Substituting (\ref{Lagr_uvw_A}) and (\ref{w_D*_sol}) in the general expression (\ref{E_L_def}), we obtain the Hamiltonian:
\be \label{E_L_D*_u_v}
E_{\cal L} = \frac{4}{3} \,\dot{u}^2 - \frac{4}{3} \frac{\dot{v}^2}{(C_5^2-C_4^2)} + V_0 \frac{v^m}{u^{m-2}} + \frac{3}{16} \frac{C_5^2 C_w^2}{(C_5^2-C_4^2)} \Sigma_0^2 \,\,\, .
\ee

Using (\ref{w_sol_A_C_hat_non_zero}), we find that the $u$ and $v$ Euler-Lagrange equations acquire the form:
\bea \label{Syst_uv_A}
\ddot{v} - \frac{3}{8} V_0 \hat{C}_0 m \frac{v^{m-1}}{u^{m-2}} &=& 0 \,\,\, , \nn \\
\ddot{u} - \frac{3}{8} V_0 (m-2) \frac{v^m}{u^{m-1}} &=& 0 \,\,\, ,
\eea
where we have denoted $\hat{C}_0 \equiv C_5^2 - C_4^2$. One can easily notice that the system (\ref{Syst_uv_A}) becomes exactly the same as (\ref{Syst_utilde_v}) under the simultaneous formal substitutions $m \rightarrow -m$ and $V_0 \rightarrow - V_0$. However, we would like to keep $V_0 > 0$, in order to have a positive-definite scalar potential. So we will view (\ref{Syst_uv_A}) as a different system, albeit quite similar to (\ref{Syst_utilde_v}). Clearly, the special choices of $m$, that simplify significantly (\ref{Syst_uv_A}), are $m = 0, 1, 2$. Let us consider them first, before turning to the generic case with $m \neq 0,1,2$\,.\footnote{Note that, for all $m \neq 0$ solutions below, the same kind of remark applies as in footnote \ref{GlobalFS}.}

\subsubsection{Special cases: $m = 0,1,2$} \label{Sec:A_special_cases}

As in Section \ref{Sec:Disk}, we begin with the simplest cases, namely $m=0$ and $m=2$. 

\vspace{0.4cm}
\noindent
$\bullet$ \hspace*{0.03cm}{\bf $m=0$ case:}

\vspace{0.2cm}
\noindent
From (\ref{Syst_uv_A}), we now have:
\bea
\ddot{v} &=& 0 \,\,\, , \nn \\
\ddot{u} + \frac{3}{4} V_0 \,u &=& 0 \,\,\, .
\eea
Hence, the solutions are:
\bea \label{uv_sol_A_m_0}
v (t) &=& C_1^v \,t + C_2^v \,\,\, , \nn \\
u (t) &=& C_1^u \sin \!\left( \frac{1}{2} \sqrt{3 V_0} \,t \right) + C_2^u \cos \!\left( \frac{1}{2} \sqrt{3 V_0} \,t \right) \,\,\, .
\eea
Substituting (\ref{uv_sol_A_m_0}) in (\ref{E_L_D*_u_v}) gives:
\be
E_{\cal L} = \left[ (C_1^u)^2 + (C_2^u)^2 \right] V_0 - \frac{4}{3} \,\frac{(C_1^v)^2}{(C_5^2 - C_4^2)} + \frac{3}{16} \frac{C_5^2 C_w^2}{(C_5^2-C_4^2)} \Sigma_0^2 \,\,\, .
\ee
Hence, to impose the constraint $E_{\cal L} = 0$, we can take for instance:
\be \label{Ham_con_A_m0_gen}
(C_1^v)^2 = \frac{3}{4} \,(C_5^2 - C_4^2) \left[ (C_1^u)^2 + (C_2^u)^2 \right] V_0 + \frac{9}{64} C_5^2 C_w^2 \Sigma_0^2 \,\,\, .
\ee 

Note that these $m=0$ solutions can have either $w = const \times v$ (single field limit) or $w \neq const \times v$ (genuine two-field case), depending on how the integration constants are chosen. In Appendix \ref{App:Annuli} we illustrate such genuine two-field solutions for certain values of the constants.

\vspace{0.4cm}
\noindent
$\bullet$ \hspace*{0.03cm}{\bf $m=2$ case:}

\vspace{0.2cm}
\noindent
In this case, (\ref{Syst_uv_A}) gives:
\bea
\ddot{v} - \frac{3}{4} V_0 \hat{C}_0 \,v &=& 0 \,\,\, , \nn \\
\ddot{u} &=& 0 \,\,\, .
\eea
Obviously, then, the solution for $u(t)$ is:
\be \label{uA_m2_s}
u(t) = C_1^u \,t + C_2^u \,\,\, .
\ee
However, unlike $C_0$ in Section \ref{Sec:Disk}, $\hat{C}_0$ here can have either sign. So we have the following two cases for $v(t)$:
\be \label{vA_m2_s1}
v (t) = C_1^v \sinh \!\left( \frac{1}{2} \sqrt{3 V_0 \hat{C}_0} \,t \right) + C_2^v \cosh \!\left( \frac{1}{2} \sqrt{3 V_0 \hat{C}_0} \,t \right) \qquad {\rm for} \qquad \hat{C}_0 > 0 \quad ,
\ee
\be \label{vA_m2_s2}
v (t) = C_1^v \sin \!\left( \frac{1}{2} \sqrt{3 V_0 |\hat{C}_0|} \,t \right) + C_2^v \cos \!\left( \frac{1}{2} \sqrt{3 V_0 |\hat{C}_0|} \,t \right) \qquad {\rm for} \qquad \hat{C}_0 < 0 \quad .
\ee
Using (\ref{uA_m2_s}), together with either (\ref{vA_m2_s1}) or (\ref{vA_m2_s2}), inside (\ref{E_L_D*_u_v}) gives:
\be
E_{\cal L} = \left[ (C_2^v)^2 - (C_1^v)^2 \right] \!V_0 + \frac{4}{3} (C_1^u)^2 + \frac{3}{16} \frac{C_5^2 C_w^2}{(C_5^2-C_4^2)} \Sigma_0^2 \,\,\, . \nn \\
\ee
Clearly, we can ensure that $E_{\cal L} = 0$ by fixing suitably one of the integration constants in the last expression.


\vspace{0.4cm}
\noindent
$\bullet$ \hspace*{0.03cm}{\bf $m=1$ case:}

\vspace{0.2cm}
\noindent
Now we obtain from (\ref{Syst_uv_A}):
\bea \label{Syst_uv_A_m_1}
\ddot{v} - \frac{3}{8} V_0 \hat{C}_0 \,u &=& 0 \,\,\, , \nn \\
\ddot{u} + \frac{3}{8} V_0 \,v &=& 0 \,\,\, .
\eea
The system (\ref{Syst_uv_A_m_1}) can be reduced to a single ODE by expressing $u$ from the first equation, namely:
\be \label{u_hatC}
u = \frac{\ddot{v}}{Q \hat{C}_0} \qquad {\rm with} \qquad Q \equiv \frac{3}{8} V_0 \,\,\, ,
\ee
and substituting this result in the second equation. One then finds:
\be \label{Eq_v_4d}
v^{(4)} + Q^2 \hat{C}_0 v = 0 \,\,\, .
\ee
Depending on the sign of $\hat{C}_0$, equation (\ref{Eq_v_4d}) has the following solutions: 
\begin{itemize}
\item[{\tiny{$\blacksquare$}}] For $\hat{C}_0 > 0$ we have:
\bea \label{v_sol_hatC_pos}
v (t) &=& C_1^v \cosh (\omega t) \cos (\omega t) + C_2^v \sinh (\omega t) \cos (\omega t) + C_3^v \cosh (\omega t) \sin (\omega t) \nn \\
&+& C_4^v \sinh (\omega t) \sin (\omega t) \,\,\, ,
\eea
where $\omega = \sqrt{\frac{1}{2} Q \hat{C}_0^{1/2}}$. Substituting (\ref{v_sol_hatC_pos}), as well as the resulting $u (t)$ from (\ref{u_hatC}), into (\ref{E_L_D*_u_v}) gives:
\be \label{E_L_v_sol_hatC_pos}
E_{\cal L} = \frac{(C_1^v C_4^v - C_2^v C_3^v)}{\sqrt{C_5^2-C_4^2}} \,V_0 + \frac{3}{16} \,\frac{C_5^2 C_w^2}{(C_5^2 - C_4^2)} \,\Sigma_0^2 \,\,\, .
\ee
\item[{\tiny{$\blacksquare$}}] For $\hat{C}_0 < 0$, the solution of (\ref{Eq_v_4d}) is:
\be \label{v_sol_hatC_neg}
v (t) = C_1^v \sin (\hat{\omega} t) + C_2^v \cos (\hat{\omega} t) + C_3^v \sinh (\hat{\omega} t) + C_4^v \cosh (\hat{\omega} t) \,\,\, ,
\ee
where $\hat{\omega} = \sqrt{Q \,|\hat{C}_0|^{1/2}}$. Now (\ref{E_L_D*_u_v}) becomes:
\be \label{E_L_v_sol_hatC_neg}
E_{\cal L} = \frac{\left[ (C_1^v)^2 + (C_2^v)^2 + (C_3^v)^2 - (C_4^v)^2 \right]}{\sqrt{C_4^2-C_5^2}} \,V_0 + \frac{3}{16} \,\frac{C_5^2 C_w^2}{(C_5^2 - C_4^2)} \,\Sigma_0^2 \,\,\, .
\ee
\end{itemize}
Clearly, one can ensure that both (\ref{E_L_v_sol_hatC_pos}) and (\ref{E_L_v_sol_hatC_neg}) satisfy the Hamiltonian constraint $E_{\cal L} = 0$ by fixing appropriately an integration constant.


\vspace{0.4cm}
\noindent
{\bf Remark on $\hat{C}_0 = 0$\,:}

\vspace{0.2cm}
\noindent
As noted in the beginning of Section \ref{Subsec:Sols_A}, when $\hat{C}_0 \equiv C_5^2 - C_4^2 = 0$, one cannot use equation (\ref{w_sol_A_C_hat_non_zero}). Instead, the $w$ equation of motion gives $\ddot{v}=0$, with the solution
\be \label{v_A_sol_m0_Ch_0_0}
v(t) = \hat{C}_1 t + \hat{C}_2
\ee
for any $m$. Then, the remaining two equations of motion acquire the form:
\bea
\ddot{w} + \frac{3}{8} V_0 m \frac{C_w C_5^2}{C_4} \frac{v^{m-1}}{u^{m-2}} &=& 0 \,\,\, . \nn \\
\ddot{u} - \frac{3}{8} V_0 (m-2) \frac{v^m}{u^{m-1}} &=& 0 \label{udd_eq_A_C_hat_zero} \,\,\, .
\eea
So we have the following $u (t)$ and $w (t)$ solutions:
\begin{itemize}
\item $m=0$: \,In this case, the solutions of (\ref{udd_eq_A_C_hat_zero}) are:
\bea \label{uw_A_sol_m0_Ch_0_0}
u &=& \hat{C}_3 \sin \left(\frac{1}{2} \sqrt{3 V_0} \,t \right) + \hat{C}_4 \cos \left(\frac{1}{2} \sqrt{3 V_0} \,t \right) \nn \\
w &=& \hat{C}_5 \,t + \hat{C}_6 \,\,\, .
\eea
Evaluating the Hamiltonian on these solutions gives:
\be
E_{\cal L} = \left( \hat{C}_3^2 + \hat{C}_4^2 \right) V_0 - \frac{3}{16} C_w^2 \Sigma_0^2 + \hat{C}_5 \Sigma_0 \,\,\, ,
\ee
where we have used that $\hat{C}_1 = \frac{3}{8} \frac{C_5^2 C_w}{C_4} \Sigma_0$ with $\Sigma_0$ being the Noether symmetry constant of motion. 
\item $m=2$: \,Now (\ref{udd_eq_A_C_hat_zero}) has the following solutions:
\bea
u(t) &=& \hat{C}_3 \,t + \hat{C}_4 \nn \\
w(t) &=& - \frac{3}{4} V_0 \frac{C_w C_5^2}{C_4} \left( \frac{1}{6} \hat{C}_1 t^3 +\frac{1}{2} \hat{C}_2 t^2 \right) + \hat{C}_5 t + \hat{C}_6 \,\,\, .
\eea
Thus, the Hamiltonian becomes:
\be
E_{\cal L} = \hat{C}_2^2 V_0 + \frac{4}{3} \hat{C}_3^2 - \frac{3}{16} C_w^2 \Sigma_0^2 + \hat{C}_5 \Sigma_0 \,\,\, , 
\ee
where again we have used $\hat{C}_1 = \frac{3}{8} \frac{C_5^2 C_w}{C_4} \Sigma_0$. 
\item $m=1$: \,In this case, the solutions of (\ref{udd_eq_A_C_hat_zero}) are given by:
\bea
u &=& - \frac{1}{16} V_0 \hat{C}_1 \,t^3 - \frac{3}{16} V_0 \hat{C}_2 \,t^2 + \hat{C}_3 \,t + \hat{C}_4 \\
w &=& \frac{3}{128} \frac{V_0 C_w C_5^2}{C_4} \left( \frac{1}{20} V_0 \hat{C}_1 t^5 + \frac{1}{4} V_0 \hat{C}_2 t^4 - \frac{8}{3} \hat{C}_3 t^3 - 8 \hat{C}_4 t^2 \right) + \hat{C}_5 t + \hat{C}_6 \,\,\, . \nn
\eea
Hence the Hamiltonian gives:
\be 
E_{\cal L} = \hat{C}_2 \hat{C}_4 V_0 + \frac{4}{3} \hat{C}_3^2 - \frac{3}{16} C_w^2 \Sigma_0^2 + \hat{C}_5 \Sigma_0 \,\,\, ,
\ee
where we have substituted $\hat{C}_1 = \frac{3}{8} \frac{C_5^2 C_w}{C_4} \Sigma_0$\,.
\end{itemize}
\noindent
Obviously, in all three special cases with $\hat{C}_0 = 0$ one can satisfy the Hamiltonian constraint $E_{\cal L} = 0$ by fixing suitably one of the integration constants. 

\subsubsection{Generic case: $m \neq 0,1,2$}

This case is similar to that considered in Subsection \ref{Sec:D_gen_m}. More precisely, for arbitrary $m$ it is not possible to find the exact solutions of equations (\ref{Syst_uv_A}) in full generality. However, just as in Section \ref{Sec:D_gen_m}, we will be able to find particular classes of exact solutions. Unlike there though, here we will find solutions for any $m \neq 0,2$.

To begin, let us observe that, together, the two equations in (\ref{Syst_uv_A}) imply the relation:
\be \label{hatC_0_uv_rel}
\hat{C}_0 \,m \,\ddot{u} \,u = (m-2) \,\ddot{v} \,v \,\,\, .
\ee
One can take (\ref{hatC_0_uv_rel}) and one of (\ref{Syst_uv_A}) as the two independent equations. An Ansatz that solves (\ref{hatC_0_uv_rel}) is given by:
\be \label{v_ans_hatC_0_u}
v \,= \,\pm \,\sqrt{\frac{\hat{C}_0 \,m}{(m-2)}} \,\, u \,\,\,\, .
\ee
To obtain real and nontrivial solutions from this Ansatz, we need that $\frac{\hat{C}_0 \,m}{(m-2)} > 0$, or in other words:
\be \label{A_part_sol_Cond_1}
\hat{C}_0 > 0 \qquad {\rm and} \qquad m \in (-\infty,0) \cup (2,\infty)
\ee
or
\be \label{hatC_neg_m_int}
\hat{C}_0 < 0 \qquad {\rm and} \qquad m \in (0,2) \,\,\, .
\ee
Substituting (\ref{v_ans_hatC_0_u}) in either equation of (\ref{Syst_uv_A}), one finds:
\be \label{uEq_gen_m_part}
\ddot{u} - \frac{3}{8} V_0 (\pm 1)^m (m-2) \left( \frac{\hat{C}_0 \,m}{m-2} \right)^{\!\frac{m}{2}} u \, = \, 0 \,\,\, .
\ee
The solutions of (\ref{uEq_gen_m_part}) depend on the sign of the combination $(\pm 1)^m (m-2)$, namely:
\be \label{A_u_part_sol_1}
u (t) = C_1^u \sinh (\hat{\omega} t) + C_2^u \cosh (\hat{\omega} t) \qquad {\rm for} \qquad (\pm 1)^m (m-2) > 0
\ee
and 
\be \label{A_u_part_sol_2}
u (t) = C_1^u \sin (\hat{\omega} t) + C_2^u \cos (\hat{\omega} t) \qquad {\rm for} \qquad (\pm 1)^m (m-2) < 0 \,\,\, ,
\ee
where 
\be
\hat{\omega} = \frac{1}{2} \sqrt{\frac{3}{2} V_0 \,|\,(\pm 1)^m \,(m-2)| \left( \frac{\hat{C}_0\,m}{m-2} \right)^{\!\frac{m}{2}}} \,\,\, .
\ee
Note that the $(\pm 1)$ in (\ref{uEq_gen_m_part}) is correlated with the $\pm$ sign in (\ref{v_ans_hatC_0_u}).

Let us now impose the Hamiltonian constraint. Substituting (\ref{v_ans_hatC_0_u}) and (\ref{A_u_part_sol_1}) into (\ref{E_L_D*_u_v}) gives:
\be \label{E_L_u_A_sol_Part_I}
E_{\cal L} = V_0 \,(\pm 1)^m \!\left( \frac{\hat{C}_0 \,m}{m-2} \right)^{\!\!\frac{m}{2}} \!\left[ (C_2^u)^2 - (C_1^u)^2 \right] + \frac{3}{16} \frac{C_5^2 C_w^2}{(C_5^2-C_4^2)} \Sigma_0^2 \,\,\, ,
\ee
whereas using (\ref{v_ans_hatC_0_u}) and (\ref{A_u_part_sol_2}) in (\ref{E_L_D*_u_v}) leads to:
\be \label{E_L_u_A_sol_Part_II}
E_{\cal L} = V_0 \,(\pm 1)^m \!\left( \frac{\hat{C}_0\,m}{m-2} \right)^{\!\!\frac{m}{2}} \!\left[ (C_2^u)^2 + (C_1^u)^2 \right] + \frac{3}{16} \frac{C_5^2 C_w^2}{(C_5^2-C_4^2)} \Sigma_0^2 \,\,\, .
\ee
Clearly, there is no problem to satisfy $E_{\cal L} = 0$ for the expression in (\ref{E_L_u_A_sol_Part_I}), by choosing suitably $C_1^u$ or $C_2^u$. On the other hand, for (\ref{E_L_u_A_sol_Part_II}) a more careful discussion is needed. Unlike in Sections \ref{Sec:Disk} and \ref{Sec:PuncDisk}, now the $\Sigma_0^2$ term can have either sign. Let us consider first $\hat{C}_0 \equiv C_5^2 - C_4^2 > 0$. In that case, either $m < 0$ or $m>2$; see (\ref{A_part_sol_Cond_1}). Only $m>2$, however, can ensure $E_{\cal L} = 0$. The reason is that, since the $\Sigma_0^2$ term in (\ref{E_L_u_A_sol_Part_II}) is positive, we need the $V_0$ term to be negative, which can only be achieved for $(\pm 1)^m = -1$. Then, the condition $(\pm 1)^m (m-2) < 0$ in (\ref{A_u_part_sol_2}) implies that $m>2$. Now let us consider $\hat{C}_0 < 0$, in which case $0<m<2$ according to (\ref{hatC_neg_m_int}). Hence the condition $(\pm 1)^m (m-2) < 0$ implies that $(\pm 1)^m = + 1$, which is exactly what is needed to have a positive $V_0$ term in (\ref{E_L_u_A_sol_Part_II}), when the $\Sigma_0^2$ term is negative. 

To summarize, the Hamiltonian constraint allows solutions with $u(t)$ as in (\ref{A_u_part_sol_2}) only in the following two parts of the parameter space:

\hspace*{0.7cm} 1) $\hat{C}_0 > 0$\,: \,$m$ odd and $m > 2$, together with the minus sign in (\ref{v_ans_hatC_0_u}).

\hspace*{0.7cm} 2) $\hat{C}_0 < 0$\,: \,$0 < m < 2$ and a plus sign in (\ref{v_ans_hatC_0_u}).

As an example of the above considerations, let us write down the particular solutions for $m=3$ and $\hat{C}_0 > 0$. In that case, the $+$ sign in (\ref{v_ans_hatC_0_u}) leads to the solution
\bea
v (t) &=& \sqrt{3 \hat{C}_0} \,u (t) \nn \\
u (t) &=& C_1^u \sinh (\hat{\omega} t) + C_2^u \cosh (\hat{\omega} t) \,\,\, ,
\eea
while the $-$ sign gives:
\bea
v (t) &=& - \sqrt{3 \hat{C}_0} \,u (t) \nn \\
u (t) &=& C_1^u \sin (\hat{\omega} t) + C_2^u \cos (\hat{\omega} t) \,\,\, ,
\eea
where $\hat{\omega} = \frac{3}{2} \sqrt{\frac{3^{1/2}}{2} V_0 \hat{C}_0^{3/2}}$\,.

As a final remark note that, among the particular solutions above, there are solutions with $m=1$, obtained for $\hat{C}_0 < 0$ as can be seen from (\ref{hatC_neg_m_int}). However, we already found the most general solution with $m=1$ and $\hat{C}_0 < 0$ in Subsection \ref{Sec:A_special_cases}, namely (\ref{v_sol_hatC_neg}) together with (\ref{u_hatC}). Hence, the latter must contain as special cases the particular $m=1$ solutions coming from (\ref{A_u_part_sol_1}) and (\ref{A_u_part_sol_2}), together with (\ref{v_ans_hatC_0_u}). One can verify that this is indeed the case, upon setting to zero either the pair $C_{1,2}^v$ or the pair $C_{3,4}^v$ of integration constants in (\ref{v_sol_hatC_neg}).

\section{Summary and discussion} \label{Sec:Discussion}
\setcounter{equation}{0}

We studied two-field cosmological $\alpha$-attractors whose scalar
manifold is any elementary hyperbolic surface. We imposed the
requirement that these models have a Noether symmetry and found those
solutions of the symmetry conditions which follow from a
separation-of-variables Ansatz. In particular, we showed that such
separated Noether symmetries exist only for a certain value of the
parameter $\alpha$. To prove these results, we rewrote the
cosmologically relevant Lagrangian in canonical form, i.e. as ${\cal
  L} (q^i,\dot{q}^i)$ in terms of generalized coordinates
$\{q^i\}=(a,\varphi,\theta)$, where $a(t)$ is the metric scale factor
and $\varphi (t)$, $\theta (t)$ are the two scalar fields. A generic
Noether symmetry generator has the form (\ref{sym_gen}), where
$\lambda_{a,\varphi,\theta} (a, \varphi, \theta)$ are functions on
configuration space such that (\ref{Lie_der}) is satisfied. With the
separation of variables Ansatz, we found that the
functions $\lambda_{a,\varphi,\theta}$ have the following form for
the elementary hyperbolic surfaces:
\begin{itemize}
\item Poincar\'e disk ($\bb{D}$):
\bea \label{D_3lambdas}
\lambda_a &=& \frac{1}{2} \sqrt{\frac{3}{2}} \,A k b_2 \,\frac{\left( C_1 \sin \theta + C_2 \cos \theta \right) \,\sinh \!\left( \sqrt{\frac{3}{8}} \,\varphi \right)}{a^{1/2}} \nn \\
\lambda_{\varphi} &=& - \frac{3}{2} A k b_2 \,\frac{\left( C_1 \sin \theta + C_2 \cos \theta \right) \,\cosh \!\left( \sqrt{\frac{3}{8}} \,\varphi \right)}{a^{3/2}} \nn \\
\lambda_{\theta} &=& - \frac{3}{4} \sqrt{\frac{3}{2}} \,A k b_2 \,\frac{\left( C_1 \cos \theta - C_2 \sin \theta \right)}{a^{3/2} \sinh \!\left( \sqrt{\frac{3}{8}} \,\varphi \right)} \quad ,
\eea
where $A,k,b_2,C_1,C_2$ are constants. Notice that this is effectively
a two-parameter family of Noether symmetries, since three of the five
parameters occur only in the combination $A k b_2$ and the latter
appears only as an overall multiplier, which thus can be factored out
of the symmetry condition (\ref{SymCond}).
\item Hyperbolic punctured disk ($\bb{D}^*$):
\bea \label{D*_3lambdas}
\lambda_a &=& \frac{1}{2} \sqrt{\frac{3}{2}} A k b_1 \,\frac{\left( C_3 \theta + \theta_0 \right) \,\exp \!\left( - \sqrt{\frac{3}{8}} \,\varphi \right)}{a^{1/2}} \nn \\
\lambda_{\varphi} &=& \frac{3}{2} A k b_1 \,\frac{\left( C_3 \theta + \theta_0 \right) \,\exp \!\left( - \sqrt{\frac{3}{8}} \,\varphi \right)}{a^{3/2}} \nn \\
\lambda_{\theta} &=& - \frac{3}{4} \sqrt{\frac{3}{2}} \,A k b_1 C_3 \,\frac{ \exp \!\left( \sqrt{\frac{3}{8}} \,\varphi \right)}{a^{3/2}} \quad ,
\eea
where $A,k,b_1,C_3,\theta_0$ are constants. The same comment as below
equations (\ref{D_3lambdas}) applies, namely (\ref{D*_3lambdas})
effectively gives a two-parameter family of symmetries.
\item Hyperbolic Annulus ($\bb{A}$):
\bea \label{A_3lambdas}
\lambda_a &=& \frac{1}{2} \sqrt{\frac{3}{2}} \,A k b_1 \,\frac{\left[ C_4 \cosh (C_R \theta) + C_5 \sinh (C_R \theta) \right] \,\cosh \!\left( \sqrt{\frac{3}{8}} \,\varphi \right) }{a^{1/2}} \nn \\
\lambda_{\varphi} &=& - \frac{3}{2} A k b_1 \,\frac{\left[ C_4 \cosh (C_R \theta) + C_5 \sinh (C_R \theta) \right] \,\sinh \!\left( \sqrt{\frac{3}{8}} \,\varphi \right)}{a^{3/2}} \nn \\
\lambda_{\theta} &=& - \frac{3}{4} \sqrt{\frac{3}{2}} \,\frac{A k b_1}{C_R} \,\frac{\left[ C_4 \sinh (C_R \theta) + C_5 \cosh (C_R \theta) \right]}{a^{3/2} \cosh \!\left(\sqrt{\frac{3}{8}} \,\varphi \right)} \quad ,
\eea
where $A,k,b_1,C_4,C_5$ are constants. Again (\ref{A_3lambdas})
is a two-parameter family of Noether symmetries.
\end{itemize}
\noindent
In (\ref{D_3lambdas})-(\ref{A_3lambdas}), we have collected the
results of (\ref{Lambda_ans}), (\ref{A123_sols}),
(\ref{Phi_solutions}), (\ref{Constants}), (\ref{Ph3_inv_Ph1}),
(\ref{Th_d_sol})-(\ref{Th3_d_sol}), (\ref{Th_1_d_star_sol}),
(\ref{Th_2_3_d_star_sol}), (\ref{Th_1_A_sol}) and
(\ref{Th_2_3_A_sol}). Note that, clearly, one can absorb the overall 
$Akb_{1,2}$ factors in (\ref{D_3lambdas})-(\ref{A_3lambdas}) inside the 
arbitrary constants $\theta_0$ and $C_i$, $i=1,...,5$; we have kept 
them explicit to facilitate tracing how the above results are obtained 
throughout Section \ref{NoetherSym}.

The requirement for the existence of a Noether symmetry restricts the
form of the scalar potential. We showed that, to be compatible with
the symmetries (\ref{D_3lambdas})-(\ref{A_3lambdas}), the Lagrangian
(\ref{Lagr_alpha}) has to have the following form:
\be \label{genLagr_form}
{\cal L} = - 3 a \dot{a}^2 + \frac{a^3 \dot{\varphi}^2}{2} + \frac{4}{3} \,a^3 \tilde{f}^{\,2}(\varphi) \,\dot{\theta}^2 - a^3 V(\varphi, \theta) \,\,\, ,
\ee
where
\be
\tilde{f}_{\bbb{D}} = \sinh \!\left( \sqrt{\frac{3}{8}} \, \varphi \right) \quad , \quad \,\tilde{f}_{\bbb{D}^*} = \exp \!\left( -\sqrt{\frac{3}{8}} \,\varphi \right) \quad , \quad \,\tilde{f}_{\bbb{A}} = C_R \,\cosh \!\left( \sqrt{\frac{3}{8}} \, \varphi \right)
\ee
and
\bea \label{3Potentials}
V_{\bbb{D}} &=& V_0 \,\cosh^2 \!\!\left( \sqrt{\frac{3}{8}} \,\varphi \right) \coth^m \!\!\left( \sqrt{\frac{3}{8}} \,\varphi \right) \left[ C_1 \cos \theta - C_2 \sin \theta \right]^{-m} \,\,\, , \nn \\
V_{\bbb{D}^*} &=& V_0 \,\exp \!\left( \!- \sqrt{\frac{3}{2}} \,\varphi + \frac{m}{2} \,e^{\sqrt{\frac{3}{2}} \,\varphi} \right) \exp \!\left[ m \,\theta \left( \frac{\theta}{2} + \frac{\theta_0}{C_3} \right) \right] \,\,\, , \\
V_{\bbb{A}} &=& V_0 \,\sinh^2 \!\!\left( \sqrt{\frac{3}{8}} \,\varphi \right) \coth^m \!\!\left( \sqrt{\frac{3}{8}} \,\varphi \right) \left[ C_4 \sinh (C_R \theta) + C_5 \cosh (C_R \theta) \right]^m \nn
\eea
with $V_0$ and $m$ being arbitrary constants. Note that the scalar
potentials (\ref{3Potentials}) depend, in each of the three cases,
only on those two parameters of the corresponding Noether symmetries
(\ref{D_3lambdas})-(\ref{A_3lambdas}), which are essential, as should
be the case.

Furthermore, we simplified the Euler-Lagrange equations of
(\ref{genLagr_form}), for each of the three elementary hyperbolic
surface cases, by transforming to generalized coordinates adapted to
the corresponding Noether symmetry. This enabled us to find many
exact solutions. For some values of the parameter $m$ (the
special cases in Sections \ref{Sec:Disk}, \ref{Sec:PuncDisk} and
\ref{Sec:Annulus}), we found the most general solutions of the
equations of motion. For the rest of the $m$-parameter space (the
generic $m$ cases), we found classes of particular
solutions.\footnote{For the punctured disk case, we
  considered only $\theta_0 = 0$ for simplicity.}

An obvious open direction to pursue further is to investigate the
physical consequences of the solutions which we have found. More precisely,
one should explore what kinds of Hubble parameter, as a function of
time, these solutions give. Furthermore, what parts of the parameter
space lead to inflationary expansion and/or to actual attractor
behavior of the solutions. It would also be interesting to understand how
the results of \cite{LWWYA} on obtaining natural inflation in
two-field $\alpha$-attractors relate to our considerations. We
already pointed out above that the relevant $\theta$-dependent part of
the scalar potential can be obtained as a special case of
$V_{\bbb{D}}$ in (\ref{3Potentials}). Hence, it is worth exploring
whether one can find a realization of hypernatural inflation which is
compatible with the Noether symmetry investigated here. In a
similar vein, it is very interesting to understand whether
considerations on primordial non-Gaussianity (along the lines of 
\cite{BBJMV}) or dark energy (along the lines of \cite{AKLV2}) in 
$\alpha$-attractor models can be compatible with our Noether symmetry. 
It is also worth investigating possible embeddings into suitable classes 
of string compactifications.\footnote{Often a useful first step in that 
direction is the successful embedding in supergravity. In that regard, the 
recent work \cite{ENNO}, on dS constructions in multi-field no-scale 
supergravity models, may be of great relevance.} In this context, the special 
value of the $\alpha$-parameter required by our separated Noether symmetry 
might play an important role. It could be related to a point of enhanced 
symmetry in some larger parameter space. Or it could be a manifestation of 
a moduli stabilization mechanism, if the $\alpha$-parameter becomes a modulus 
in the underlying compactification.

Another important problem (on which we plan to report in the near
future) is to find more general solutions to the Noether symmetry
conditions that do not rely on the separation of variables
Ansatz. Indications are that such solutions have an elegant
mathematical theory, though only a subclass of them restricts the
value of the $\alpha$-parameter (equivalently, the Gaussian curvature)
of the scalar manifold. It would be interesting to compare the
solutions of the equations of motion in the presence of such more
general symmetries to the solutions of the field equations that we
obtained here. This might uncover some characteristic features of
cosmological behavior which arise in the presence of separated Noether
symmetries when compared to more general symmetries.

A different line of investigation is to extend the study of Noether
symmetries to two-field models defined on arbitrary hyperbolic
surfaces and to general multifield models and to explore their
description in the Hamiltonian approach. A proper formulation of this
problem requires the geometric approach to Noether symmetries provided
by the jet bundle formalism.

While the Noether approach requires the Lagrangian formulation
discussed in the present paper, we should mention that classical
cosmological dynamics can also be studied using the formulation used
in \cite{LS, BL1, BL2, BL3}, which is obtained by solving the
Friedmann equation in order to eliminate the cosmological scale factor
$a(t)$. As explained in those references, this leads to a geometric
system of non-linear second order ODEs which involves only the scalar
fields $\phi^I$. In fact, the Friedmann equation provides an energy
shell constraint which {\em must} be imposed on the Lagrangian system
described by \eqref{Lagr_alpha} in order to isolate those solutions of
the Euler-Lagrange equations which are of actual cosmological
relevance. Due to the non-holonomic character of this
equation, the resulting geometric system of ODEs for $\phi^I$ does
not generally admit a non-constrained Lagrangian formulation. This
system of ODEs defines a dissipative geometric dynamical system on the
tangent bundle of the scalar manifold, which can be studied with the
methods of dynamical systems theory \cite{dyn}. In particular,
symmetries of the cosmological model could be studied directly at this
level using Lie's theory of symmetries of systems of ODEs, which in
this setting has an elegant geometric formulation. We hope to address
this topic in the future.

\section*{Acknowledgements}
L.A. would like to thank the Stony Brook Simons Workshop in
Mathematics and Physics and the Mainz Institute for Theoretical
Physics for hospitality during the completion of this work. L.A. has
received partial support from the Bulgarian NSF grant DN 08/3 and the
bilateral grant STC/Bulgaria-France 01/6. E.M.B. has been supported
mainly by the Romanian Ministry of Research and Innovation, grant PN
18 09 01 01/2018. The work of L.A. and E.M.B has also been partially
supported by the ICTP - SEENET-MTP project NT-03 Cosmology - Classical
and Quantum Challenges. The work of C. I. L. was supported by grant
IBS-R003-D1.

\appendix

\section{Elementary hyperbolic surfaces} \label{ElHypSs}
\setcounter{equation}{0}

Any smooth and complete hyperbolic surface is isometric to a quotient
of the hyperbolic plane $\bb{H}$ (the open upper half plane of the
complex plane endowed with the Poincar\'e metric) by a discrete
subgroup of its group of isometries ${\rm PSL}(2,\bb{R})$. Elements of
${\rm PSL}(2,\bb{R})$ are classified according to their fixed
points. Elliptic elements have a single fixed point located in
$\bb{H}$, parabolic elements have a fixed point on the conformal
boundary\footnote{The hyperbolic plane does not have a boundary in the
sense of manifold theory.  However, one can define a conformal
boundary for $\bb{H}$ (`a boundary at infinity', which is $\pd \bb{H}
= \bb{R} \cup \{\infty \}$) by using the conformal structure of the
hyperbolic metric, in the same vein as for the Penrose conformal
boundary in general relativity. See \cite{LS} and references
therein for details and generalization.} $\pd \bb{H}$ of $\bb{H}$ 
and hyperbolic elements have two
distinct fixed points on $\pd \bb{H}$. A complete hyperbolic surface
is called {\em elementary} if it is isometric with $\bb{H}$ or with a
quotient of $\bb{H}$ by a cyclic subgroup of ${\rm PSL}(2,\R)$ (i.e. a group 
generated by a single element), which is of parabolic or of
hyperbolic type. There are three types of elementary hyperbolic
surfaces: the hyperbolic disk $\mD$ (also called the Poincar\'e disk, since
it is isometric with $\bb{H}$), the hyperbolic punctured disk
$\mD^\ast$ and the hyperbolic annuli $\mA(R)$. The hyperbolic disk and
hyperbolic punctured disk are unique up to isometry, while the
isometry class of a hyperbolic annulus depends on a real modulus
$R>1$. We briefly discuss these hyperbolic surfaces in turn, referring
the reader to \cite{BL1} for more detail:

\begin{itemize}
\item {\em Poincar\'e disk}:\\
The Poincar\'e disk $\mD$ is the open subset of the complex plane $\C$
defined by the condition 
\be
|z| < 1 \,\,\, , 
\ee
endowed with the complete
hyperbolic metric:
\be \label{ds2_D}
ds^2 = \frac{4}{\left( 1- z \bar{z} \right)^2} \,dz d\bar{z} \,\,\, .
\ee
For various reasons, some going as far back as \cite{CSF}, in the
literature on cosmological $\alpha$-attractors this metric appears in
the scalar kinetic terms with a different overall constant factor. One
can transform (\ref{ds2_D}) to polar coordinates $\rho$ and $\theta$,
determined via $z \equiv \rho e^{i \theta}$ with $\rho \in [0,1)$, and
then, by changing suitably the radial variable, to semi-geodesic
coordinates (see \cite{BL1}). This is what is achieved with the
redefinition (\ref{rho_redef}) that maps the action (\ref{S_Pd}) into
the form (\ref{S_alpha}).
\item {\em Hyperbolic punctured disk}:\\
The hyperbolic punctured disk $\mD^\ast$ is the open subset of $\C$
defined by 
\be
0<|z|<1 \,\,\, , 
\ee
endowed with the complete hyperbolic metric:
\be
ds^2 = \frac{1}{\left( \rho \ln \rho \right)^2} \left( d\rho^2 + \rho^2 d \theta^2 \right) \,\,\, ,
\ee
where $\rho = |z|$ and $\theta = \arg(z)$ are polar coordinates on the
complex plane. As explained in \cite{BL1}, one can transform this
metric to semi-geodesic coordinates, i.e. to the form $ds^2 = d
\varphi^2 + f (\varphi) d\theta^2$ using a certain change of variables
$\varphi = \varphi (\rho)$. This is what the transformation
(\ref{rho_redef_D*}) amounts to.
\item {\em Hyperbolic annuli}:\\ The hyperbolic annulus $\mA(\hat{R})$ is the
open domain in the complex plane defined through:
\be \label{Annulus_domain}
\qquad \frac{1}{\hat{R}}<|z|< \hat{R} \quad {\rm where} \quad \hat{R}>1 \quad ,
\ee
endowed with the complete hyperbolic metric (in polar coordinates):
\be \label{ds2_A}
ds^2 = \frac{C_R^2}{\left[ \rho \cos \left( C_R \ln \!\rho \right) \right]^2} \left( d\rho^2 + \rho^2 d \theta^2 \right) \quad {\rm where} \quad C_R \equiv \frac{\pi}{2 \ln \hat{R}} \quad .
\ee
The transformation (\ref{rho_redef_A}), modulo an overall numerical factor, maps the metric (\ref{ds2_A}) to the form $ds^2 = d \varphi^2 + f (\varphi) d\theta^2$, where $\varphi \in (-\infty , \infty)$. Note that $\varphi<0$ corresponds to $\frac{1}{\hat{R}}<\rho<1$, while $\varphi>0$ corresponds to $1<\rho<\hat{R}$.
\end{itemize}
We refer the reader to \cite{BL1} for more detail on the geometry of
elementary hyperbolic surfaces.

\section{Nontrivial trajectories for $m=0$} \label{App:Traj_m0}
\setcounter{equation}{0}

In this appendix we illustrate some of the exact solutions we have obtained in Sections \ref{Sec:Disk}, \ref{Sec:PuncDisk} and \ref{Sec:Annulus}. A comprehensive investigation of the phenomenological implications of all new solutions, in their entire parameter spaces, is a rather laborious effort that we leave for the future. Nevertheless, here we will illustrate, in a certain corner of parameter space, the existence of nontrivial two-field trajectories among our solutions for $m=0$, in each of the three elementary hyperbolic surface cases.\footnote{The possibility of having nontrivial multi-field trajectories, even for a potential without angular dependence, was already shown in \cite{AB,MM}.} We will also consider the behavior of the Hubble parameters in the three cases, for the relevant parts of parameter space.

\subsection{Poincar\'e disk} \label{App:Disk}

In Section \ref{Sec:Disk} we pointed out that, for the Poincare disk case, the single field limit is obtained when $m=0$ and $w = const \times v$. Indeed, for $m=0$ the scalar potential becomes:
\be \label{V_disk_m0}
V (\varphi , \theta) = V_0 \,\cosh^2 \!\left( \sqrt{\frac{3}{8}}\,\varphi \right) \,\, ,
\ee
as can be seen from (\ref{3Potentials}), while $w = const \times v$ implies $\theta = const$, as is evident from (\ref{InvCoordTr}). However, by choosing suitably the integration constants in (\ref{w_D_sol}) and (\ref{SingleF-D-sol}), one can have $w \neq conts \times v$ even for $m=0$. Thus, one can obtain nontrivial $(\varphi , \theta)$ trajectories, even though the potential has no angular dependence.

We will illustrate these trajectories in a certain part of parameter space. To underline their dependence on the parameters, we will explore how the trajectories change as we vary two of the integration constants, namely $C_1^u$ and $C_2^u$, while keeping the rest fixed. Let us make the following convenient choices: 
\be \label{Constants_D}
C_1 = \frac{1}{\sqrt{3}} \,\, , \,\, C_2 = 0 \,\, , \,\, C_w = 1 \,\, , \,\, V_0 = 3 \,\, , \,\, \Sigma_0 = 2 \,\, , \,\, C_0^w = 1 \,\, , \,\, C_2^v = 0 \,\, .
\ee
Recall that the constant $C_1^v$ is determined from the Hamiltonian constraint (\ref{HamCon_m0_D}).
To be able to solve the letter, one needs $C_2^u \neq 0$ and even $|C_2^u| > |C_1^u|$. We also have to take $|C_1^u| > 1$ for the choices in (\ref{Constants_D}), in order to ensure a real and positive scale factor $a(t)$ for any $t\ge0$\,. This can be understood by noting that $a(t)|_{t=0} = \left[ (C_1^u)^2 - (C_0^w)^2 - 3 (C_2^v)^2 \right]^{\frac{1}{3}}$. Thus, if $(C_1^u)^2 - (C_0^w)^2 - 3 (C_2^v)^2 < 0$\,, then $a(t)$ becomes complex in a neighborhood of $t=0$. So, to recapitulate, we need to take:
\be
1 < |C_1^u| < |C_2^u| \,\, .
\ee
\begin{figure}[t]
\begin{center}
\hspace*{-0.2cm}
\includegraphics[scale=0.32]{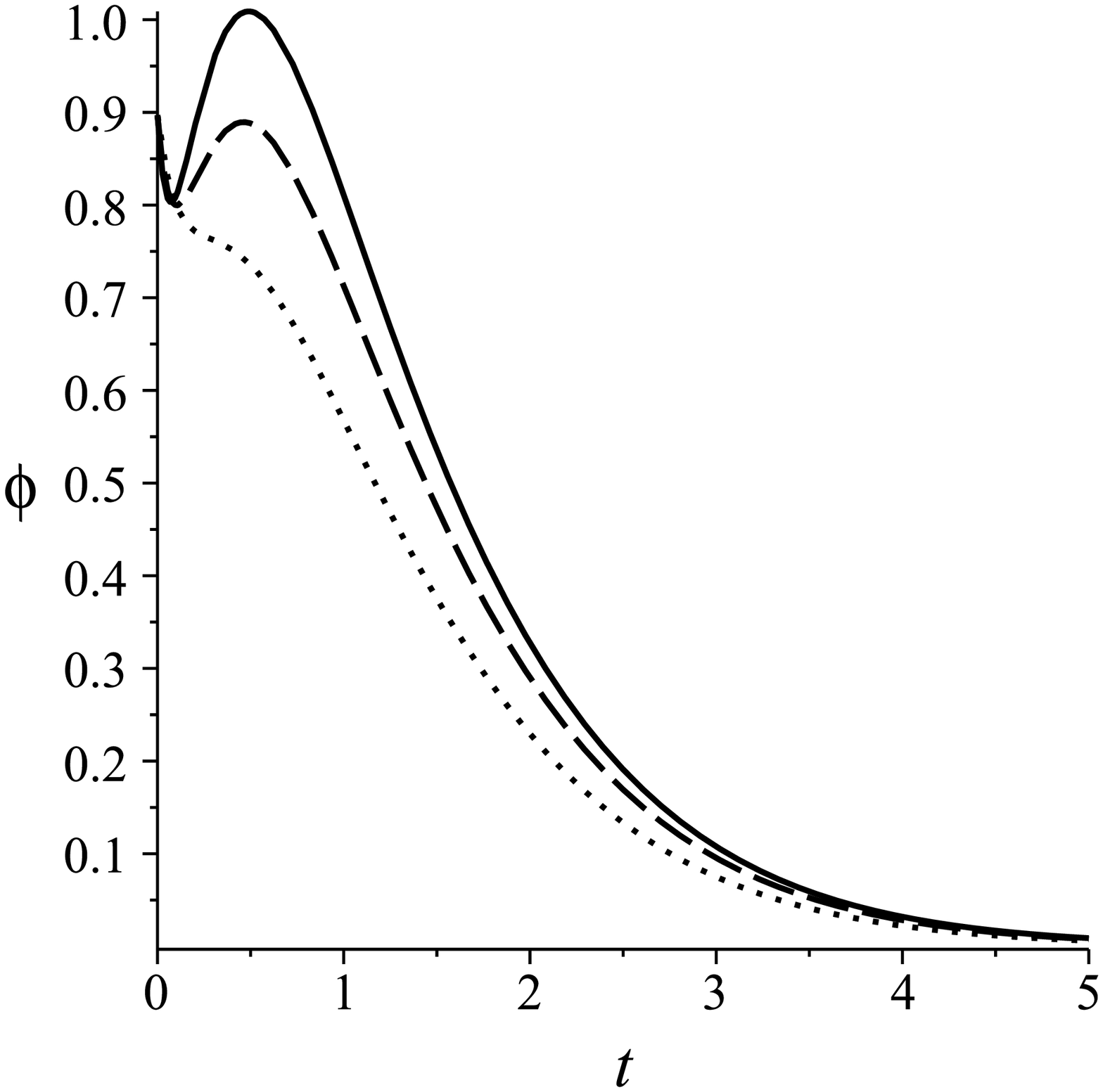}
\hspace*{0.5cm}
\includegraphics[scale=0.32]{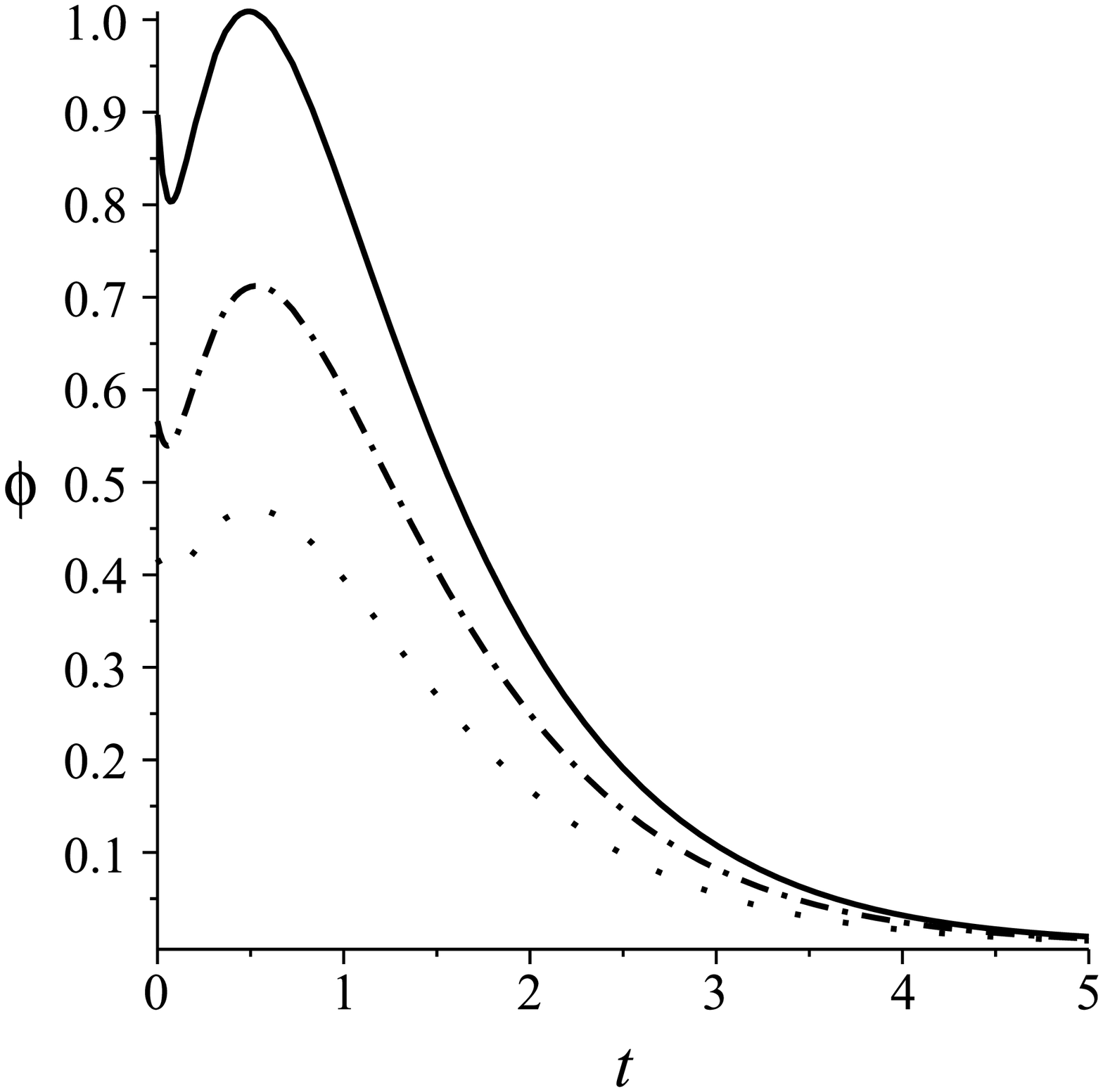}
\end{center}
\vspace{-0.7cm}
\caption{{\small Plots of $\varphi (t)$ for different values of the constants $C_{1,2}^u$\,. {\it On the left}\,, $C_1^u = 2$ and $C^u_2$ takes the following values: $C_2^u = 5$ (solid line), $C_2^u = 4$ (dashed line) and $C_2^u = 3$ (dotted line). {\it On the right}\,, $C_2^u = 5$ and $C^u_1$ takes the following values: $C_1^u = 2$ (solid line), $C_1^u = 3$ (dash-dotted line) and $C_1^u = 4$ (space-dotted line). Note that the solid lines on the left and right sides are the same curve.}}
\label{PhiD}
\vspace{0.1cm}
\end{figure}
\indent
Now we are ready to investigate numerically the $m=0$ solutions, obtained from substituting (\ref{w_D_sol}) and (\ref{SingleF-D-sol}), together with (\ref{utilde_def}) and (\ref{Constants_D}), into (\ref{InvCoordTr}). On Figure \ref{PhiD} we have plotted the scalar $\varphi (t)$ for different choices of $C_{1,2}^u$. On the left $C_1^u = const$, while $C_2^u$ varies. In this case, the initial value of $\varphi$ at $t=0$ stays the same, although the shape of the function $\varphi (t)$ changes. In particular, increasing $C_2^u$ increases $\varphi$. On the right of Figure \ref{PhiD}, $C_2^u = const$ and $C_1^u$ varies. Clearly, now the initial value of $\varphi$ also changes. However, increasing $C_1^u$ decreases $\varphi$. In all of the cases on Figure \ref{PhiD}, $\varphi$ starts at a finite value at $t=0$ and $\varphi \rightarrow 0$ as $t \rightarrow \infty$. Note that $\varphi = 0$ is precisely the minimum of the potential (\ref{V_disk_m0}).

\begin{figure}[t]
\begin{center}
\hspace*{-0.2cm}
\includegraphics[scale=0.32]{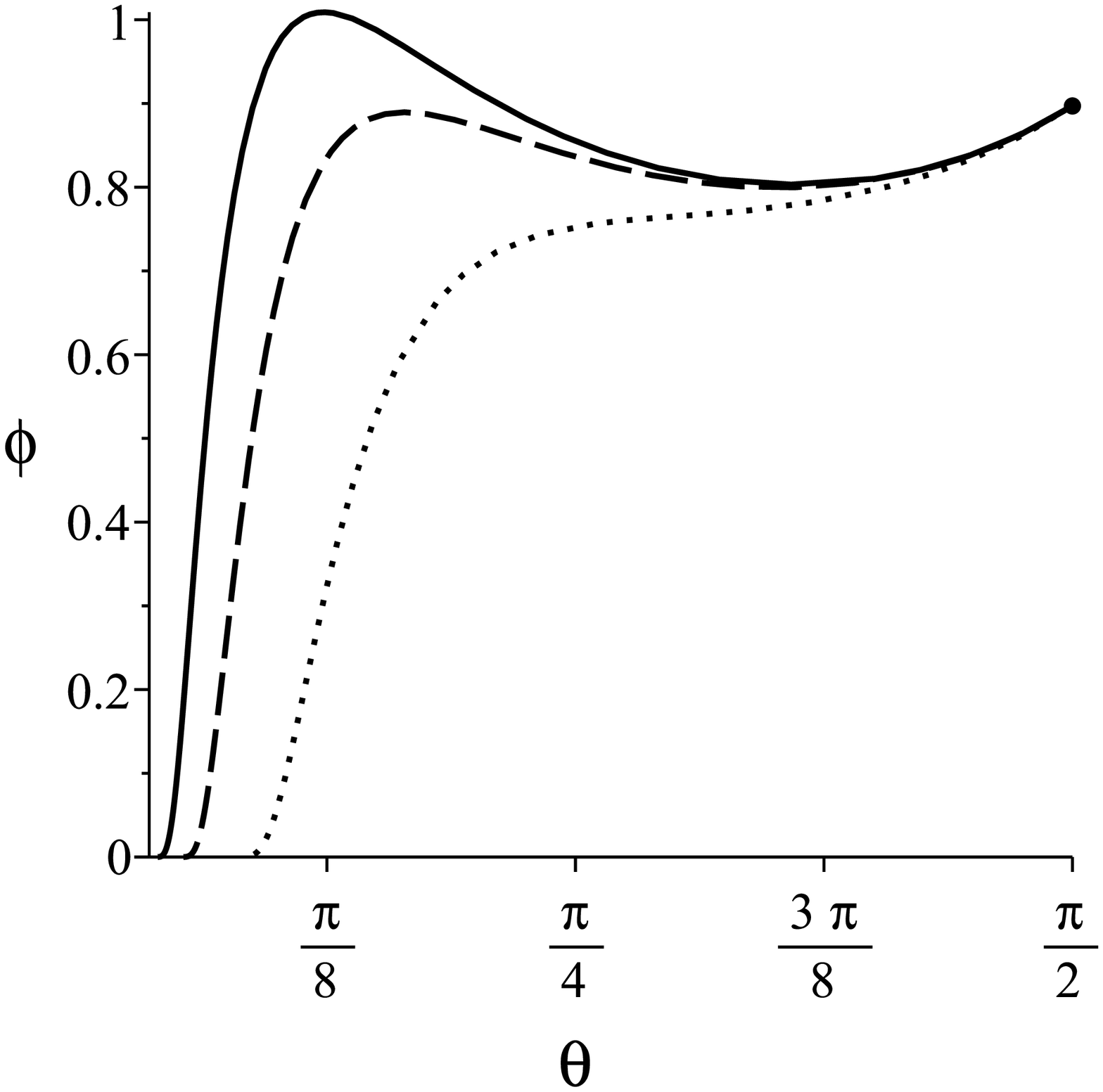}
\hspace*{0.5cm}
\includegraphics[scale=0.32]{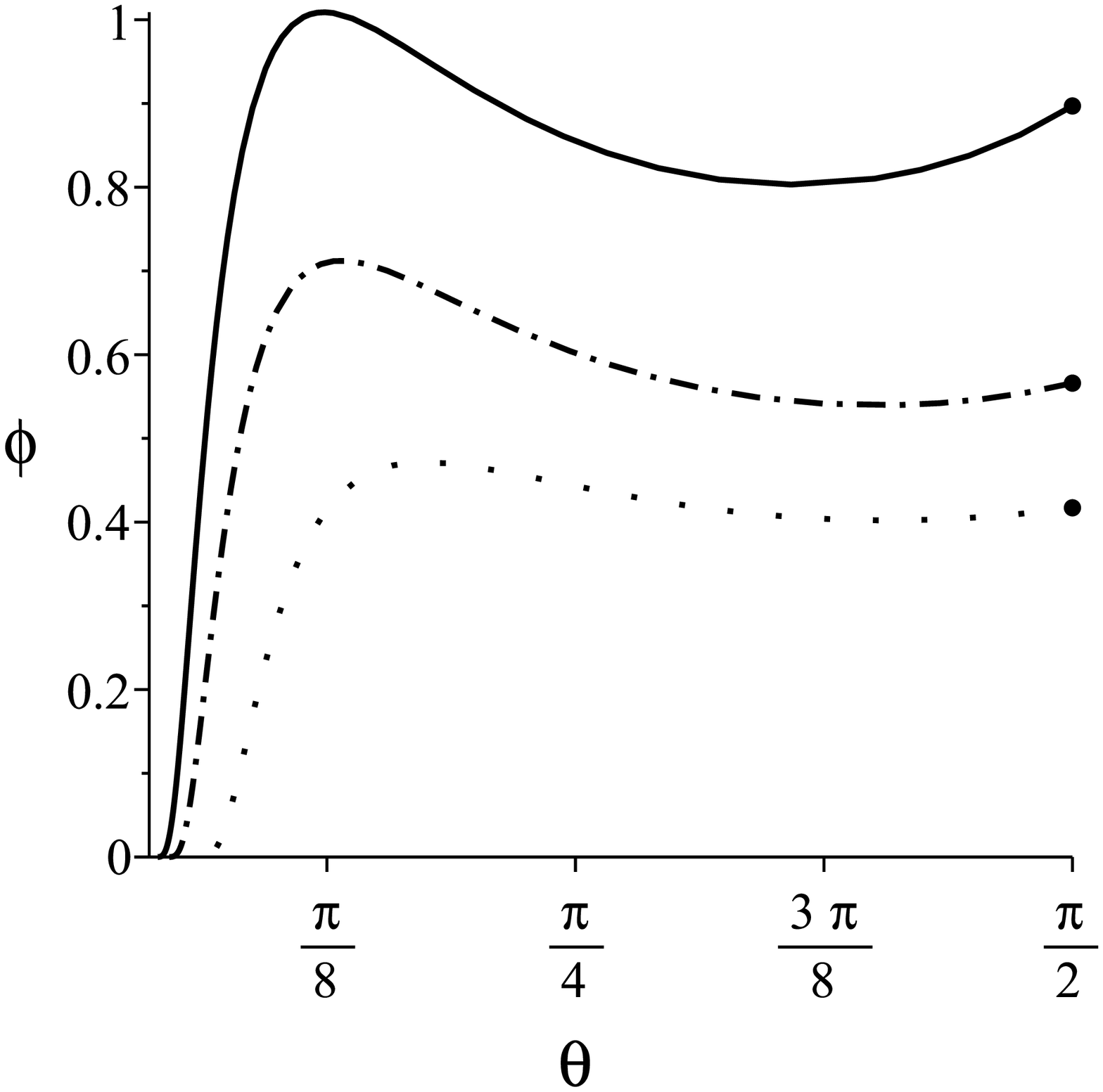}
\end{center}
\vspace{-0.7cm}
\caption{{\small The trajectories $\big( \varphi (t) , \theta (t) \big)$ for the same values of the constants as in Figure \ref{PhiD}. The dot at one end of a trajectory denotes its starting point at $t=0$.}}
\label{TrajectoriesD}
\vspace{0.3cm}
\end{figure}

\begin{figure}[h!]
\begin{center}
\hspace*{-0.2cm}
\includegraphics[scale=0.33]{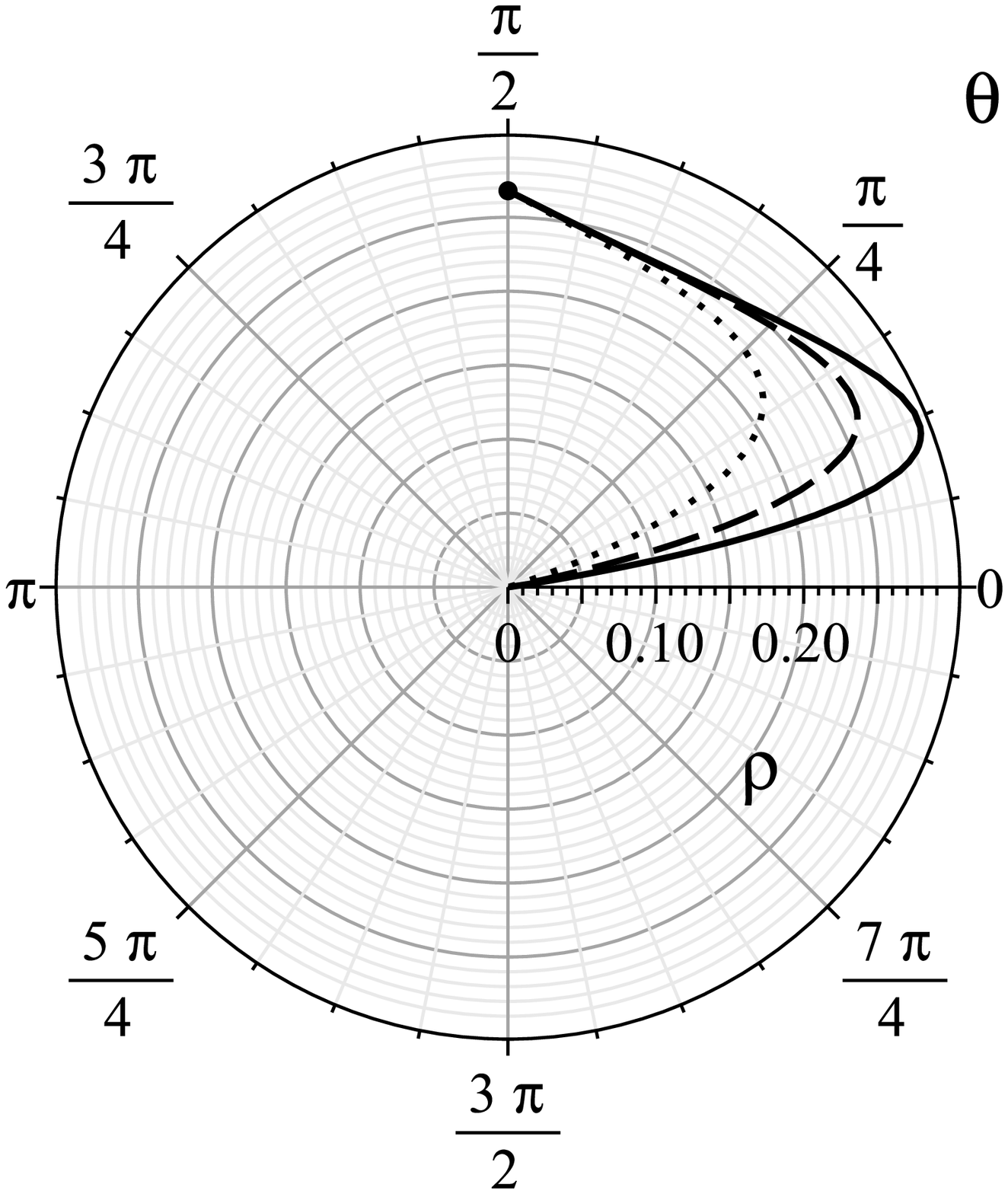}
\hspace*{0.5cm}
\includegraphics[scale=0.33]{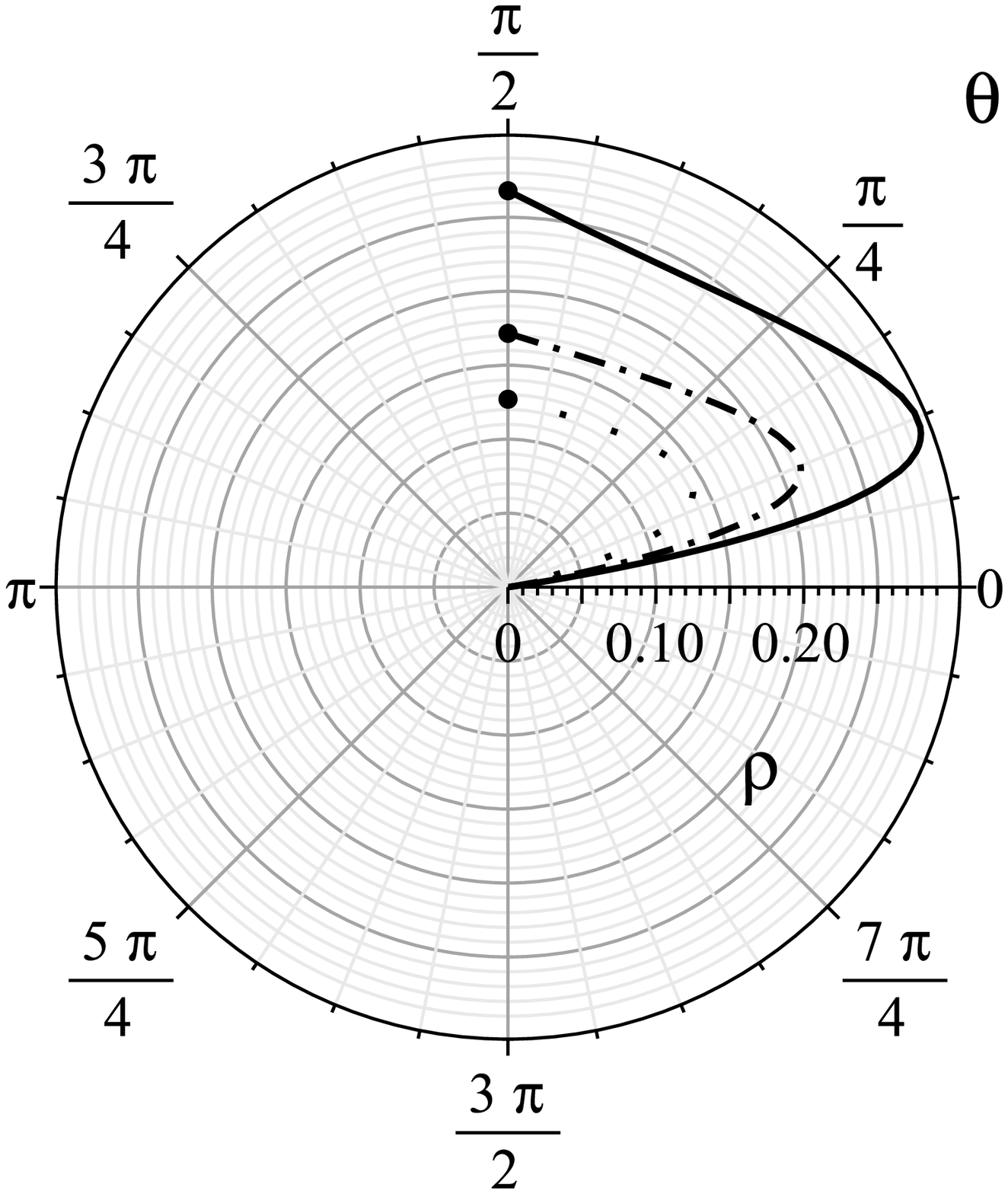}
\end{center}
\vspace{-0.7cm}
\caption{{\small The trajectories $\big( \rho (t) , \theta (t) \big)$, with $\rho$ being the radial variable on the unit disk, for the same values of the constants as in Figure \ref{PhiD}.}}
\label{TrajectoriesDR}
\vspace{0.1cm}
\end{figure}

On Figure \ref{TrajectoriesD} we have plotted the trajectories $\big( \varphi (t) , \theta (t) \big)$ obtained for the same values of the constants $C_{1,2}^u$ as in Figure \ref{PhiD}. At $t=0$ these trajectories start at $\theta = \frac{\pi}{2}$, while as $t \rightarrow \infty$ they tend to $\varphi = 0$. In fact, it is more illuminating to plot them in polar coordinates. For easier comparison with the punctured disk and annuli cases, on Figure \ref{TrajectoriesDR} we plot these trajectories in terms of the canonical radial variable of the Poincar\'e disk $\rho \in [0,1)$\,, which is related to $\varphi$ via (\ref{rho_redef}) .\footnote{Note that for the ranges of $\varphi$ and $\rho$ relevant here, relation (\ref{rho_redef}) becomes $\rho \approx \frac{\sqrt{6}}{8} \varphi$\,. So in polar ($\varphi$,$\theta$) coordinates, the trajectories are the same as on Figure \ref{TrajectoriesDR}, up to a rescaling of the radial direction.} 
\hspace*{-0.35cm}
Clearly, when $C_1^u = const$ and $C_2^u$ varies, the starting point at $t=0$ remains the same, although the shape of the trajectory changes. When $C_2^u = const$ and $C_1^u$ varies, the starting point changes as well. In both cases, though, the trajectories start at $t=0$ at a finite $\rho$ and as $t \rightarrow \infty$ they tend to $\rho = 0$\,, or equivalently $\varphi = 0$\,, which is the minimum of the potential (\ref{V_disk_m0}).

\begin{figure}[t]
\begin{center}
\hspace*{-0.2cm}
\includegraphics[scale=0.32]{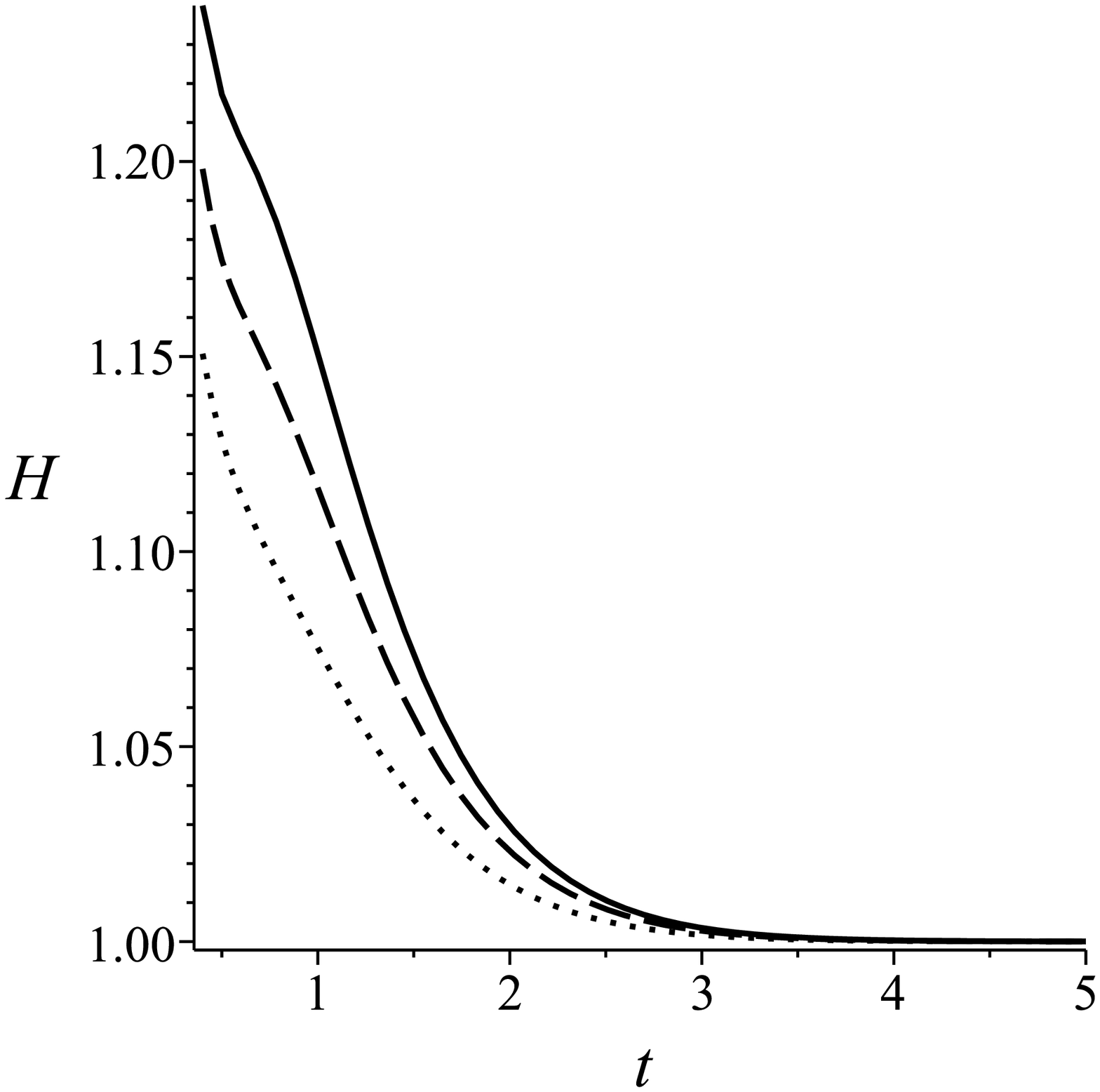}
\hspace*{0.5cm}
\includegraphics[scale=0.32]{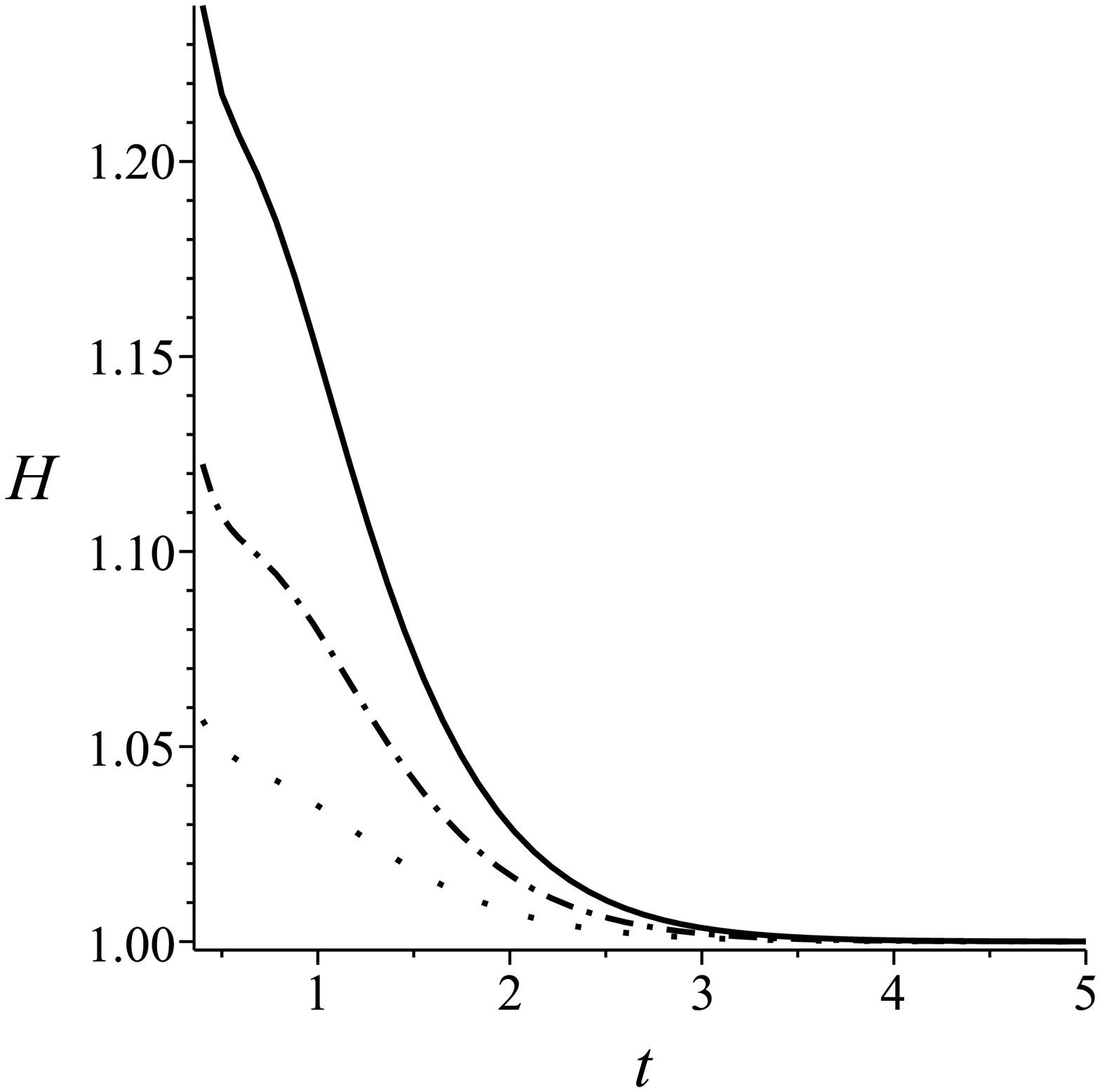}
\end{center}
\vspace{-0.7cm}
\caption{{\small The Hubble parameters $H (t)$ for the same values of the constants as in Figure \ref{PhiD}.}}
\label{HubbleD}
\vspace{0.2cm}
\end{figure}

Finally, on Figure \ref{HubbleD} we plot the Hubble parameters $H(t) = \frac{\dot{a} (t)}{a(t)}$ for the same trajectories studied above. In all cases, we have $H(t) \rightarrow 1$ as $t \rightarrow \infty$. So the spacetimes, corresponding to these solutions, asymptote to dS space. Note that the horizontal axis starts at $t=0.4$ only for better visibility of the distinctions between the graphs. In each case, $H(0)$ is finite. For example, $H(0) = 3.2$ for the solid line that is common for the left and right sides.

\subsection{Hyperbolic punctured disk} \label{App:PuncD}

The $m=0$ potential for the hyperbolic punctured disk case is:
\be \label{V_mo_D*}
V (\varphi , \theta) = V_0 \,\exp \!\left( -\sqrt{\frac{3}{2}}\,\varphi \right) \, ,
\ee
as one can see from (\ref{3Potentials}). To obtain the single-field limit, we also need $w = const \times u$ (implying that $\theta = const$), as discussed in Section \ref{Sec:PuncDisk}. However, by appropriately choosing the integration constants in (\ref{Punct_d_sol_w}) and (\ref{Punct_d_uv_sol_m_0}), we can have $w \neq const \times u$ although $m=0$. So, in this case too, there are nontrivial two-field trajectories, even when the scalar potential does not depend on $\theta$. Before turning to their numerical investigation, it will be useful to write down explicitly the inverse of (\ref{rho_redef_D*}). Substituting $\alpha = \frac{4}{3}$, according to (\ref{Constants}), gives:
\be \label{rho_phi_trans}
\rho = \exp \!\left( -e^{\frac{\sqrt{6}}{4}\,\varphi} \right) \, ,
\ee
where we have also used that by definition $\rho < 1$ (see Appendix \ref{ElHypSs}).

We will explore, again, the dependence of the $(\varphi , \theta)$ trajectories on the two integration constants characterizing $u(t)$, namely $C_1^*$ and $C_2^*$, while keeping all the other constants fixed. In the process, a certain complementarity between the two constants in $u(t)$ will become even more apparent. It is convenient to take:
\be \label{Const_m0_D*}
C_w = 1 \,\, , \,\, V_0 = 3 \,\, , \,\, \Sigma_* = 2 \,\, , \,\, C_0^w = 0 \,\, , \,\, C_4^* = 1 \,\, ,
\ee
while solving the constraint (\ref{Ham_Con_m0_D*}) for $C_3^*$. To ensure, with the choices (\ref{Const_m0_D*}), that the scale factor $a(t) > 0$ for every $t \ge 0$ and that (\ref{Ham_Con_m0_D*}) can be solved, we need:
\be
C_2^* > 0 \qquad {\rm and} \qquad C_1^* C_3^* > 2 \,\, .
\ee
\begin{figure}[t]
\begin{center}
\hspace*{-0.2cm}
\includegraphics[scale=0.32]{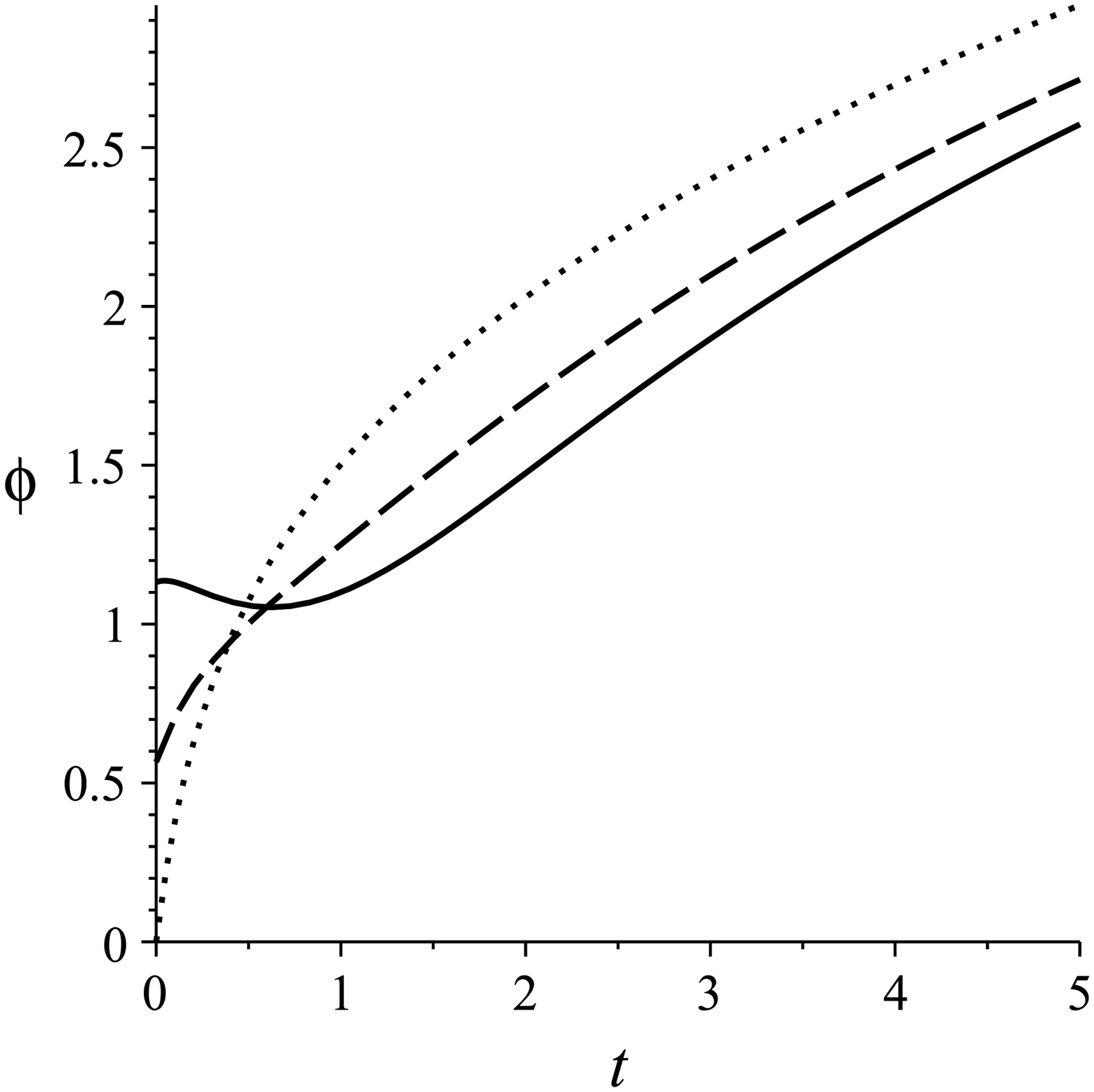}
\hspace*{0.5cm}
\includegraphics[scale=0.32]{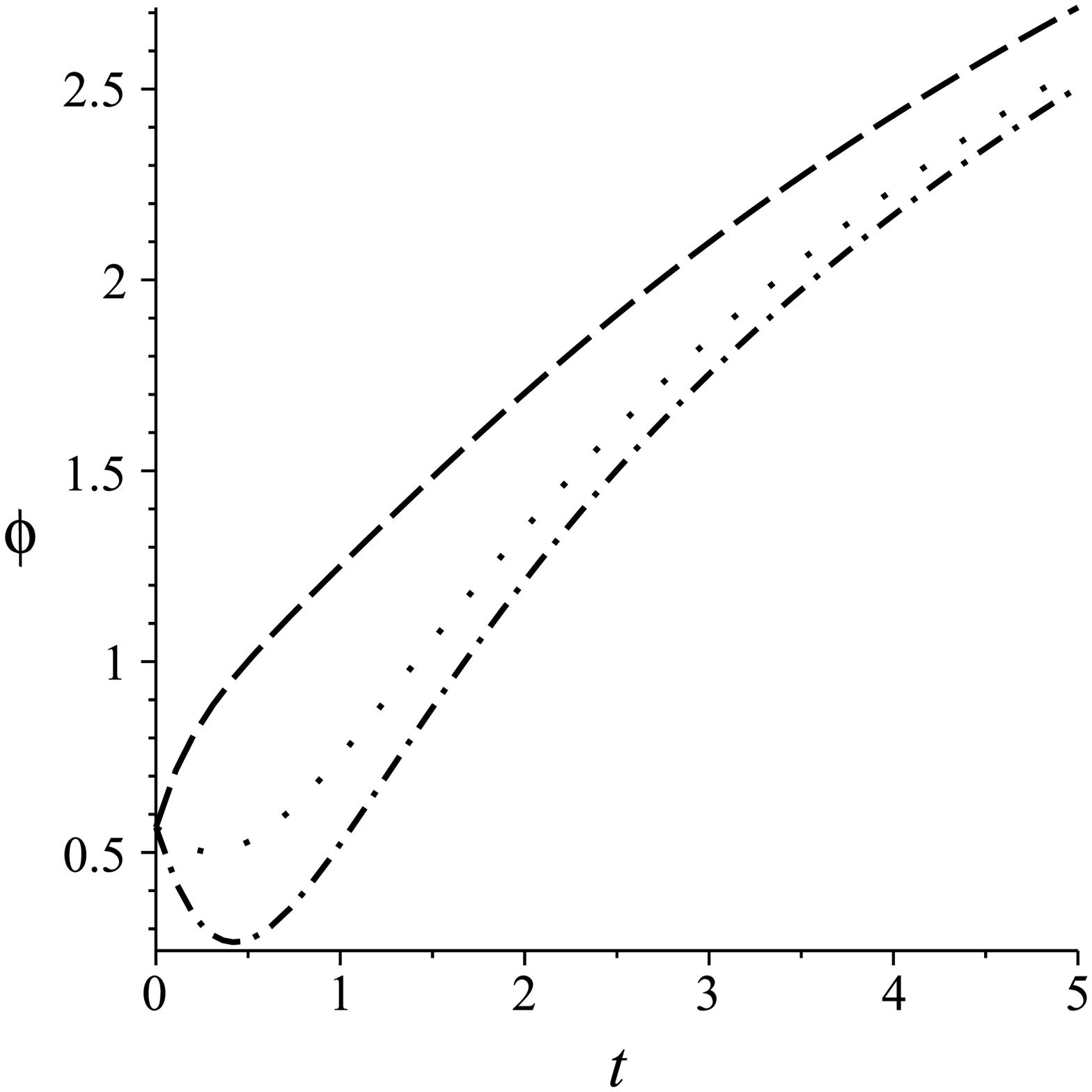}
\end{center}
\vspace{-0.6cm}
\caption{{\small Plots of $\varphi (t)$ for different values of $C_{1,2}^*$\,. {\it On the left}\,, $C_1^* = 1$ and $C^*_2$ takes the values: $C_2^* = \frac{1}{2}$ (solid line), $C_2^* = 1$ (dashed line) and $C_2^* = 2$ (dotted line). {\it On the right}\,, $C_2^* = 1$ and $C^*_1$ takes the values: $C_1^* = 1$ (dashed line), $C_1^* = 2$ (space-dotted line) and $C_1^* = 3$ (dash-dotted line). Note that the dashed lines on the left and right sides are the same curve.}}
\label{phi_t_puncD}
\vspace{-0.4cm}
\end{figure}
\begin{figure}[h!]
\begin{center}
\hspace*{-0.2cm}
\includegraphics[scale=0.32]{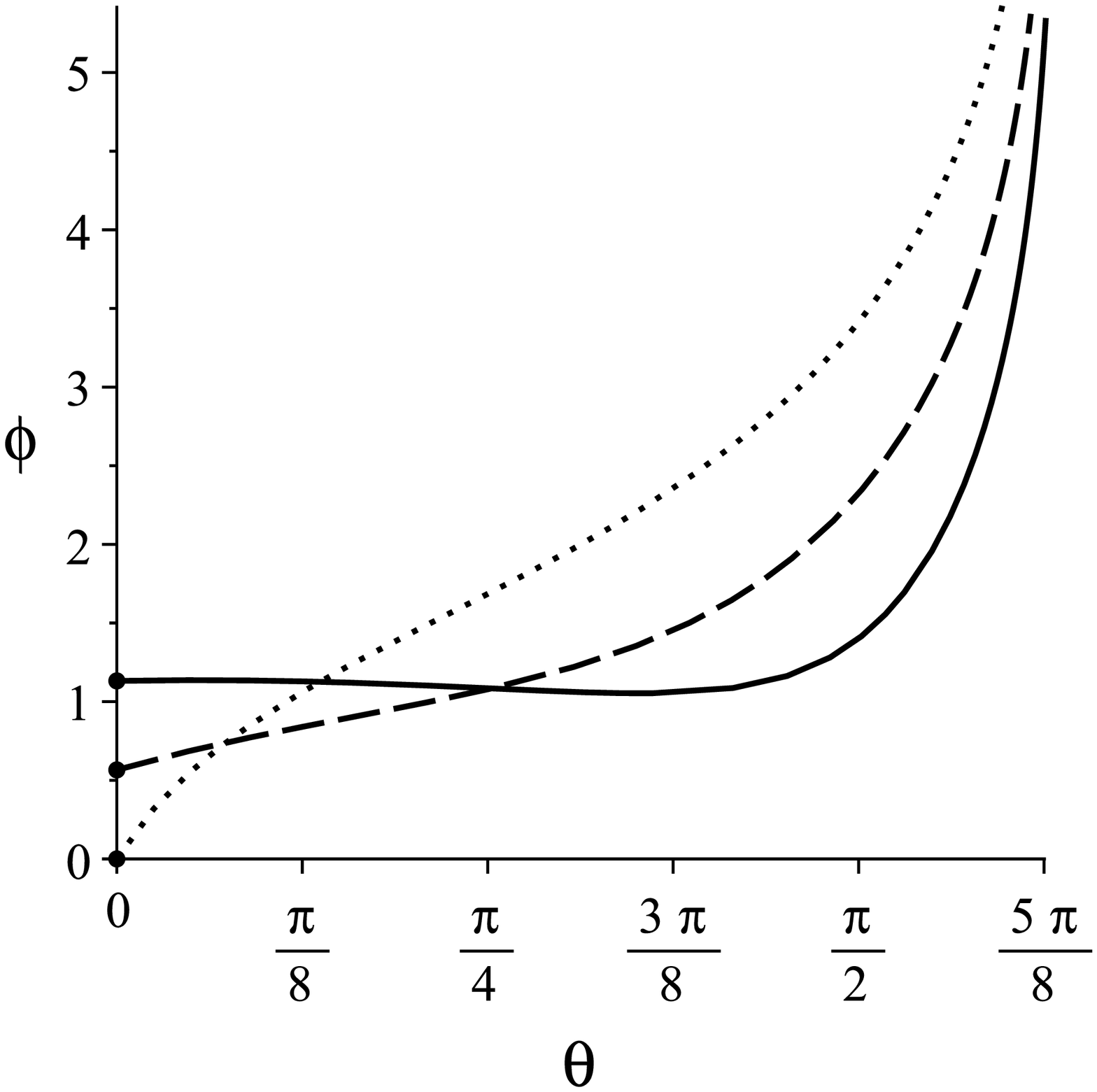}
\hspace*{0.5cm}
\includegraphics[scale=0.32]{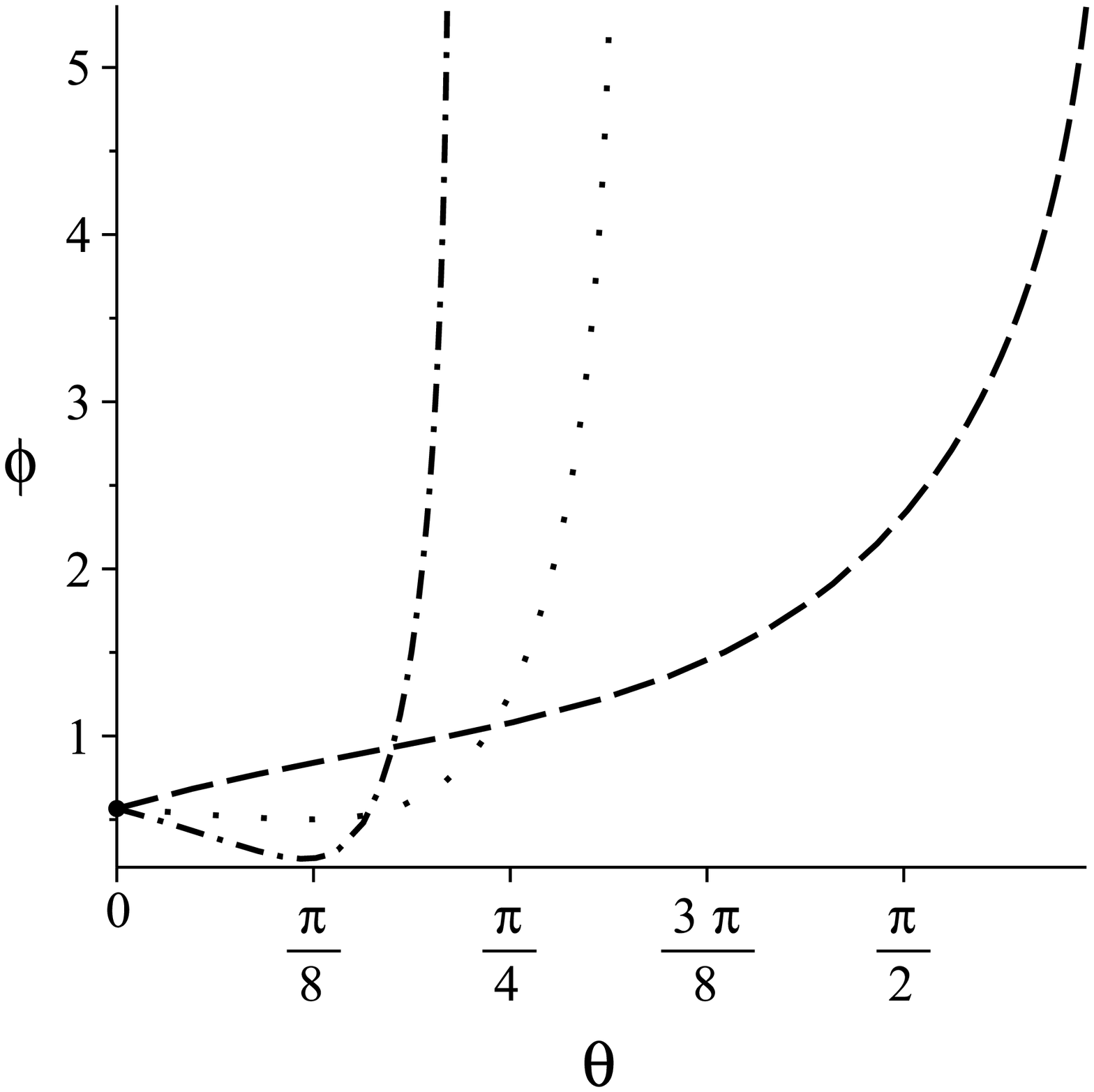}
\end{center}
\vspace{-0.6cm}
\caption{{\small The trajectories $\big( \varphi (t) , \theta (t) \big)$ for the same values of the constants as in Figure \ref{phi_t_puncD}. The dot at one end of a trajectory denotes its starting point at $t=0$.}}
\label{Trajectories_puncD}
\vspace{0.1cm}
\end{figure}

Let us now turn to the numerical investigation of the solutions, obtained by substituting (\ref{Punct_d_sol_w}) and (\ref{Punct_d_uv_sol_m_0}), together with (\ref{v_new_var_hat}) and (\ref{Const_m0_D*}), into (\ref{InvCoordTr_D*}). On Figure \ref{phi_t_puncD} we plot $\varphi (t)$; on the left $C_1^* = const$ and $C_2^*$ varies, while on the right $C_2^* = const$ and $C_1^*$ varies. In all cases $\varphi \rightarrow \infty$ as $t \rightarrow \infty$. This is in perfect agreement with the fact that the minimum of the potential (\ref{V_mo_D*}) is achieved for $\varphi \rightarrow \infty$. Note that, due to (\ref{rho_phi_trans}), $\varphi \rightarrow \infty$ corresponds to $\rho \rightarrow 0$.
\begin{figure}[t]
\begin{center}
\hspace*{-0.2cm}
\includegraphics[scale=0.33]{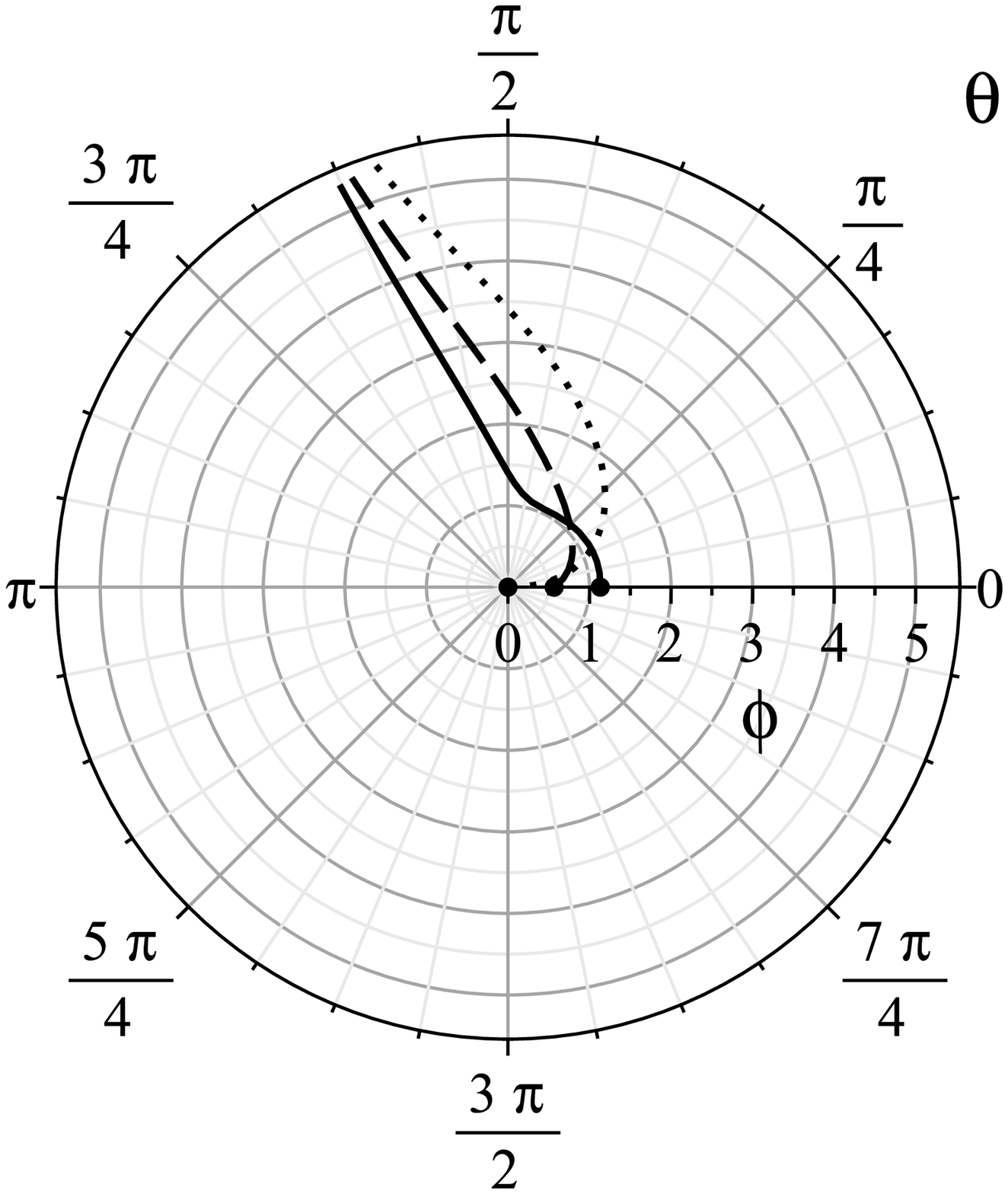}
\hspace*{0.5cm}
\includegraphics[scale=0.33]{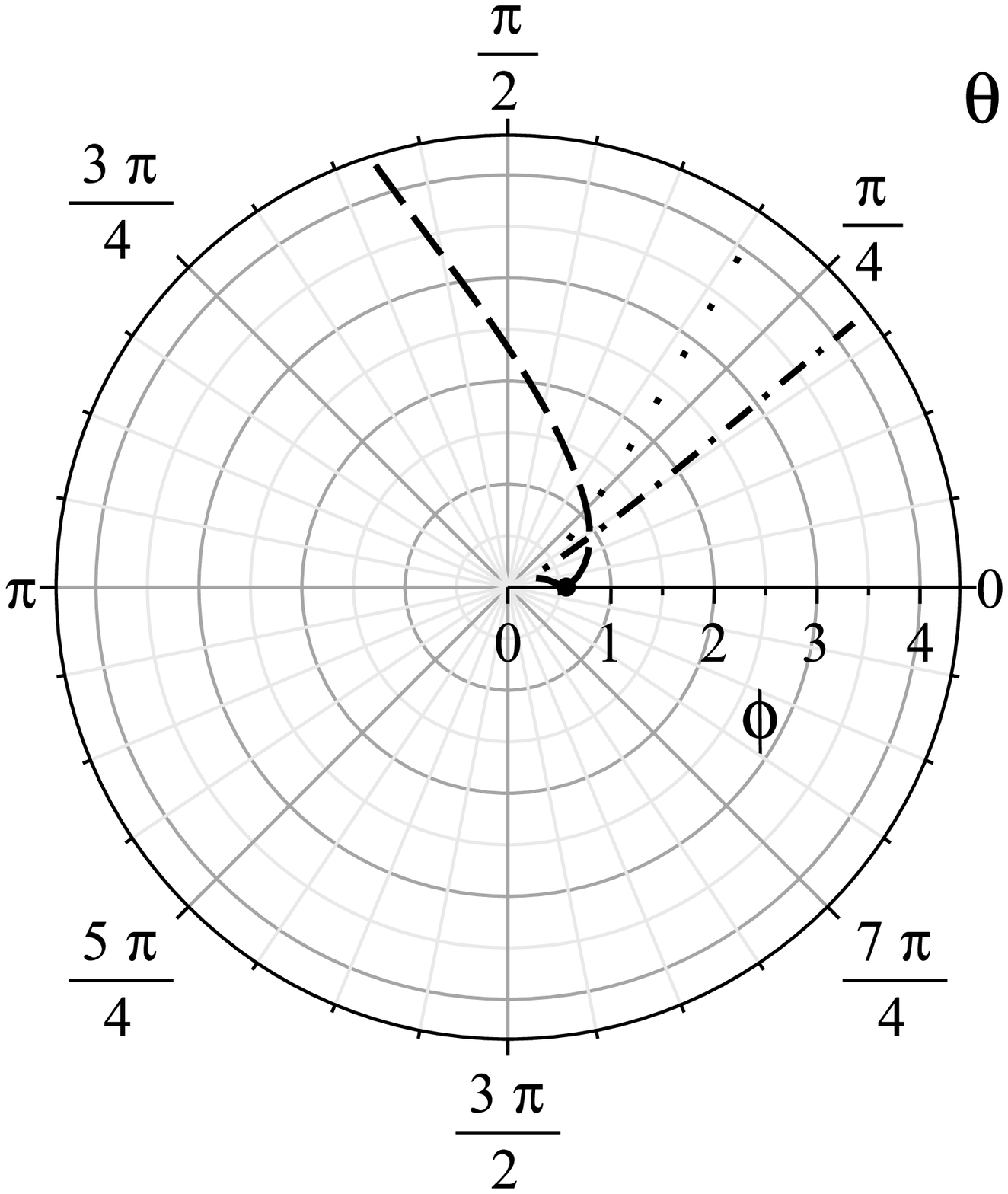}
\end{center}
\vspace{-0.6cm}
\caption{{\small The trajectories of Figure \ref{Trajectories_puncD} in polar $(\varphi , \theta)$ coordinates.}}
\label{Trajectories_th_ph_pol_puncD}
\vspace{0.35cm}
\end{figure}
\begin{figure}[h!]
\begin{center}
\hspace*{-0.2cm}
\includegraphics[scale=0.33]{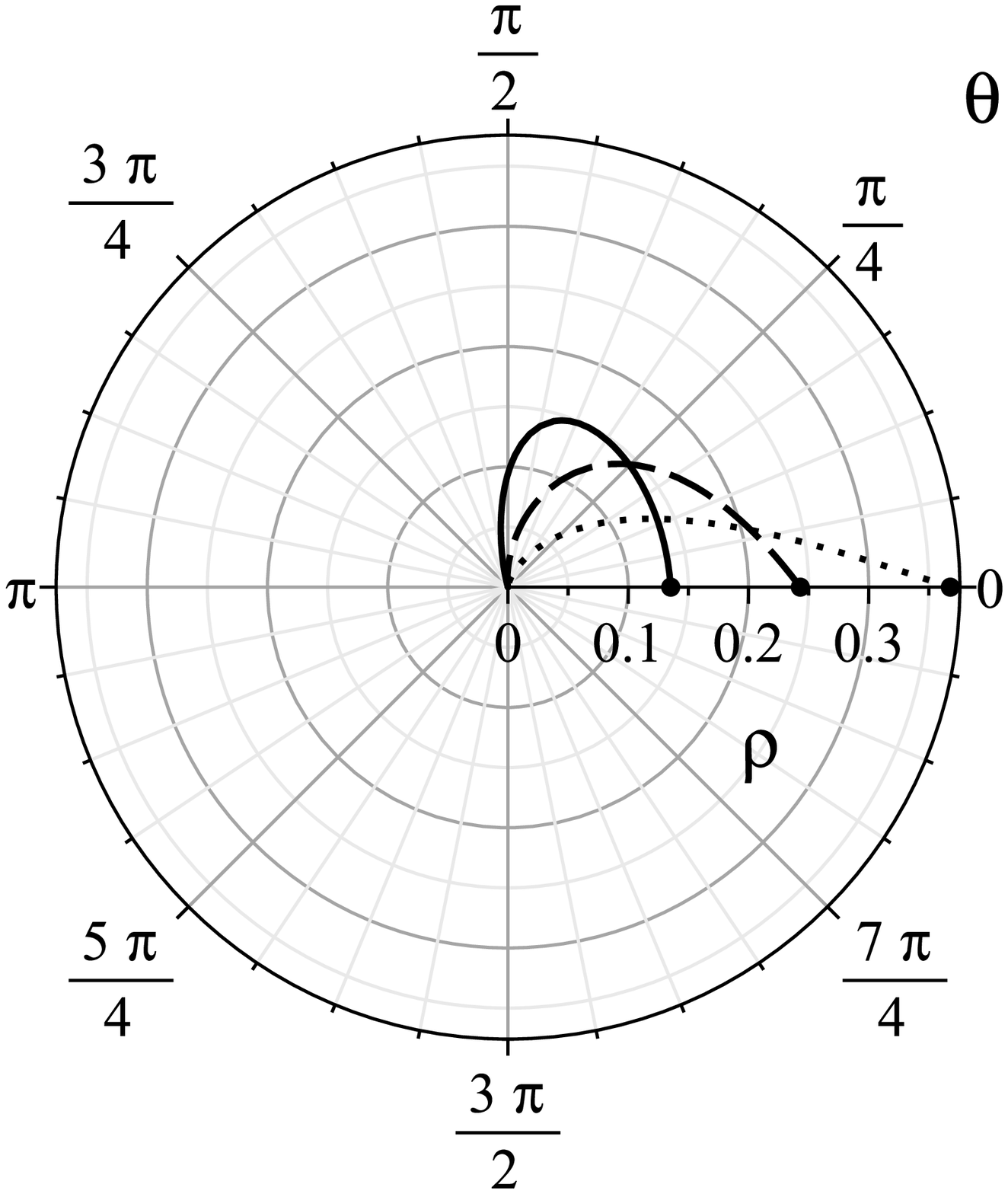}
\hspace*{0.5cm}
\includegraphics[scale=0.33]{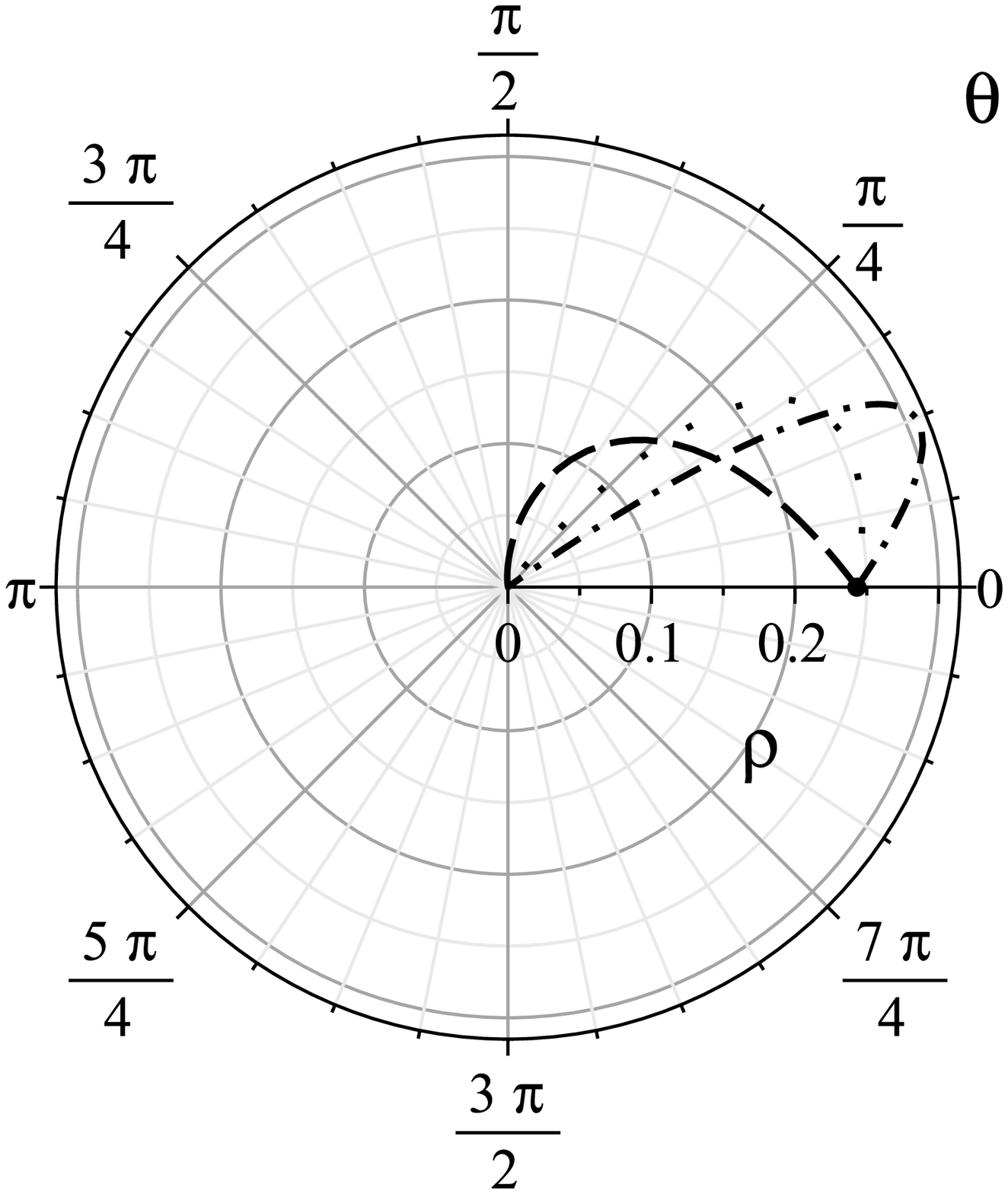}
\end{center}
\vspace{-0.6cm}
\caption{{\small The trajectories $\big( \rho (t) , \theta (t) \big)$ for the same values of the constants as in Figure \ref{phi_t_puncD}.}}
\label{th_rh_puncD}
\vspace{0.1cm}
\end{figure}
\hspace*{-0.5cm}
On Figure \ref{Trajectories_puncD} we plot the trajectories $\big( \varphi (t) , \theta (t) \big)$ for the same values of the constants as in Figure \ref{phi_t_puncD}. On the left, for different choices of $C_2^*$ (with $C_1^*$ fixed) the trajectories start at $t=0$ at different values of $\varphi$, while they all tend to $\varphi \rightarrow \infty$ and $\theta = \frac{5\pi}{8}$ as $t \rightarrow \infty$. On the right, for different values of $C_1^*$ (with $C_2^*$ fixed) all trajectories start at the same point, while for $t \rightarrow \infty$ they tend to different values of $\theta$. This is even more clear in polar $(\varphi , \theta)$ coordinates; see Figure \ref{Trajectories_th_ph_pol_puncD}. For easier comparison with the disk and annuli cases, on Figure \ref{th_rh_puncD} we also plot the same trajectories in polar $(\rho , \theta)$ coordinates, with $\rho \in (0,1)$ being the canonical radial variable of the hyperbolic punctured disk. Note that at $t=0$, the different trajectories start at different $\rho$, but as $t \rightarrow \infty$ they all tend to $\rho = 0$, which corresponds to the minimum of the scalar potential.

\begin{figure}[t]
\begin{center}
\hspace*{-0.2cm}
\includegraphics[scale=0.32]{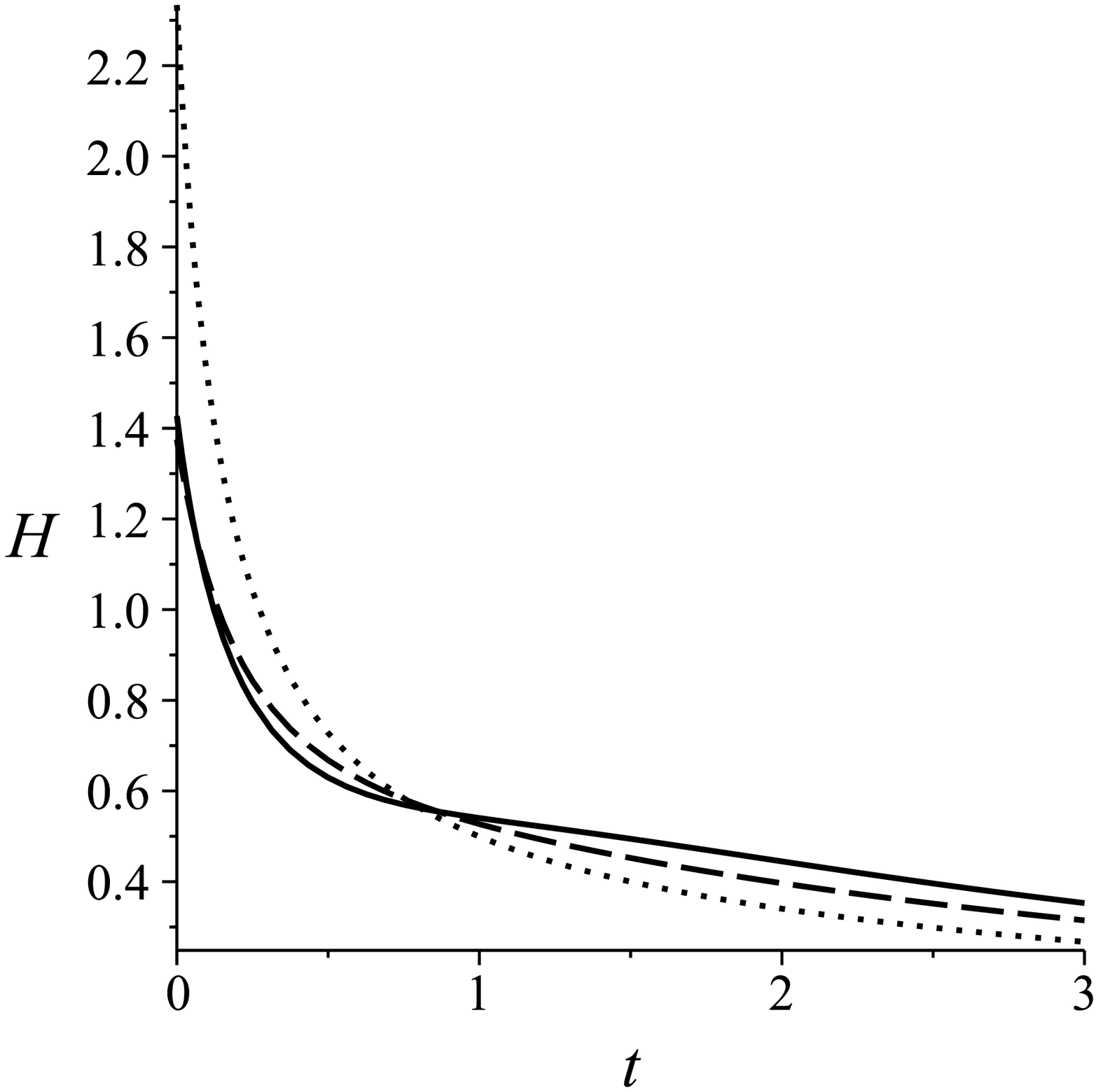}
\hspace*{0.5cm}
\includegraphics[scale=0.32]{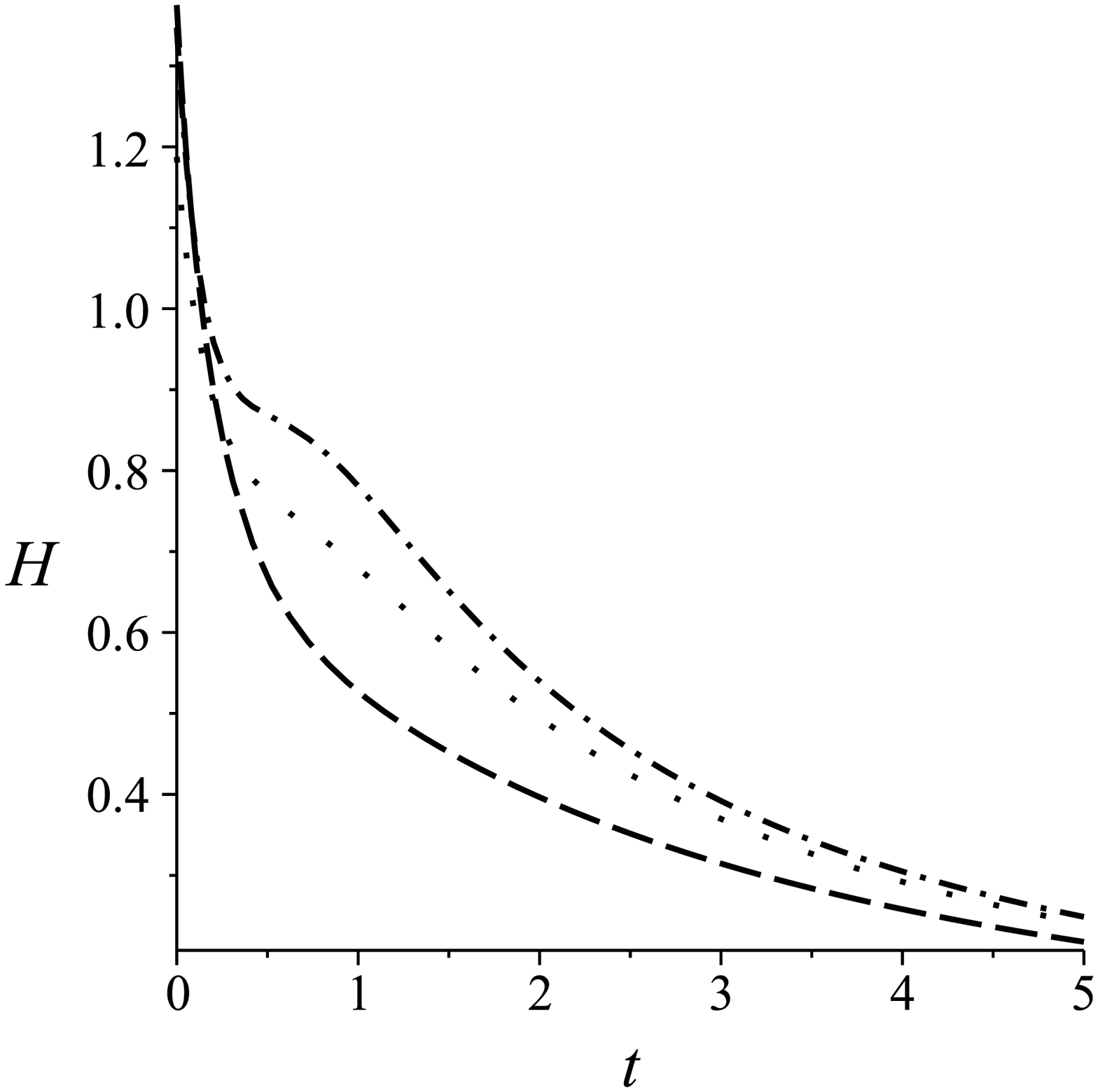}
\end{center}
\vspace{-0.6cm}
\caption{{\small The Hubble parameters $H (t)$ for the same values of the constants as in Figure \ref{phi_t_puncD}.}}
\label{H_puncD}
\vspace{0.1cm}
\end{figure}

Finally, on Figure \ref{H_puncD} we plot the Hubble parameters corresponding to the trajectories considered above. In all cases, $H(t)|_{t=0}$ is finite and $H (t) \rightarrow 0$ as $t \rightarrow \infty$. This is in accordance with the fact that, for large $t$, the scalar $\varphi \rightarrow \infty$ and thus the potential (\ref{V_mo_D*}), i.e. the effective cosmological constant, tends to zero. So the spacetimes of these solutions tend to Minkowski space. This may represent a natural mechanism for relaxation of the cosmological constant. Or it may indicate that this class of models has to be considered only in a finite time-range, assuming that at later times a different effective description (for example, containing new fields) would become more appropriate.

\subsection{Hyperbolic Annuli} \label{App:Annuli}

For the hyperbolic annuli case, the $m=0$ potential is:
\be \label{V_annuli_m0}
V (\varphi , \theta) = V_0 \,\sinh^2 \!\left( \sqrt{\frac{3}{8}}\,\varphi \right) \,\, ,
\ee
according to (\ref{3Potentials}). From Section \ref{Sec:Annulus}, it is clear that the single-field limit is obtained when, in addition, one has $w = const \times v$, which implies $\theta = const$. However, just like in Appendices \ref{App:Disk} and \ref{App:PuncD}, one can have $w \neq const \times v$ even when $m=0$, for suitable choices of the integration constants in (\ref{w_D*_sol}) and (\ref{uv_sol_A_m_0}).\footnote{In this Appendix, we will focus on the generic $\hat{C}_0 \neq 0$ case in Section \ref{Sec:Annulus}. Note, however, that for $m=0$, the solutions in the degenerate $\hat{C}_0 = 0$ case are of the same form as for $\hat{C}_0 \neq 0$, as can be seen easily by comparing (\ref{w_D*_sol}) and (\ref{uv_sol_A_m_0}) to (\ref{v_A_sol_m0_Ch_0_0}) and (\ref{uw_A_sol_m0_Ch_0_0}), although the $m=1$ and $m=2$ solutions in the two cases differ significantly.} So, again, one can have nontrivial $(\varphi , \theta)$ trajectories, even though the potential is independent of $\theta$. To study numerically those trajectories, it will be convenient to use the canonical radial variable $\rho$ of the hyperbolic annuli, which is related to $\varphi$ via (\ref{rho_redef_A}). Note that the inverse transformation (with $\alpha = \frac{4}{3}$ substituted) is:
\be \label{rho_phi_trans_A}
\ln \rho = \frac{2}{C_R} \arctan \!\left[ \tanh \!\left( \frac{\sqrt{6}}{8} \,\varphi \right) \right] \, ,
\ee
where $\varphi \in (-\infty , \infty)$, with $\varphi < 0$ corresponding to $\rho < 1$ and $\varphi > 0$ corresponding to $\rho > 1$.

As before, we will study numerically the dependence of the nontrivial two-field trajectories on the integration constants in $u(t)$, i.e. on $C_1^u$ and $C_2^u$, with all other constants fixed. For convenience, let us take the following values:
\be \label{Constants_A_m0}
C_4 = 0 \,\, , \,\, C_5 = \frac{1}{\sqrt{3}} \,\, , \,\, C_w = 30 \,\, , \,\, V_0 = 3 \,\, , \,\, \Sigma_0 = 2 \,\, , \,\, C_0^w = 1 \,\, , \,\, C_2^v = 3 \,\, , \,\, \hat{R}=2 \,\, ,
\ee
with $C_1^v$ determined from the Hamiltonian constraint (\ref{Ham_con_A_m0_gen}). This, in particular, means that we are considering the annulus given by:
\be
\frac{1}{2} < \rho < 2 \,\, .
\ee
Note that,
with the choices (\ref{Constants_A_m0}), we need to have: 
\be
(C_2^u)^2 < 23 \,\, ,
\ee
in order to ensure that $a(t)>0$ for any $t \ge 0$\,. Finally, unlike in Appendices \ref{App:Disk} and \ref{App:PuncD}, the Hamiltonian constraint in this case does not impose any restriction on the choices of $C_1^u$ and $C_2^u$.

\begin{figure}[t]
\begin{center}
\hspace*{-0.2cm}
\includegraphics[scale=0.32]{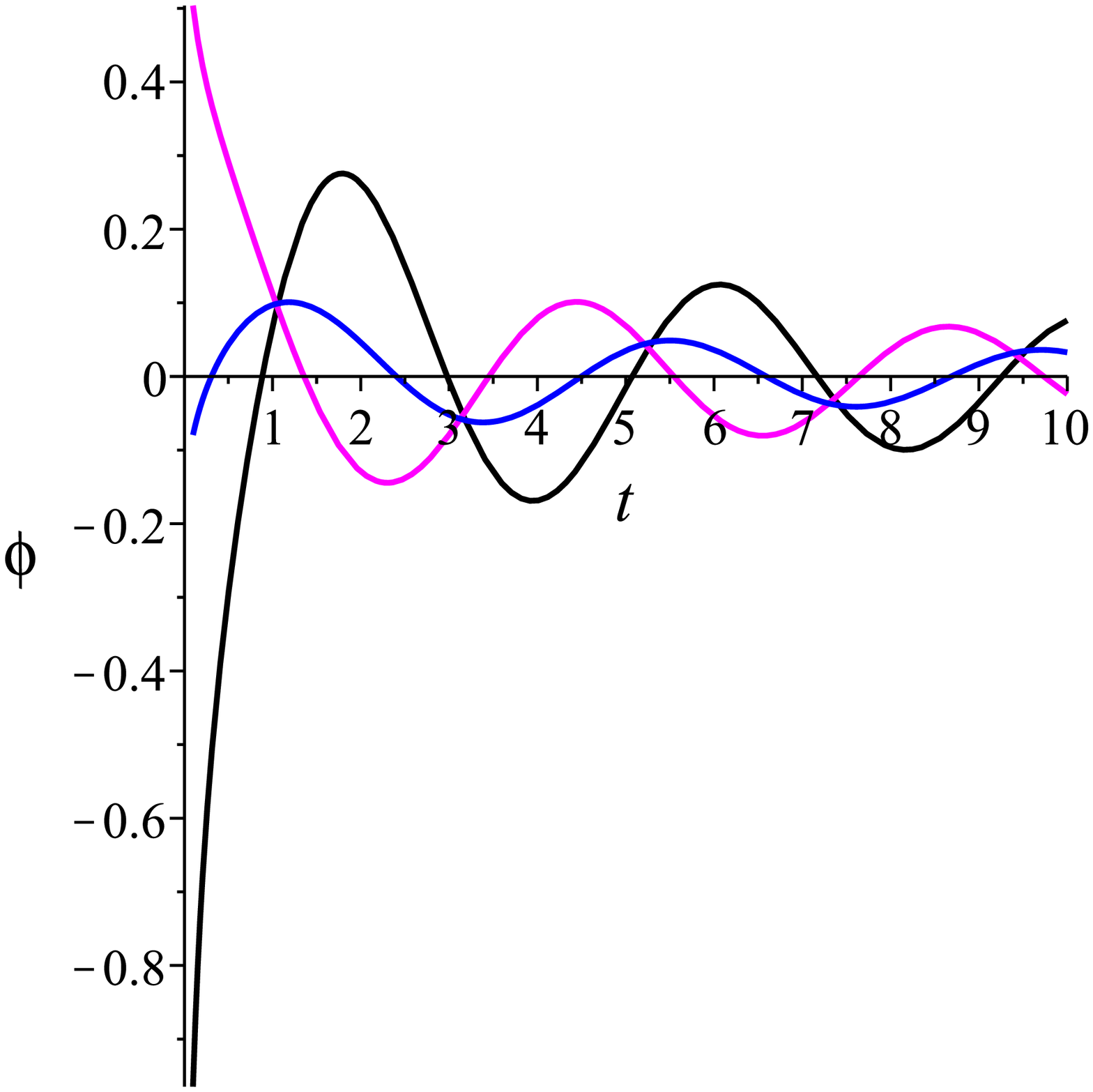}
\hspace*{0.5cm}
\includegraphics[scale=0.32]{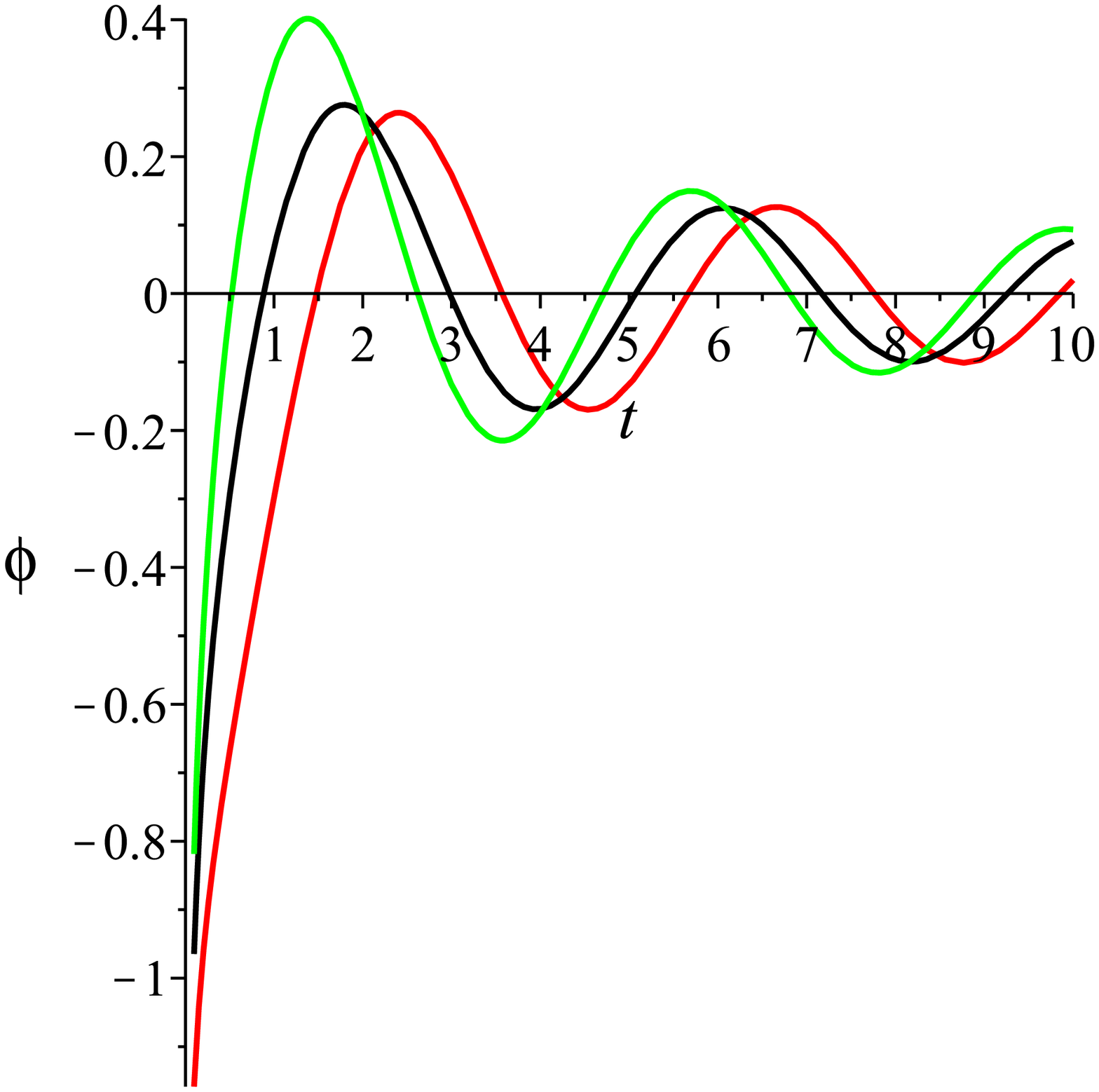}
\end{center}
\vspace{-0.6cm}
\caption{{\small Plots of $\varphi (t)$ for several values of $C_{1,2}^u$\,. {\it On the left}\,, $C_1^u = 1$ and $C^u_2$ takes the values: $C_2^u = - 4$ (black line), $C_2^u = -\frac{1}{2}$ (blue line) and $C_2^u = 2$ (magenta line). {\it On the right}\,, $C_2^u = -4$ and $C^u_1$ takes the values: $C_1^u = -3$ (red line), $C_1^u = 1$ (black line) and $C_1^u = 4$ (green line). Note that the black lines on the left and right sides are the same curve.}}
\label{phi_t_annulus}
\vspace{0.1cm}
\end{figure}
\begin{figure}[h!]
\begin{center}
\hspace*{-0.2cm}
\includegraphics[scale=0.32]{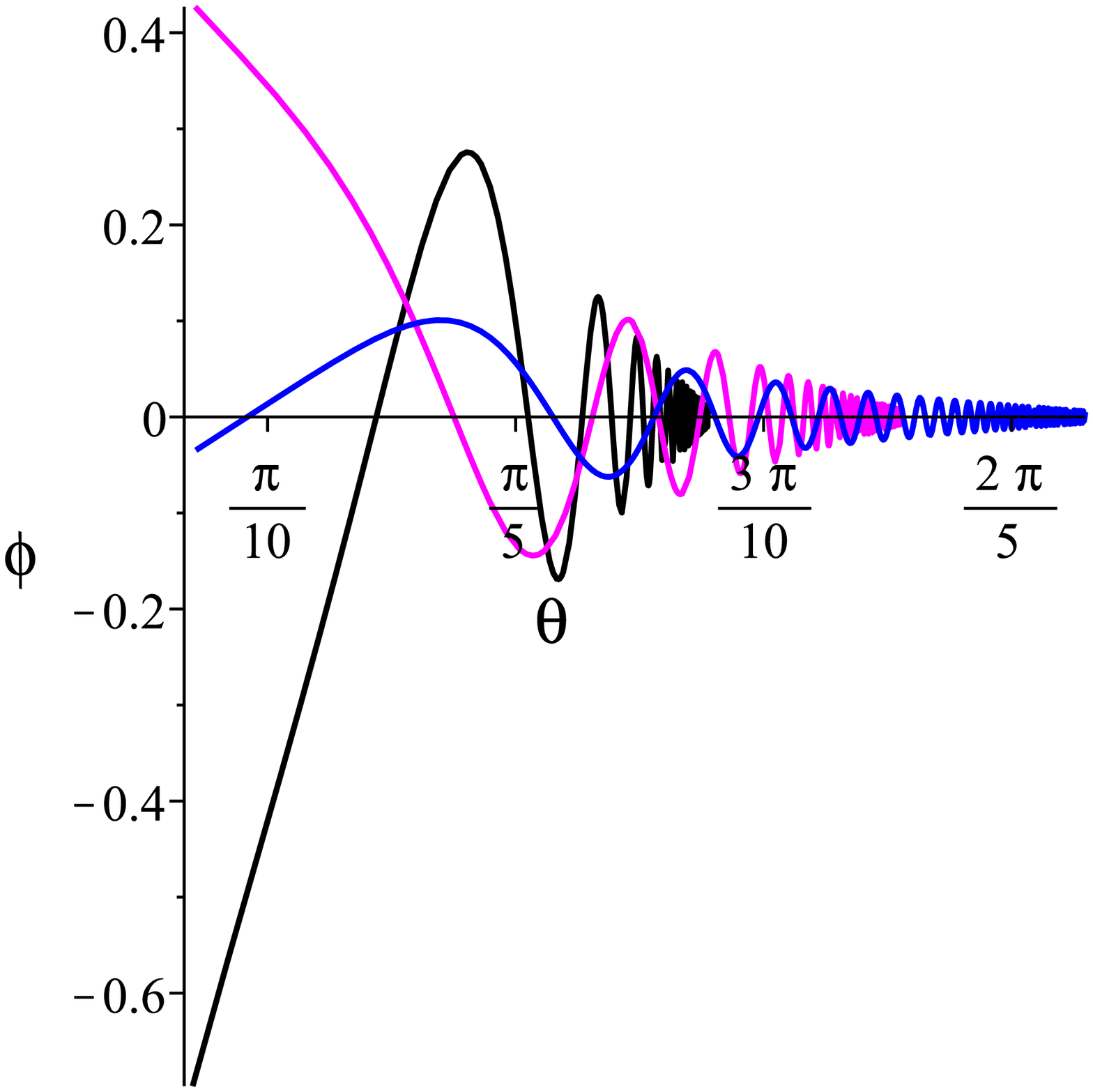}
\hspace*{0.5cm}
\includegraphics[scale=0.32]{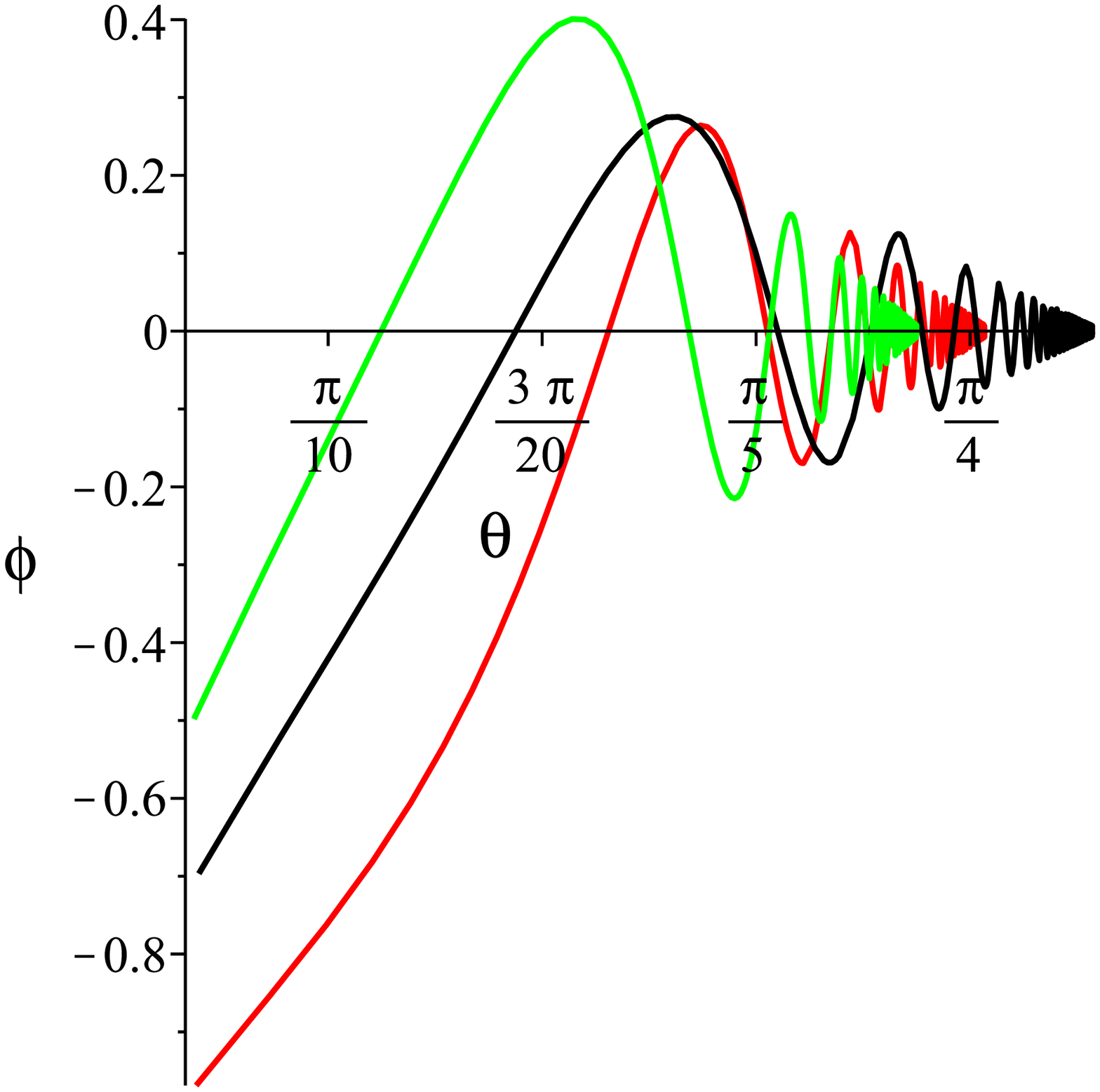}
\end{center}
\vspace{-0.6cm}
\caption{{\small The trajectories $\big( \varphi (t) , \theta (t) \big)$ for the same values of the constants as in Figure \ref{phi_t_annulus}.}}
\label{phi_th_annulus}
\vspace{0.1cm}
\end{figure}
\begin{figure}[h!]
\begin{center}
\hspace*{-0.2cm}
\includegraphics[scale=0.32]{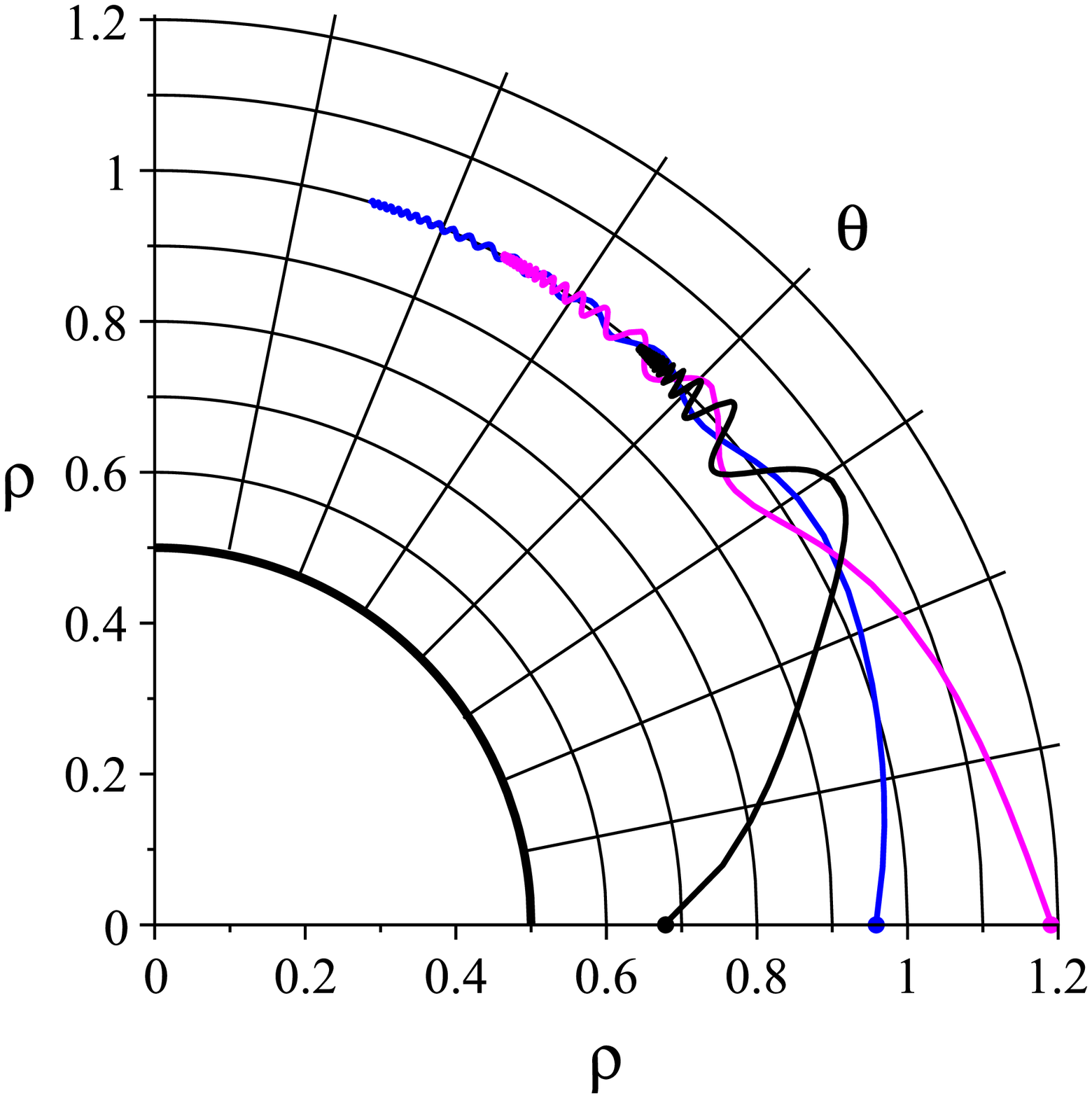}
\hspace*{0.5cm}
\includegraphics[scale=0.32]{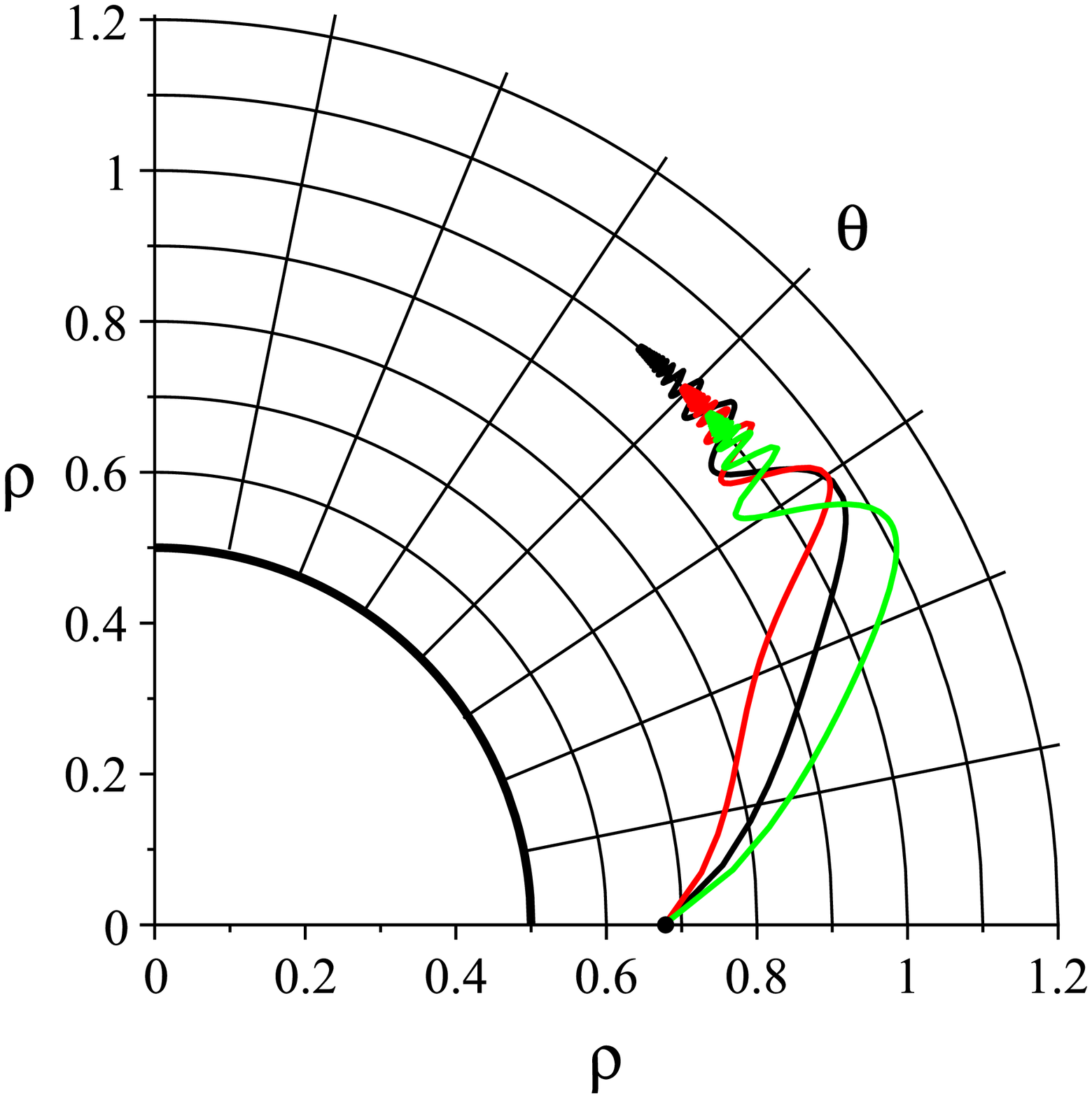}
\end{center}
\vspace{-0.6cm}
\caption{{\small The trajectories $\big( \rho (t) , \theta (t) \big)$ for the same values of the constants as in Figure \ref{phi_t_annulus}. The dot at one end of a trajectory denotes its starting point at $t=0$.}}
\label{rh_th_annulus}
\vspace{0.1cm}
\end{figure}
\begin{figure}[h!]
\begin{center}
\hspace*{-0.2cm}
\includegraphics[scale=0.32]{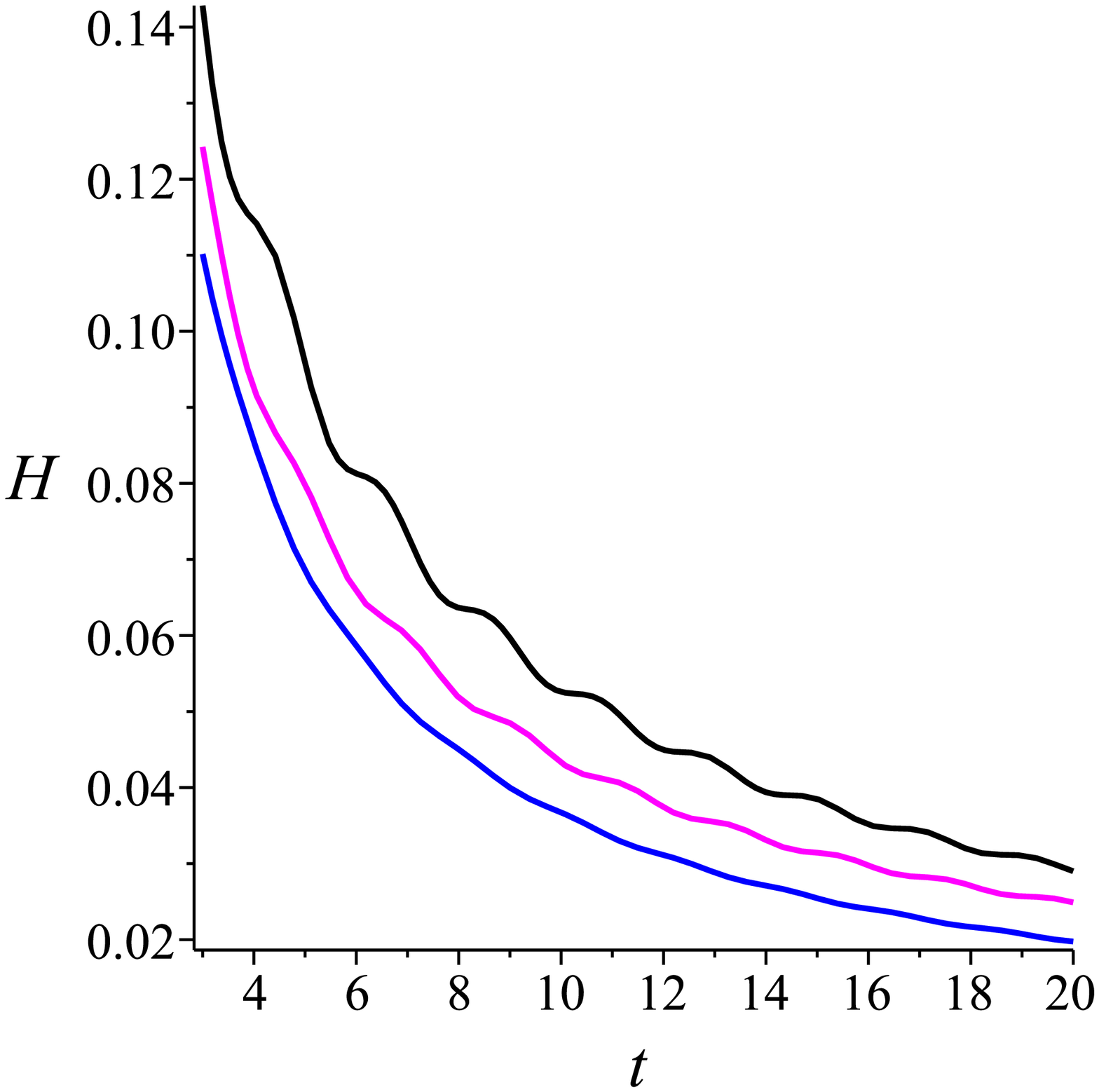}
\hspace*{0.5cm}
\includegraphics[scale=0.32]{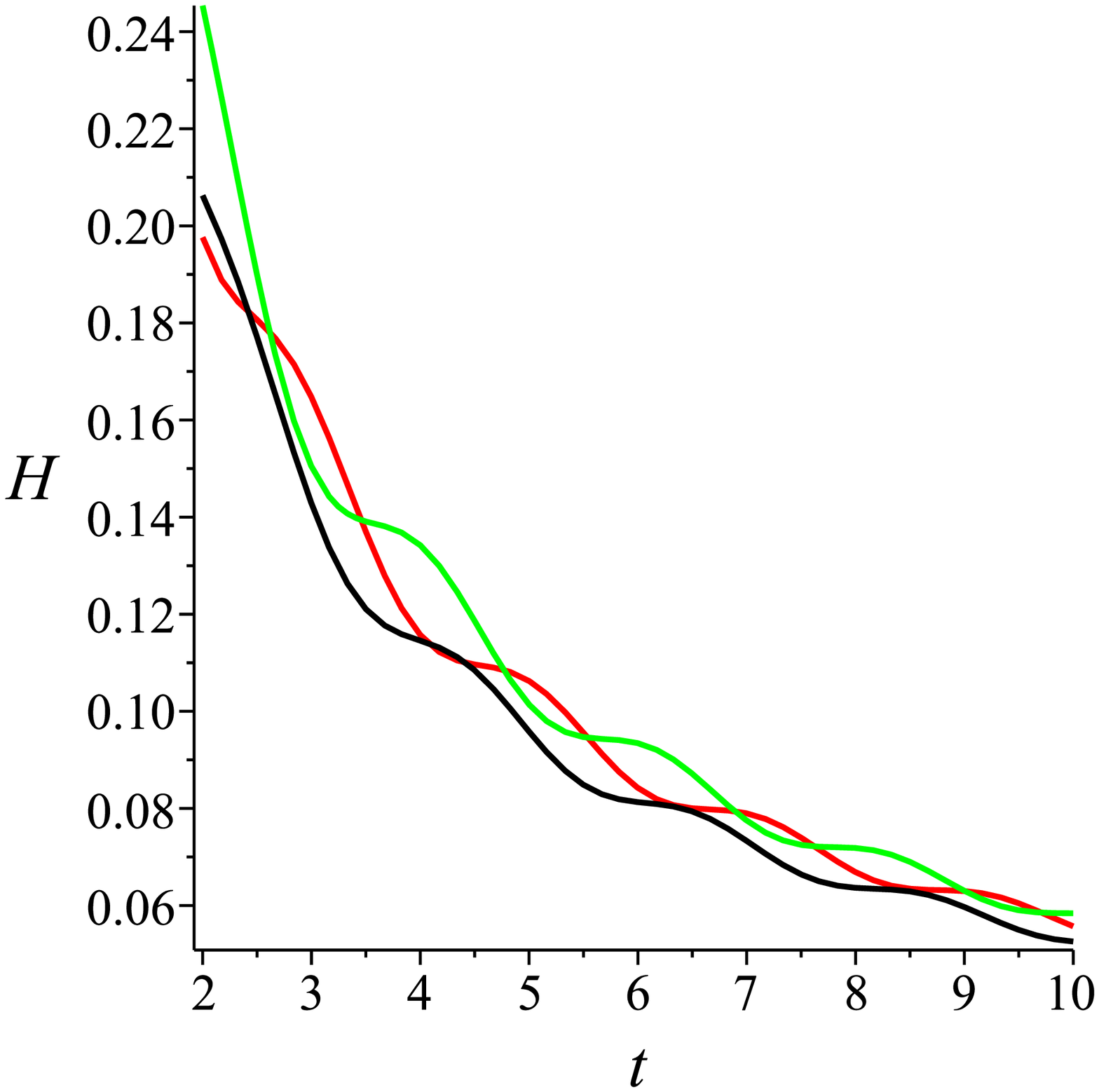}
\end{center}
\vspace{-0.6cm}
\caption{{\small The Hubble parameters $H (t)$ for the same values of the constants as in Figure \ref{phi_t_annulus}.}}
\label{H_t_annulus}
\vspace{0.1cm}
\end{figure}
Now we turn to studying numerically the $m=0$ solutions, obtained from substituting (\ref{w_D*_sol}) and (\ref{uv_sol_A_m_0}), together with (\ref{Constants_A_m0}), into (\ref{InvCoordTr_A}). On Figure \ref{phi_t_annulus} we plot $\varphi (t)$; on the left $C_1^u = const$ and $C_2^u$ changes, while on the right $C_2^u = const$ and $C_1^u$ changes. Note that in all cases $\varphi (0)$ is finite; this is not obvious, because we have started the plots at $t=0.1$ in order to make the overall features of the graphs better visible. Also, on the right side $\varphi (0) = -1.67$ for all three graphs. Notice that in all cases $\varphi (t)$ oscillates around $\varphi = 0$ with an ever decreasing amplitude. Eventually, as $t \rightarrow \infty$, the scalar $\varphi (t)$ settles at $\varphi = 0$, which is the minimum of the potential (\ref{V_annuli_m0}). This is even more clear on Figure \ref{phi_th_annulus}, where we plot the trajectories $\big( \varphi(t) , \theta(t) \big)$ for the same values of the constants as in Figure \ref{phi_t_annulus}. The plots on Figure \ref{phi_th_annulus} start at $t=0.2$\,, again for better visibility of the features of the graphs at large $t$. (They end at $t=140$.) This obscures the fact that all trajectories on the right side start at the same point. To illustrate clearly the entirety of the trajectories, it is most useful to change variables from $\varphi$ to the radial coordinate $\rho \in (\frac{1}{2} , 2)$ of the hyperbolic annulus. On Figure \ref{rh_th_annulus} we plot the trajectories $\big( \rho (t) , \theta (t) \big)$ for the same values of the constants as in Figure \ref{phi_t_annulus}. We have restricted the plot to the segment with $\theta \in [0 , \frac{\pi}{2}]$\,, in order to make the graphs better visible. Clearly, they all oscillate around $\rho = 1$ with decreasing amplitudes and, as $t \rightarrow \infty$, they settle at $\rho = 1$ and different values of $\theta$. Note that, due to (\ref{rho_phi_trans_A}), $\rho = 1$ corresponds precisely to $\varphi = 0$, which is the minimum of the scalar potential (\ref{V_annuli_m0}). It is also interesting to observe that trajectories, which start closer to $\rho = 1$, reach greater values of $\theta$ as $t \rightarrow \infty$, although trajectories starting further from $\rho = 1$ have greater amplitudes early on. 

Finally, on Figure \ref{H_t_annulus} we plot the Hubble parameters for the same trajectories as in Figures \ref{phi_t_annulus} - \ref{rh_th_annulus}. We have restricted the range of $t$ from below only to make the distinctions between the graphs, as well as their features, visible. For each curve, $H(0)$ is finite and $H(t) \rightarrow 0$ as $t \rightarrow \infty$. This is in agreement with the fact that at late times $\varphi$ settles at $\varphi = 0$ and so the potential (\ref{V_annuli_m0}) vanishes. This conclusion is similar to the one at the end of Appendix \ref{App:PuncD}. However, the present case has the rather peculiar feature that $H(t)$ exhibits a damped oscillations pattern. Thus, this class of models describes a kind of a cascading spacetime evolution. It would be interesting to explore whether, considered in a finite-time range, a transient stage of this kind (at the time of horizon exit of the largest observable CMB scales) might be helpful for explaining low multipole-moment anomalies in the CMB.

\end{document}